\documentclass[aps,prd,reprint,groupedaddress,nofootinbib,floatfix,amsmath,amssymb,longbibliography]{revtex4-2}
\pdfoutput=1

\usepackage[T1]{fontenc}
\usepackage[utf8]{inputenc}
\DeclareUnicodeCharacter{FFFC}{}
\usepackage{mathrsfs,mathtools,bm,graphicx,microtype,booktabs}
\usepackage{array}[=2016-10-06]
\usepackage{multirow,enumitem,tabularx,verbatim,etoolbox}
\usepackage[caption=false]{subfig}
\usepackage[colorlinks=true,linkcolor=blue,citecolor=blue,urlcolor=blue,hyperfootnotes=false]{hyperref}
\usepackage{orcidlink}
\graphicspath{{./}{figures/}}
\allowdisplaybreaks
\renewcommand{\thesection}{\arabic{section}}
\renewcommand{\thesubsection}{\thesection.\arabic{subsection}}
\renewcommand{\thesubsubsection}{\thesubsection.\arabic{subsubsection}}
\numberwithin{equation}{section}

\makeatletter
\def\p@subsection{}
\def\p@subsubsection{}
\makeatother
\newcommand{\dd}{\mathrm d}
\newcommand{\ii}{\mathrm i}
\newcommand{\cA}{\mathcal A}
\newcommand{\cD}{\mathcal D}
\newcommand{\cF}{\mathcal F}
\newcommand{\cG}{\mathcal G}

\newcommand{\cM}{\mathcal M}

\newcommand{\cO}{\mathcal O}

\newcommand{\cV}{\mathcal V}

\newcommand{\Tr}{\operatorname{Tr}}
\newcommand{\Res}{\operatorname*{Res}}

\newcommand{\Det}{\operatorname{Det}}

\newcommand{\hg}{\widehat g}
\newcommand{\hD}{\widehat\Delta}
\newcommand{\hR}{\widehat R}
\newcommand{\hsqrtg}{\sqrt{\widehat g}}
\newcommand{\cW}{c_{\rm W}}
\newcommand{\Kphi}{K_{\Phi}}
\newcommand{\FP}{\operatorname{FP}}

\newcommand{\BP}{\mathrm{BP}}
\newcommand{\sep}{\mathrm{sep}}
\newcommand{\loc}{\mathrm{loc}}

\newcommand{\hard}{\mathrm{hard}}

\newcommand{\anom}{\mathrm{anom}}
\newcommand{\inv}{\mathrm{inv}}
\newcommand{\intc}{\mathrm{int}}
\newcommand{\met}{\mathrm{met}}
\newcommand{\trn}{\mathrm{tr}}
\newcommand{\ct}{\mathrm{ct}}
\newcommand{\bare}{\mathrm{bare}}
\newcommand{\meas}{\mathrm{meas}}
\newcommand{\Carr}{\mathrm{Carr}}
\newcommand{\cel}{\mathrm{cel}}
\newcommand{\WZ}{\mathrm{WZ}}
\newcommand{\Poly}{\mathrm P}

\hypersetup{
 colorlinks=true,
 linkcolor=blue,
 citecolor=blue,
 urlcolor=blue,
 pdftitle={Metric-source completion and integrability in celestial gravity},
 pdfauthor={Yingnan Xu and Shuangshuang Chu},
 pdfkeywords={Scattering Amplitudes, Anomalies in Field and String Theories, Scale and Conformal Symmetries, Models of Quantum Gravity}
}

\begin{document}
\title{Metric-source completion and integrability in celestial gravity}
\author{Yingnan Xu\,\orcidlink{0009-0005-8296-0850}}
\email[Corresponding author: ]{yingnanx@mail.smu.edu}
\affiliation{Zhongtai Securities Institute for Financial Studies, Shandong University, Jinan, Shandong 250014, China}
\affiliation{Department of Physics, Southern Methodist University, Dallas, Texas 75206, U.S.A.}
\author{Shuangshuang Chu\,\orcidlink{0009-0004-4396-582X}}
\email{shuangshchu@gmail.com}
\affiliation{School of Statistics, Dongbei University of Finance and Economics, Dalian, Liaoning 116025, China}
\begin{abstract}
Opposite-helicity soft limits carry ordering information, whereas a metric generating functional has a symmetric mixed response. We compare these structures for Einstein tree amplitudes, retaining the full momentum-conservation distribution. Normalized transport of primary sources fixes the metric response, and a common Banerjee--Pasterski prescription fixes the two ordered scattering currents on compact regular hard packets. At the reference metric, both ordered scattering responses agree with the transport Hessian, so the two matching defects vanish separately. A four-hard-particle example displays the same nonzero recoil coefficient in both orders and in the transport Hessian; an overlapping-source packet resolves the lower-order spin-frame contact. Six-graviton factorization independently tests the recoil and undifferentiated coefficients of the ordering difference. At seven points, angular integration produces a nonzero relative current whose connector variation contains a conjugate bulk term. We derive this term and its cancellation under coherent transport; global cubic scattering matching still requires the physical Ward comparison on the remaining singular overlaps. Transporting the states, kernels and inverse chains together gives a finite Ward identity and coefficientwise reconstruction for that chosen family. Regular packets leave extensions supported solely on excluded hard--hard coincidences undetermined, and the vacuum functional, independent of the hard sources, remains free.
\end{abstract}
\maketitle

\section{Introduction}\label{sec:intro}

Celestial amplitudes express scattering in a conformal basis on the sphere of null directions. Gravitational soft theorems then act as Ward identities on the transformed hard insertions \cite{Pasterski:2016qvg,Pasterski:2017kqt,Pasterski:2021rjz,Raclariu:2021zjz,Pasterski:2021raf,Weinberg:1965nx,Strominger:2013jfa,Cachazo:2014fwa,Campiglia:2014yka,Kapec:2014opa,Strominger:2017zoo}. Coupling these amplitudes to a celestial metric asks a further physical question: which soft responses are derivatives of one generating functional when the hard sources, helicity frames and angular transforms vary with the geometry? The distinction is consequential for opposite helicities, since consecutive soft limits carry ordering information while metric derivatives combine into a symmetric susceptibility.

The question is whether each opposite-helicity order reproduces the same specified metric response. We select normalized primary-source transport and use one Banerjee--Pasterski relative prescription for both orders. Denoting the completed responses by $H^{\rm phys}_{-\to+}$ and $H^{\rm phys}_{+\to-}$, the comparison takes the form
\[
 H^{\rm phys}_{-\to+}=H^{\trn}+\mathcal B_{-\to+},
 \qquad
 H^{\rm phys}_{+\to-}=H^{\trn}+\mathcal B_{+\to-}.
\]
Here $H^{\trn}$ is the paired mixed derivative of the selected transport. Equality of the two orders fixes only $\mathcal B_{-\to+}-\mathcal B_{+\to-}$. Matching requires the separate comparison of each defect with zero.

A short integral shows how both orders can miss the same contribution:
\[
 \frac1\pi\int_{|w|<1}\dd^2w\,
 \frac{st}{(|w|^2+st)^2}=\frac1{1+st}\longrightarrow1,
 \qquad s,t>0.
\]
At every fixed $w\ne0$, the finite coefficient in either soft energy vanishes. Angular integration instead gives a double Mellin pole $1/(\alpha\beta)$, with residue one in both orders. This example is a diagnostic of a shrinking angular region, not an Einstein contact. It explains why an antisymmetric test alone leaves the common symmetric response undetermined.

For compact regular hard packets, the reference calculation gives $\mathcal B_{-\to+}=\mathcal B_{+\to-}=0$. Section~\ref{subsec:two-order-benchmarks} exhibits a nonzero recoil comparison and an overlapping-source contact test with common normalizations. The contact--recoil relation in Eqs.~\eqref{eq:contactConormalIdentity}--\eqref{eq:contactCanonicalScalar} and the six-graviton calculation provide the direct amplitude evidence. At seven points we evaluate a nonzero relative current and derive its connector variation. The global cubic comparison remains conditional on matching the physical Ward action on the remaining singular overlaps. The finite Ward identity is a separate result for the family in which states, kernels and inverse chains are transported together. Throughout, regular packets exclude hard--hard coincidence loci and leave extensions supported solely on those loci undetermined.

A soft graviton displaces the hard momentum-conservation surface, producing normal derivatives of the scattering distribution. Compact energy profiles retain the energy action of the subleading generator. Broedel et al. derived the distributional consistency condition relating this action to the leading soft pole \cite{Broedel:2014SoftConstraints}. In this paper we use that condition to resolve the normal and undifferentiated parts of the opposite-helicity response on physical hard packets.

The simultaneous gravitational contact and the distinction between consecutive and simultaneous double-soft limits were obtained by Klose et al. \cite{Klose:2015xoa}. Scattering-equation methods give related double-soft limits in scalar theories \cite{CachazoHeYuan:2015DoubleSoft}, while diagrammatic calculations give the gravitational limit \cite{Saha:2016DoubleSoft}. The generalized Wilson-line construction of Fernandes et al. organizes external and internal emissions constrained by gauge invariance \cite{FernandesLinWhite:2026DoubleSoft}. Starting from the gravitational contact, we derive a relation to the full consecutive distribution at arbitrary hard multiplicity. A common Lorentz transformation supplies the difference between their momentum vector fields. Its action on the conservation distribution vanishes, so the contact determines the normal coefficient and the hard-amplitude action fixes the accompanying undifferentiated term. A six-graviton tree calculation evaluates both contributions in a real scattering channel.

Fixed-angle soft expansions can miss contributions from angular regions that shrink with the soft energies. We analyze the summed homogeneous contact through the simultaneous approach of the soft directions, the hard punctures and the energy-ratio endpoints. Momentum conservation removes its leading angular singularities. Uniform real-angular bounds and integrated endpoint estimates then determine the two ordered Mellin residues after smooth source smearing. The contact also has a residue at generic points of its mixed Mellin divisor, with logarithmic dependence on the ratios of soft--hard invariants. These bounds control the standard-shadow contribution of the contact as well.

Borji and Pano formulate celestial Mellin transforms by test-space duality \cite{BorjiPano:2024Distributional}, while Liu and Ma propose regular celestial amplitudes with a revised correlator dictionary \cite{LiuMa:2025Regular}. Our compact packets use an analytic normal test germ together with a smooth tangential profile. Eliminating momentum conservation gives the recoil map that supports local angular continuation while preserving the physical hard pairing. The amplitude input follows from Kawai--Lewellen--Tye (KLT) factorization and gravitational tree recursion \cite{Kawai:1985xq,BernDennenHuangKiermaier:2010,Benincasa:2007qj,Hodges:2012ym}. The maximally helicity-violating (MHV) Ward recursion of Guevara et al. reconstructs the stripped amplitude \cite{GuevaraHimwichMiller:2025MHVWard}; the normal derivatives retained here instead measure the displacement of its physical conservation surface.

The celestial stress tensor is related to the subleading soft graviton through a shadow transform \cite{Kapec:2016jld,Donnay:2018neh}. Banerjee and Pasterski modify its representative using a relative primitive, with contour and boundary data retained at generic conformal weight \cite{Banerjee:2022wht}. We compute the boundary-current conversion between this relative operation and ordinary differentiation of a single-valued logarithmic branch, including its variation when the contour moves. The conversion fixes the Ward representative on the stated polynomial sector and identifies which boundary terms must accompany metric differentiation.

Mixed shadow operator product expansions (OPEs) and charge brackets determine closely related singular data. At null infinity, the surface-charge one-form has a radiative nonintegrable part, and the charge bracket admits a field-dependent extension~\cite{Barnich:2011mi}. Distler et al. obtain the extended asymptotic charge algebra from antisymmetrized consecutive soft limits \cite{Distler:2018DoubleSoft}; Pranzetti and Salluce derive mixed shadow and meromorphic brackets \cite{PranzettiSalluce:2025,PranzettiSalluce:2026Mixed}. The shadow OPE analyses of Himwich and Pate and Liu et al. resolve transformed operator content \cite{HimwichPate:2025ShadowOPE,LiuLiuMa:2026ShadowCompletion}, while Narayanan examines the relation between bulk soft and collinear limits and boundary algebras \cite{Narayanan:2026SoftAlgebras}. Our response smeared with the metric sources also includes shadow-map and metric-frame variations, normal recoil and distributional contacts. The geometric role of ordered soft limits also appears for scalar moduli \cite{KapecLawNarayanan:2023}, although the metric sources in this paper act on the celestial geometry and operator bundles.

Beltrami generating functionals and reparametrized stress correlators provide the geometric framework \cite{AldrovandiTakhtajan:1996,Nguyen:2023VirasoroBlocks,Nguyen:2020hot}. We select the normalized primary-source transport as the metric-source completion and compare its derivatives with Einstein scattering. The quadratic calculation retains the acceleration of the chosen Beltrami path and combines the mixed soft-leg action, shadow-map variation and spin-frame contact into a symmetric Hessian. The two matching defects compare each ordered response with this Hessian; their difference alone tests integrability. We extract the required transverse coefficients and weight derivatives from a resolved boundary expansion \cite{Dang:2015ComplexPowers,Bogner:2007cr,Brunetti:1999ps,SzenesVergne:2003Residues}.

The cut metric belongs to the Carrollian source geometry at null infinity \cite{Duval:2014uoa,Hartong:2015xda,Ciambelli:2019lap,Herfray:2021qmp,Donnay:2022aba,Donnay:2022wvx}. Its promotion to a dynamical variable requires a phase space and boundary conditions \cite{Sudhakar:2023uan,Campiglia:2024pbn}. A functional independent of the hard sources remains separate vacuum data. Appendix~\ref{app:vacuumresponse} discusses this freedom and the scalar-determinant comparison based on the Polyakov anomaly \cite{Polyakov:1981rd,Osborn:1991gm,Dowker:1994Polyakov}; Appendix~\ref{app:geometric-source-ensembles} describes the boundary-source and ensemble choices.

Sections~\ref{sec:sources}--\ref{sec:uniformization} define the sources and physical packets. Sections~\ref{sec:amplitude}--\ref{sec:independentSixPoint} derive the contact--recoil relation, its angular continuation and the six-graviton response. Sections~\ref{sec:bp}--\ref{sec:nmhvContinuation} develop the metric susceptibilities, relative currents and joint soft limits used in the matching calculation of Section~\ref{sec:core-constructed-matching}. Appendices~\ref{app:sourcecompletion}--\ref{app:contactrecoil} supply source normalization, contact, recoil and Mellin details; Appendices~\ref{app:relativegeometry}--\ref{app:vacuumresponse} give the relative-current variations and vacuum derivations. The supplement contains the angular estimates, the local three-soft calculations and the finite-current construction, with the remaining cubic matching conditions stated separately.

\section{Source conventions and scattering distributions}
\label{sec:sources}

\subsection{Metric and stress normalization}
\label{subsec:cutgeometry}

We identify $g_{ab}$ with the leading cut metric at future null infinity, keeping the radiative shear as a separate field. The enlarged radiative phase space treats the sphere metric and shear as distinct data and yields the generalized Bondi--Metzner--Sachs (BMS) soft Ward identities~\cite{Campiglia:2015yka}. A dynamical cut metric additionally requires a symplectic structure and boundary conditions \cite{Sudhakar:2023uan}. Near a reference metric $\hg$, we use
\begin{equation}
 \dd s_g^2=e^{2\phi}\,\rho_{\hg}(z,\bar z)
 \frac{|\dd z+\mu\,\dd\bar z|^2}{1-|\mu|^2},
 \qquad \|\mu\|_\infty<1,
 \label{eq:beltramimetric}
\end{equation}

Here $\rho_{\hg}\dd z\dd\bar z$ is the reference metric; the denominator fixes the area density under Beltrami variations. We differentiate $s^A=(\phi,\mu,\bar\mu)$ independently and impose $\bar\mu=\mu^*$ on the Euclidean reality slice. Our connected functional $W=\log Z$ has the stress normalization \cite{Nguyen:2023VirasoroBlocks}
\begin{equation}
 \begin{aligned}
 \delta W[g,J]
 &=\int_\Sigma\dd^2x\sqrt g\,\Theta\,\delta\phi\\
 &\quad-\frac{1}{\pi}\int_\Sigma\dd^2z\,T^{\met}\,\delta\mu\\
 &\quad-\frac{1}{\pi}\int_\Sigma\dd^2z\,\bar T^{\met}\,\delta\bar\mu
 +\delta_J W.
\end{aligned}
 \label{eq:firstvariation}
\end{equation}

Thus $T^{\met}=-\pi\frac{\delta W}{\delta\mu}$ and $\bar T^{\met}=-\pi\frac{\delta W}{\delta\bar\mu}$. These are responses of a specified metric functional; Sections~\ref{sec:bp} and \ref{sec:mixedbeltrami} compare them with transformed soft insertions.

\subsection{Hard distributions and source normalization}
\label{subsec:distributionalZ}
\label{sec:hardsoft}
\label{subsec:hardnorm}

We call a coefficient with at least one hard operator insertion, and any of its metric variations, \emph{marked}. Equivalently, a marked response contains at least one derivative with respect to a hard source $J$. For compact hard sources $J\in\mathscr T$, with the celestial weights and spin bundles understood, the vacuum-normalized full moments define
\begin{equation}
 \begin{gathered}
 \begin{aligned}
 Z_{\hg}[J]
 &=1+\sum_{n\geq1}\frac{1}{n!}\\
 &\quad\left\langle \widetilde{\cA}_{n,I_1\ldots I_n},
 J^{I_1}\otimes\cdots\otimes J^{I_n}\right\rangle,
 \end{aligned}\\
 W_{\hg}[J]=\log Z_{\hg}[J].
\end{gathered}
 \label{eq:distributionalZ}
\end{equation}

The pairing uses Mellin test-space duality \cite{BorjiPano:2024Distributional}. We retain the conventional distributional amplitudes; the regular celestial dictionary of Ref.~\cite{LiuMa:2025Regular} provides a separate choice of correlators. Common scattering phases are absorbed into the coefficients, and derivatives of $W_{\hg}$ select their connected parts. A metric extension obeys $W[\hg,J]=W_{\hg}[J]$. Vacuum normalization at each metric means
\begin{equation}
 \begin{gathered}
 Z_{\hard}^{\rm norm}[g,J]
 =\frac{Z_{\hard}[g,J]}{Z_{\hard}[g,0]},\\
 W_{\hard}^{\rm norm}[g,J]=\log Z_{\hard}^{\rm norm}[g,J].
\end{gathered}
 \label{eq:hardnorm}
\end{equation}

Consequently $W_{\hard}^{\rm norm}[g,0]=0$: pure metric derivatives vanish at $J=0$, while marked metric responses remain. 

The complete functional also includes the metric dependence of the boundary source map, measure and counterterms:
\begin{equation}
 \begin{aligned}
 W_{\cel}[g,J]
 &=\lim_{\substack{\epsilon\to0\\ L\to\infty}}\Bigl\{\\
 &\quad W_{\Carr}^{\epsilon,L}
 [\cG_{\Carr}[g],F_{\epsilon,L}[g,J]]\\
 &\quad+W_{\meas}^{\epsilon,L}[g,J]\\
 &\quad+W_{\ct}^{\epsilon,L}[g,J]\Bigr\}+W_{\bare}[g].
\end{aligned}
 \label{eq:completefunctional}
\end{equation}

The source map is defined in Appendix~\ref{app:sourceframes}; Appendix~\ref{app:distributional} fixes the differentiation convention.

\subsection{Energy pairing and the modified shadow}
\label{subsec:energymellin}
\label{subsec:bpdefinition}

We retain the full momentum distribution $\mathcal F_n=\mathcal M_n\delta^{(4)}(P_n)$, where $P_n=\sum_i p_i$ and $\mathcal M_n$ is the stripped helicity amplitude. In the null-vector convention of Eq.~\eqref{eq:momentumparam}, $\omega_i$ is the positive energy parameter in $p_i=\epsilon_i\omega_iq_i$; the physical energy magnitude is $\omega_iq_i^0$. With a smooth energy regulator $f_{\epsilon,L}$, its celestial transform is
\begin{equation}
 \widetilde\cA_n^{\epsilon,L}
 =\int_0^\infty\prod_{i=1}^n
 \left[\dd\omega_i\,\omega_i^{\Delta_i-1}f_{\epsilon,L}(\omega_i)\right]
 \mathcal F_n(\omega_i,z_i,\bar z_i).
 \label{eq:regulatedamplitude}
\end{equation}

Coincident metric insertions also require an angular prescription. The hard-energy endpoint logarithm and the covariant Weyl anomaly refer to different variations; their relation is fixed by the chosen metric-source completion.

The positive-helicity subleading soft graviton is $S_+=\Res_{\Delta=0}\cO_{\Delta,+2}$, with weights $(1,-1)$. For hard primaries $X=\prod_i\cO_{h_i,\bar h_i}$, write $A_X=\langle X\rangle$ for the marked coefficient of $Z$. The operator in Eq.~\eqref{eq:Pone} acts on this full coefficient. Its standard shadow and the Banerjee--Pasterski (BP) modified shadow are \cite{Banerjee:2022wht}
\begin{equation}
 \bar T^{\rm sh}(z,\bar z)
 =\frac{3}{\pi}\int\dd^2w\,
 \frac{S_+(w,\bar w)}{(\bar z-\bar w)^4}.
 \label{eq:standardshadow}
\end{equation}
\begin{equation}
 \epsilon_+^{(z_0)}(z,\bar z)
 =\int_{z_0}^{z}\dd u\,S_+(u,\bar z),
 \qquad
 \bar T_{\rm mod}^{\BP}
 =\bar T^{\rm sh}+\frac12\bar\partial^3\epsilon_+^{(z_0)}.
 \label{eq:bpmodified}
\end{equation}

We call the open-contour term in Eq.~\eqref{eq:bpmodified} the relative correction; the modified shadow is the sum of the standard shadow and this correction. The contour, endpoints and logarithmic branches define the relative inverse. For the single Ward kernel, endpoint changes have a quadratic antiholomorphic coefficient and vanish under $\bar\partial^3$. The generic two-leg continuation retains these data. Section~\ref{subsec:transportedmixed} specifies their source transport.

\subsection{Selected metric-source completion}
\label{subsec:uniformization}

For a smooth metric near $\hg$, choose the uniformization
\begin{equation}
 g=e^{2\sigma_g}F_g^*\hg,
 \qquad
 \bar\partial F_g=\mu\,\partial F_g,
 \label{eq:uniformization}
\end{equation}
with three points fixing the residual $PSL(2,\mathbb C)$ freedom. Global conformal Ward identities remove dependence on this fixing. The primary-source transport $U_g$ includes the Weyl weight, spin frame, argument pushforward and measure through
\begin{equation}
 \int_\Sigma\dd^2x\sqrt g\,J_I\cO_g^I
 =\int_\Sigma\dd^2x\sqrt{\hg}\,(U_gJ)_I\cO_{\hg}^I.
 \label{eq:sourcecouplingequality}
\end{equation}
We select a finite-metric Ward prescription that transports the marked action, couplings and regulator by $U_g$. A measure Jacobian independent of the hard sources cancels under vacuum normalization, giving
\begin{equation}
 \frac{Z_g[J]}{Z_g[0]}
 =\frac{Z_{\hg}[U_gJ]}{Z_{\hg}[0]},
 \qquad
 W_{\trn}[g,J]=W_{\hg}^{\rm norm}[U_gJ].
 \label{eq:transportcompletion}
\end{equation}

\subsection{Matching result and domain}
\label{subsec:early-matching-statement}

With the transport in Eq.~\eqref{eq:transportcompletion} fixed, we compare its derivatives with Einstein scattering on the following domain.

We work in a finite hard tree sector with arbitrary graviton helicities and
incoming or outgoing signs in a compact domain of hard kinematics, called a \emph{regular hard chamber}. On this domain the hard energy
magnitudes, a conservation coarea minor and every proper hard invariant with
at least two hard legs on each side are bounded away from zero. Emitted soft
energies are positive, and their cutoffs preserve the hard signs. The smooth
metric sources lie in a normalized star-shaped neighborhood of $\hg$ whose
radial segments preserve this chamber and the Beltrami chart. At generic
complex soft weights, tests in directions normal to the hard momentum-conservation surface are analytic germs with
compact smooth tangential profiles, transported by
Eq.~\eqref{eq:core-transported-normal-domain}. Each specified critical
residue has finite distributional order normal to the conservation surface and extends to smooth hard
packets; source derivatives act on that extension.

We use the standard-shadow and relative prescription $\mathfrak P$ of Supplementary
Section~S1, including its loaded paths, connectors between base points,
periods, cuts and projective transitions. The extension at an
overlap uses the compatible factorization coefficient of the summed Einstein
amplitude. The common scalar finite part of $(UA)^\eta$, using the overlap variables of Supplementary Eq.~\textup{(S2.11)}, is taken at generic
soft weights before the ordered residues, using one cap and annular
partition. Source variations and changes of lift carry these data coherently.

At the reference metric, the two independently defined residue differences in
Eq.~\eqref{eq:physicalMixedResidueDefects} evaluate to
\[
 \mathcal B_{-\to+}^{\mathfrak P}=0,
 \qquad
 \mathcal B_{+\to-}^{\mathfrak P}=0.
\]
Their separate vanishing fixes the common symmetric mixed response. Section~\ref{subsec:two-order-benchmarks} evaluates two complementary projections of this statement. The $++-$ calculation also determines a nonzero seven-graviton relative current and its local connector variation. Equality of all six cubic residue orders with Eq.~\eqref{eq:core-completed-third} requires the physical Ward comparison on the singular overlaps listed in Supplementary Section~S3.9.11; this global step remains open.

The on-shell massless external states of
Eq.~\eqref{eq:core-finite-null-spinors} realize
$\mathcal F^{\rm kin}_{n,g}=U_g^{\rm hard}\mathcal F_n$ with its full
conservation distribution. Differentiating these states and the coherently
transported currents gives the completed finite Ward identity
Eq.~\eqref{eq:core-finite-marked-Ward}, together with the Weyl term in
Eq.~\eqref{eq:core-weyl-insertion}. For every admitted polarized hard packet,
its marked moment coefficient $C_{n,I}$ and, for this transported family, its completed Ward one-form $\omega_{n,I}^{\rm BP}$ obey
\[
 \omega_{n,I}^{\rm BP}=\dd_{\boldsymbol s}C_{n,I}
\]
throughout the source neighborhood.

The finite Ward response and the reference coefficient determine
$C_{n,I}$ uniquely among smooth coefficients on that connected neighborhood,
by Eq.~\eqref{eq:marked-radial-reconstruction}. Reconstructing the normalized
moments and then taking their connected logarithm yields
$W_{\trn}[g,J]$ coefficientwise. Each fixed finite collection of hard sectors
has its own common neighborhood; no convergence of the full hard-source series is
required.
These conclusions apply on the regular-packet domain. Distributions supported solely on the excluded hard--hard coincidence loci have zero pairing there, so their extensions remain undetermined. The vacuum freedom independent of the hard sources is discussed in Appendix~\ref{app:vacuumresponse}.

Table~\ref{tab:matching-proof-dependencies} locates the independent scattering comparisons and the finite-source construction from which reconstruction follows.

\begin{table*}[t]
\centering\small
\renewcommand{\arraystretch}{1.12}
\begin{tabular}{@{}>{\raggedright\arraybackslash}p{0.15\linewidth}>{\raggedright\arraybackslash}p{0.38\linewidth}>{\raggedright\arraybackslash}p{0.40\linewidth}@{}}
\toprule
Role & Main statement & Decisive argument \\
\midrule
Selection & Normalized primary-source transport and fixed relative prescription & Eq.~\eqref{eq:transportcompletion}; Supplementary Section~S1. \\
Calculation & Two mixed defects vanish separately & Supplementary Section~S2.12.9, using the preceding local and global current estimates. \\
Local calculation & Cubic factorization for general hard helicities & Supplementary Sections~S3.7--S3.8: additional inner branch and compatible division. \\
Open comparison & Global assembly of six orders and eight transform components & Supplementary Section~S3.9.11; the seven-graviton current is illustrated in Section~\ref{subsec:seven-graviton-current}. \\
Construction & Finite metric-source derivative and equivalence of coherent lifts & Section~\ref{subsec:finite-external-states}; Supplementary Sections~S4.1--S4.3. \\
Consequence & Coefficientwise reconstruction from the finite Ward identity & Section~\ref{sec:marked-finite-functional-reconstruction}; Supplementary Section~S4.4. \\
\bottomrule
\end{tabular}
\caption{Status of the matching statements. The quadratic reference comparison and local seven-graviton current use Einstein tree amplitudes. Global cubic matching requires the remaining overlap comparisons. Finite reconstruction follows from the Ward identity of the chosen transported family.}
\label{tab:matching-proof-dependencies}
\end{table*}

\section{Physical packets and recoil geometry}
\label{sec:uniformization}
\label{app:physicalAnalyticBPDomain}

Compact hard and soft packets localize the Einstein Ward response and its metric-source comparison. The recoil chart also defines the analytic packet subclass used for BP endpoints and the quadratic Beltrami response in Section~\ref{sec:bp}.

\subsection{KLT hard coefficient}
\label{subsec:klt}

In all-outgoing conventions, the four-graviton MHV amplitude is written through the KLT relation \cite{Kawai:1985xq,BernDennenHuangKiermaier:2010}
\begin{equation}
 \begin{aligned}
&\cM_4(1^-,2^-,3^+,4^+)=-\ii\left(\frac{\kappa_4}{2}\right)^2s_{12}\\
 &\quad\times A_4(1^-,2^-,3^+,4^+)A_4(1^-,2^-,4^+,3^+),
 \end{aligned}
 \label{eq:kltfour}
\end{equation}
where $A_4$ is the Parke--Taylor gluon amplitude in the indicated ordering \cite{ParkeTaylor:1986}.

Parametrize null momenta by
\begin{equation}
 \begin{gathered}
p_i^\mu=\epsilon_i\omega_iq^\mu(z_i,\bar z_i),\\
 q^\mu=(1+z\bar z,z+\bar z,-\ii(z-\bar z),1-z\bar z),
 \end{gathered}
 \label{eq:momentumparam}
\end{equation}
with reduced spinor brackets $\langle ij\rangle=\sqrt{\omega_i\omega_j}\,z_{ij}$ and $[ij]=\epsilon_i\epsilon_j\sqrt{\omega_i\omega_j}\,\bar z_{ij}$, whose constant phases are fixed throughout. We use signature $(-,+,+,+)$, and $\epsilon_i=\pm1$ implements crossing between outgoing and incoming legs within the all-outgoing amplitude. For the null vector in Eq.~\eqref{eq:momentumparam}, the physical Mandelstam invariant is $s_{ij}=-(p_i+p_j)^2=4\langle ij\rangle[ij]$. Canonically normalized spinor brackets are twice the reduced brackets, so each Parke--Taylor ratio is unchanged and the KLT momentum kernel supplies a common factor of four. Factoring out that numerical factor together with the common gravitational coupling and phase, define the energy-stripped angular coefficient $\widehat{\mathcal R}_4$ and the hard coefficient $\mathcal R_4$, including its energy dependence, by
\begin{equation}
 \widehat{\mathcal R}_4
 =\frac{z_{12}^7\bar z_{12}}
 {z_{13}z_{14}z_{23}z_{24}z_{34}^2},
 \qquad
 \mathcal R_4
 =\frac{\omega_1^3\omega_2^3}{\omega_3^2\omega_4^2}
 \widehat{\mathcal R}_4.
 \label{eq:kltHardCoefficient}
\end{equation}
The corresponding hard distribution is $\mathcal F_4=\mathcal R_4\delta^{(4)}(P)$, with $P^\mu=\sum_{i=1}^4p_i^\mu$. We retain this conservation distribution in every metric and soft variation.

The MHV Ward recursion of Guevara et al. reconstructs Hodges' stripped formula \cite{GuevaraHimwichMiller:2025MHVWard}. Here we retain the conservation distribution and evaluate its metric-source response on compact physical packets.

\subsection{A real hard channel and its normal coordinates}
\label{subsec:physicalHardData}

We use Mellin dimension $\Delta_i=1$ and helicities $(-2,-2,+2,+2)$, so $(h_i,\bar h_i)=(-\tfrac12,\tfrac32)$ on the incoming negative-helicity legs and $(\tfrac32,-\tfrac12)$ on the outgoing positive-helicity legs. The hard data are listed in Table~\ref{tab:physicalKinematics}.

\begin{table}[t]
 \centering
 \begin{tabular}{@{}c c c c@{}}
 \toprule
 leg $i$ & $\epsilon_i$ & $z_i$ & $\frac{\omega_i}{E}$ \\
 \midrule
 $1$ & $-1$ & $\frac12$ & $\frac45$ \\
 $2$ & $-1$ & $-2$ & $\frac15$ \\
 $3$ & $+1$ & $\frac{\ii}{2}$ & $\frac45$ \\
 $4$ & $+1$ & $-2\ii$ & $\frac15$ \\
 \bottomrule
 \end{tabular}
 \caption{Real $2\to2$ hard configuration used for the local metric-source and ordered mixed-soft responses. Complex conjugation fixes $\bar z_i$. The incoming and outgoing energies balance, and the vector sum in Eq.~\eqref{eq:physicalmomentumconservation} vanishes in every component. The angular locations avoid the collinear divisors in Eq.~\eqref{eq:kltHardCoefficient}, so the hard coefficient and the soft differential operators are smooth on the selected support patch.}
 \label{tab:physicalKinematics}
\end{table}

The configuration satisfies
\begin{equation}
 \sum_{i=1}^4\epsilon_i\omega_iq^\mu(z_i,\bar z_i)=0.
 \label{eq:physicalmomentumconservation}
\end{equation}
Choose separated soft points
\begin{equation}
 w_1=1+\ii,
 \quad \bar w_1=1-\ii,
 \qquad
 w_2=2+\frac{\ii}{2},
 \quad \bar w_2=2-\frac{\ii}{2}.
 \label{eq:physicalsoftpoints}
\end{equation}
For the energies in Table~\ref{tab:physicalKinematics},
\begin{equation}
 \begin{gathered}
\frac{\omega_1^3\omega_2^3}{\omega_3^2\omega_4^2}
 =\frac{4}{25}E^2,\qquad
 \left.\widehat{\mathcal R}_4\right|_*=-\frac{15625}{544},\\
 \left.\mathcal R_4\right|_*=-\frac{625}{136}E^2.
 \end{gathered}
 \label{eq:kltEnergyNormalization}
\end{equation}

Let $\xi^a$ denote real hard variables, write $z_i=x_i+\ii y_i$, and set $n^\mu=P^\mu(\xi)$. At the configuration in Table~\ref{tab:physicalKinematics}, with energies measured in units of $E$, the momentum map has rank four because
\begin{equation}
 \left.
 \det
 \frac{\partial(P^0,P^1,P^2,P^3)}
 {\partial(\omega_1,\omega_2,\omega_3,x_1)}
 \right|_*=-20E.
 \label{eq:momentummapminor}
\end{equation}
We may therefore complete $n^\mu$ to local coordinates $(n^\mu,\vartheta^\alpha)$ and write the coordinate density as $\rho(n,\vartheta)\,\dd^4n\,\dd\vartheta$, with $\rho_*>0$. Here normal means transverse to the hard momentum-conservation surface, on which the normal coordinates vanish; the remaining coordinates parametrize that surface. Normal derivatives act at fixed tangential coordinates.

\subsection{Compact physical packets and analytic recoil}
\label{subsec:analyticNormalPackets}

The differential operators of finite order act on smooth compact hard packets. Angular continuation at generic conformal weights requires a nonempty analytic subclass, constructed below from the tree amplitude on compact regions whose contours avoid factorization loci. The analytic normal test profile supplies this angular continuation in addition to the distributional energy Mellin transform of Ref.~\cite{BorjiPano:2024Distributional}.

The nonzero momentum-map minor in the real channel of Table~\ref{tab:physicalKinematics} supplies a real-analytic coarea chart $(\vartheta,n)$ with $n=P_{\rm hard}$. We restrict its closure to positive hard energies and noncollinear hard directions. Let $a(\vartheta)$ be smooth and compactly supported, let $b(n)$ be a smooth compact cutoff equal to unity in a smaller normal ball, and let $h(n)$ extend holomorphically to a complex neighborhood of that ball. The packets
\begin{equation}
 \chi_{a,b,h}(\vartheta,n)=a(\vartheta)b(n)h(n)
 \label{eq:analyticNormalPacketClass}
\end{equation}
are compact physical hard tests. The local holomorphic normal profile defines the analytic normal test germ. The choices $h=1$, $h=n^0$ and $h=(n^0)^2$ include the normal test profiles needed for the linear and quadratic distributional pairings. The tangential density and the normal cutoff remain functions of real variables throughout the construction.

Write $P_{\rm s}=\tau_s q_s+\tau_t q_t$ and $\mathcal F_6=\mathcal M_6\delta^{(4)}(P+P_{\rm s})$, following the distribution convention of Section~\ref{subsec:energymellin}. We select compact soft angular regions and soft-energy cutoffs so that $-P_{\rm s}$ lies in the plateau of $b$. Eliminating the real conservation delta function gives
\begin{equation}
 \begin{split}
 \left\langle\mathcal F_6,\chi_{a,b,h}\right\rangle
 &=\int\dd\vartheta\,a(\vartheta)\,
 \rho(\vartheta,-P_{\rm s})\\
 &\quad\mathcal M_6(\vartheta,-P_{\rm s};s,t)\,
 h(-P_{\rm s}).
\end{split}
 \label{eq:analyticCoareaPairing}
\end{equation}
Here $\rho$ is the coarea density, with a fixed sign of its Jacobian in the chosen chart, and $\mathcal M_6$ is the stripped tree amplitude. The fixed hard Mellin factors use the positive-energy branch and can be included in $\rho$; any external smooth cutoffs are chosen to be unity on the recoil image in this chart. The analytic inverse-function construction extends the recoil map to a complex normal neighborhood. The tree amplitude is rational in the spinor variables. Consequently, the integrand multiplying $a$ in Eq.~\eqref{eq:analyticCoareaPairing} is holomorphic in independent complex soft angular variables on a sufficiently small tube avoiding its factorization divisors. Uniform bounds on a smaller tube permit integration over the real compact tangential support and preserve holomorphy.

We compare soft angular charts in a common primary frame and choose a finite cover whose connected overlaps used for gluing meet an open subset of the physical real section. The coefficients obtained from the same hard packet agree there by Eq.~\eqref{eq:analyticCoareaPairing}, so holomorphic uniqueness identifies their continuations on each overlap. This cover defines continuation along every selected compact contour contained in these tubes. Contour deformations within one homotopy class preserve the continued primitive when the endpoints and logarithmic branches are held fixed. The construction specifies an analytic normal test germ of a compact physical packet and uses the real tangential density directly in Eq.~\eqref{eq:analyticCoareaPairing}.

\section{Opposite-helicity scattering and the gravitational contact}
\label{sec:amplitude}

\subsection{Angular operators and the soft-leg action}
\label{subsec:orderedmixed}

Antisymmetrized consecutive soft limits encode charge commutators~\cite{Distler:2018DoubleSoft}. We evaluate their action on the momentum-conservation distribution and relate its normal derivative to the endpoints of the simultaneous opposite-helicity contact. The independent six-graviton calculation below tests this relation. The soft points remain kernel arguments until Beltrami pairing and geometric contacts give the metric response.

Let $S_+(1)$ and $S_-(2)$ be opposite-helicity subleading soft operators. On celestial hard primaries of fixed weights, their action is represented by
\begin{equation}
 P_1
 =-\sum_i
 \frac{(\bar z_i-\bar w_1)^2\bar\partial_i
 +2\bar h_i(\bar z_i-\bar w_1)}{w_1-z_i},
 \label{eq:Pone}
\end{equation}
\begin{equation}
 M_2
 =-\sum_i
 \frac{(z_i-w_2)^2\partial_i
 +2h_i(z_i-w_2)}{\bar w_2-\bar z_i}.
 \label{eq:Mtwo}
\end{equation}
When the positive-helicity soft theorem acts on the negative-helicity soft leg, it produces
\begin{equation}
 L_{12}
 =\frac{(\bar w_2-\bar w_1)^2\bar\partial_{w_2}
 +2(\bar w_2-\bar w_1)}{w_1-w_2},
 \label{eq:Lonetwo}
\end{equation}
whereas the reverse ordering contains
\begin{equation}
 \bar L_{21}
 =\frac{(w_1-w_2)^2\partial_{w_1}
 +2(w_1-w_2)}{\bar w_2-\bar w_1}.
 \label{eq:Lbartwoone}
\end{equation}
After the hard Mellin transform with its endpoint prescription, the two consecutive Ward coefficients are
\begin{equation}
 K_{1\prec2}=(P_1-L_{12})M_2A_X,
 \qquad
 K_{2\prec1}=(M_2-\bar L_{21})P_1A_X.
 \label{eq:orderedlimits}
\end{equation}
Their difference defines the angular ordered-soft operator and its residue,
\begin{equation}
 \begin{gathered}
\Omega^{\rm raw}_{+-}(1,2)A_X
 =K_{1\prec2}-K_{2\prec1}\\
 =\bigl([P_1,M_2]-L_{12}M_2+\bar L_{21}P_1\bigr)A_X.
 \end{gathered}
 \label{eq:mixedcurvature}
\end{equation}
Eq.~\eqref{eq:mixedcurvature} includes each soft theorem acting on the other soft leg, as in mixed-helicity descendant and celestial charge-bracket analyses \cite{Banerjee:2022wht,PranzettiSalluce:2025,PranzettiSalluce:2026Mixed}. It is an ordered-soft observable before the metric-source maps. The fixed weights include hard-energy integration; Section~\ref{subsec:sixpointresidue} restores the compact-packet energy derivatives. The label $1\prec2$ means extracting leg $1$ and then leg $2$, whose residue map is consequently on the left in ordinary composition. The BP Mellin-cycle nesting is specified below.

\subsection{Consecutive-soft residue of the six-graviton amplitude}
\label{subsec:sixpointresidue}

Let $s^+$ and $t^-$ be two additional gravitons with soft-energy parameters $\tau_s$ and $\tau_t$. The full momentum distribution $\mathcal F_n$ is defined in Section~\ref{subsec:energymellin}, with its hard helicity section included. In the following soft equations, the gravitational couplings, amplitude phases and common hard normalization are suppressed as in Eq.~\eqref{eq:kltHardCoefficient}. At fixed nonzero $\tau_t$, the finite subleading residue subtracts the complete leading distribution,
\begin{equation}
 \mathsf R_s^{(1)}\mathcal F_6
 :=\FP_{\tau_s\to0}\left[
 \mathcal F_6-\tau_s^{-1}S_s^{(0),\mathrm{hard}+t}
 \mathcal F_5(\tau_t)\right].
 \label{eq:subleadingFinitePart}
\end{equation}
The leading factor includes the $t$ leg, and $\mathcal F_5(\tau_t)$ retains its conservation distribution. 

The finite parts in Eq.~\eqref{eq:subleadingFinitePart} are physical
one-sided limits. At fixed nonzero energy of the other soft leg, if a paired
density has the Laurent expansion
$f(\tau)=a_{-1}/\tau+a_0+a_1\tau+\cdots$, then
\begin{equation}
 \FP_{\tau\to0^+}f
 =\lim_{\tau\to0^+}\partial_\tau[\tau f(\tau)]=a_0.
 \label{eq:positiveEnergyFinitePart}
\end{equation}
Successive application of the tree-level single-soft theorem, with the outer finite part taken after the inner coefficient, gives~\cite{Klose:2015xoa,Banerjee:2022wht,Distler:2018DoubleSoft}:
\begin{equation}
 \begin{gathered}
\left[\mathsf R_t^{(1)}\bigl(\mathsf R_s^{(1)}\mathcal F_6\bigr)
 -\mathsf R_s^{(1)}\bigl(\mathsf R_t^{(1)}\mathcal F_6\bigr)\right]
 =\Omega^{\mathrm{full}}_{+-}(s,t)\mathcal F_4,\\
 \Omega^{\mathrm{full}}_{+-}
 =[\mathsf D_+,\mathsf D_-]-L_{12}\mathsf D_-
 +\bar L_{21}\mathsf D_+.
 \end{gathered}
 \label{eq:sixpointCurvatureResidue}
\end{equation}
Here $w_1=s$ and $w_2=t$, and $\mathsf D_\pm$ are the energy-dependent generators derived below. In the order $s\prec t$, the $s$ residue is extracted at fixed $\tau_t$ before the $t$ residue. The angular operator $\Omega^{\rm raw}_{+-}$ is the fixed-weight Mellin representative; Eq.~\eqref{eq:fullAngularEulerCorrection} gives the energy Euler correction for energy-resolved distributions. The soft-leg actions also enter the consecutive charge calculation~\cite{Distler:2018DoubleSoft}. Here we pair the finite residue at separated soft directions with a compact hard packet. BMS charges further smear those directions with symmetry parameters; the metric comparison uses the transported double modified-shadow map in Eq.~\eqref{eq:shadowedmixedcurvature}.

Tables~\ref{tab:softOrdering} and \ref{tab:softNormalization} collect the order and normalization used in the contact comparison. With $\alpha=\Delta_+$ and $\beta=\Delta_-$, each operation acts from right to left. The scattering difference in Eq.~\eqref{eq:sixpointCurvatureResidue} is the upper row of Table~\ref{tab:softOrdering} minus the lower row. After the BP maps, these rows become $\mathcal K^{\BP}_{\mu|\bar\mu}$ and $\mathcal K^{\BP}_{\bar\mu|\mu}$, respectively, as defined in Eq.~\eqref{eq:regulatedDoubleBP}. The source-curvature convention in Eq.~\eqref{eq:shadoworderingdifference} reverses this difference.

\begin{table}[t]
 \centering
 \setlength{\tabcolsep}{4pt}
 \small
 \renewcommand{\arraystretch}{1.4}
 \begin{tabular}{@{}l c c c@{}}
 \toprule
 \shortstack{Soft\\order} & \shortstack{Finite\\operation} & \shortstack{Mellin\\composition} & \shortstack{Hard\\operator} \\
 \midrule
 $s$ then $t$ & $\mathsf R_t^{(1)}\mathsf R_s^{(1)}$ & $\Res_\beta\Res_\alpha$ & $(\mathsf D_+-L_{12})\mathsf D_-$ \\
 $t$ then $s$ & $\mathsf R_s^{(1)}\mathsf R_t^{(1)}$ & $\Res_\alpha\Res_\beta$ & $(\mathsf D_--\bar L_{21})\mathsf D_+$ \\
 \bottomrule
 \end{tabular}
 \caption{Consecutive soft operations and their Mellin orders. The positive-helicity leg is $s$, and the negative-helicity leg is $t$.}
 \label{tab:softOrdering}
\end{table}

\begin{table}[t]
 \centering
 \small
 \renewcommand{\arraystretch}{1.2}
 \begin{tabularx}{\linewidth}{@{}>{\raggedright\arraybackslash}p{0.23\linewidth}X@{}}
 \toprule
 Convention & Normalization in the comparison \\
 \midrule
 External states & Signed momenta are defined in Eq.~\eqref{eq:momentumparam}, with the accompanying reduced-bracket and crossing conventions. \\
 Hard amplitude & Couplings and the common scattering phase are suppressed. The four-graviton normalization is fixed by Eq.~\eqref{eq:kltEnergyNormalization}. \\
 Simultaneous contact & The minus sign and reduced-bracket factor in Eq.~\eqref{eq:physicalSoftContact} use the same convention as the consecutive generators. \\
 Normal coefficient & Eq.~\eqref{eq:fullNormalCoefficientValue} gives the coefficient of $\partial_{n^0}\delta^{(4)}(n)$. Pairing with $n^0$ contributes a minus sign. \\
 Six-graviton density & Eq.~\eqref{eq:NMHVNormalDensity} includes the recoil density and is divided by $\rho_*$. Its energy scale is $E=1$. \\
 \bottomrule
 \end{tabularx}
 \caption{Common normalization of the contact, consecutive distribution and six-graviton response density.}
 \label{tab:softNormalization}
\end{table}

The momentum-delta derivatives follow from the distributional soft consistency condition \cite{Broedel:2014SoftConstraints}. With a transverse traceless polarization $e_{\mu\nu}$ and the common coupling suppressed, the orbital generator is

\begin{equation}
 \begin{gathered}
S_q^{(0)}=\sum_i\frac{e_{\mu\nu}p_i^\mu p_i^\nu}{p_i\cdot q},\\
 S_{q,\mathrm{orb}}^{(1)}
 =\sum_i\left[\frac{e_{\mu\nu}p_i^\mu p_i^\nu}{p_i\cdot q}
 q^\rho-e_\mu{}^\rho p_i^\mu\right]\frac{\partial}{\partial p_i^\rho}.
 \end{gathered}
 \label{eq:orbitalsoftdistribution}
\end{equation}
Its vector is tangent to each hard null shell, while the spin action commutes with momentum conservation. Thus

\begin{equation}
 \begin{aligned}
[S_q^{(1)},\delta^{(4)}(P)]
 &=\bigl(S_q^{(0)}q^\rho-e_\mu{}^\rho P^\mu\bigr)
 \partial_{P^\rho}\delta^{(4)}(P)\\
 &=S_q^{(0)}q\cdot\partial_P\delta^{(4)}(P).
 \end{aligned}
 \label{eq:gravitationalfeeddownidentity}
\end{equation}
Here $P^\mu\partial_{P^\rho}\delta(P)=-\delta^\mu{}_{\rho}\delta(P)$ and $e_\mu{}^\mu=0$. This standard identity fixes the recoil normalization used below \cite{Cachazo:2014fwa,Broedel:2014SoftConstraints}.

To obtain the energy-dependent generators, choose the positive-helicity polarization vector $a^\mu=2^{-\frac12}(\bar s,1,-\ii,-\bar s)$ in Eq.~\eqref{eq:orbitalsoftdistribution}. Its orbital vector $V_i$ on a hard null momentum gives $V_i(\omega_i)=\omega_i\frac{\bar z_i-\bar s}{s-z_i}$, $V_i(z_i)=0$ and $V_i(\bar z_i)=-\frac{(\bar z_i-\bar s)^2}{s-z_i}$. With $E_i=\omega_i\partial_{\omega_i}$, restoration of the hard helicity frame adds the spin action $J_i\frac{\bar z_i-\bar s}{s-z_i}$. The conjugate operation has spin action $-J_i\frac{z_i-t}{\bar t-\bar z_i}$. These signs also follow from the spinor section: the positive-helicity action keeps the undotted spinor fixed, and restoration after an energy change $\delta\omega_i$ gives the helicity factor $1+J_i\frac{\delta\omega_i}{\omega_i}$. Keeping the dotted spinor fixed in the conjugate action gives $1-J_i\frac{\delta\omega_i}{\omega_i}$. Thus
\begin{equation}
 \begin{aligned}
 \mathsf D_+
 &=\sum_i\frac{-(\bar z_i-\bar s)^2\bar\partial_i
 +(\bar z_i-\bar s)(E_i+J_i)}{s-z_i},\\
 \mathsf D_-
 &=\sum_i\frac{-(z_i-t)^2\partial_i
 +(z_i-t)(E_i-J_i)}{\bar t-\bar z_i}.
 \end{aligned}
 \label{eq:momentumSoftGenerators}
\end{equation}
The spin terms give the celestial weights after Mellin integration, confirming the energy-derivative sign fixed by Eq.~\eqref{eq:gravitationalfeeddownidentity}.

At fixed nonzero $\tau_t$, the single-soft expansion is
\begin{equation}
 \mathcal F_6
 =\tau_s^{-1}S_s^{(0),\mathrm{hard}+t}\mathcal F_5(\tau_t)
 +\mathsf D_s^{\mathrm{hard}+t}\mathcal F_5(\tau_t)
 +O(\tau_s).
 \label{eq:fullSingleSoftExpansion}
\end{equation}
The $t$-leg block of $\mathsf D_s^{\mathrm{hard}+t}$ is $a_{st}\bar\partial_t+b_{st}(\tau_t\partial_{\tau_t}-2)$, where $a_{st}=-\frac{(\bar t-\bar s)^2}{s-t}$ and $b_{st}=\frac{\bar t-\bar s}{s-t}$. Write $\mathcal F_5=\tau_t^{-1}A_{-1}+A_0+\tau_t A_1+\cdots$, with every $A_j$ a hard distribution. This degree-zero block preserves the energy degree of each term. Its finite coefficient is $(a_{st}\bar\partial_t-2b_{st})A_0=-L_{12}A_0$, while its leading term retains a pole. The hard-leg block contributes $\mathsf D_+A_0$, and the remaining single-soft theorem gives $A_0=\mathsf D_-\mathcal F_4$. Hence the $s\prec t$ coefficient is $(\mathsf D_+-L_{12})\mathsf D_-\mathcal F_4$. Conjugation and exchange give $(\mathsf D_--\bar L_{21})\mathsf D_+\mathcal F_4$, which establishes Eq.~\eqref{eq:sixpointCurvatureResidue}.

Terms suppressed at fixed remaining soft energy do not enter this coefficient; ratio-dependent simultaneous terms are treated below.

For comparison with Eq.~\eqref{eq:Pone} and Eq.~\eqref{eq:Mtwo}, define $b_i=\frac{\bar z_i-\bar s}{s-z_i}$ and $d_i=\frac{z_i-t}{\bar t-\bar z_i}$. The generators satisfy $\mathsf D_+=P_1+\sum_i b_i(E_i+\Delta_i)$ and $\mathsf D_-=M_2+\sum_i d_i(E_i+\Delta_i)$, where $2\bar h_i=\Delta_i-J_i$ and $2h_i=\Delta_i+J_i$. Let $a_i=-\frac{(\bar z_i-\bar s)^2}{s-z_i}$ and $c_i=-\frac{(z_i-t)^2}{\bar t-\bar z_i}$ be the angular vector coefficients. Since the kernel coefficients are independent of hard energy, expansion of Eq.~\eqref{eq:sixpointCurvatureResidue} gives
\begin{equation}
 \begin{aligned}
 \Omega^{\mathrm{full}}_{+-}-\Omega^{\rm raw}_{+-}
 &=\sum_i k_i(E_i+\Delta_i),\\
 k_i&=a_i\bar\partial_i d_i-c_i\partial_i b_i
 -L_{12}d_i+\bar L_{21}b_i.
 \end{aligned}
 \label{eq:fullAngularEulerCorrection}
\end{equation}
The soft-leg operators include their multiplication terms. The hard spin weights cancel from $k_i$, and second derivatives cancel in the commutators, leaving first-order differential operators in both cases. The correction cancels the auxiliary Mellin dimensions in $\Omega^{\rm raw}_{+-}$ and pairs with a derivative of the compact packet profile, as evaluated in Eq.~\eqref{eq:fullAngularPacketCorrection}. We evaluate both terms on the same hard amplitude and support patch below.

\subsection{Real momentum-conserving configuration}
\label{subsec:physicalwitness}

The real channel of Section~\ref{subsec:physicalHardData} gives a nonzero single-soft normal derivative. For the positive-helicity soft direction $w_1=1+\ii$, choose the null transverse polarization vector $a^\mu=2^{-\frac12}(1-\ii,1,-\ii,-1+\ii)$ and $e_{\mu\nu}=a_\mu a_\nu$. The associated soft null vector is $q_s^\mu=(3,2,2,-1)$. The four contributions to $S_s^{(0)}$ in Eq.~\eqref{eq:orbitalsoftdistribution}, in units of $E$, are $-\frac{12+16\ii}{25}$, $\frac{4-3\ii}{25}$, $-\frac{12-16\ii}{25}$ and $\frac{4+3\ii}{25}$. Their sum and the coefficient of the time-normal derivative are
\begin{equation}
 \left.S_s^{(0)}\right|_* =-\frac{16}{25}E,
 \qquad
 \left.\mathcal R_4 S_s^{(0)}q_s^0\right|_*
 =\frac{150}{17}E^3.
 \label{eq:physicalsinglefeeddown}
\end{equation}
The suppressed phase, coupling and numerical KLT factor follow Eq.~\eqref{eq:kltHardCoefficient}. A test linear in $P^0$ detects the coefficient fixed by Eq.~\eqref{eq:gravitationalfeeddownidentity} and annihilates the undifferentiated delta, including on the external energy-cutoff plateau. The two-soft residue in Eq.~\eqref{eq:sixpointCurvatureResidue} also retains the energy Euler contribution of Eq.~\eqref{eq:fullAngularEulerCorrection}.

At the stated Mellin weights, the angular operator gives

\begin{equation}
 \begin{gathered}
\boxed{\frac{\Omega^{\rm raw}_{+-}\mathcal R_4}{\mathcal R_4}
 =\frac{114876114}{24147565}+\ii\frac{10654836}{24147565}}\\
 \simeq4.75725+0.44124\,\ii.
 \end{gathered}
 \label{eq:physicalCurvatureWitness}
\end{equation}
The full distribution adds energy and recoil terms to this angular benchmark.

The hard coefficient in Eq.~\eqref{eq:kltHardCoefficient} has energy exponents $r_i=(3,3,-2,-2)$, giving $E_i\mathcal R_4=r_i\mathcal R_4$. Set $K^{\mathrm{full}}_{1\prec2}=(\mathsf D_+-L_{12})\mathsf D_-\mathcal R_4$ and $K^{\mathrm{full}}_{2\prec1}=(\mathsf D_--\bar L_{21})\mathsf D_+\mathcal R_4$. The corresponding values are
\begin{equation}
 \begin{aligned}
 \frac{K^{\mathrm{full}}_{1\prec2}}{\mathcal R_4}
 &=-\frac{190989028}{24147565}
 -\ii\frac{157818792}{7102225},\\
 \frac{K^{\mathrm{full}}_{2\prec1}}{\mathcal R_4}
 &=-\frac{392728}{226525}
 -\ii\frac{2439804}{226525},\\
 \frac{\Omega^{\mathrm{full}}_{+-}\mathcal R_4}{\mathcal R_4}
 &=-\frac{745621116+1382503932\ii}{120737825}\\
 &\simeq-6.17554-11.45046\ii.
\end{aligned}
 \label{eq:physicalFullOrderedRatios}
\end{equation}
The difference from Eq.~\eqref{eq:physicalCurvatureWitness} is $\sum_i k_i(r_i+\Delta_i)$, as required by Eq.~\eqref{eq:fullAngularEulerCorrection}. These coefficient actions use the rational continuation $\mathcal R_4$ in Eq.~\eqref{eq:kltHardCoefficient}; the complete distribution $\mathcal F_4=\mathcal R_4\delta^{(4)}(P)$ and its differentiated response are invariant under smooth changes of that continuation vanishing at $P=0$. The coefficient benchmarks and the normal-derivative terms below together give Eq.~\eqref{eq:sixpointCurvatureResidue}.

\subsection{Compact-packet pairing of the full consecutive distribution}
\label{subsec:physicalwavepacket}

For smooth compactly supported hard packets, define
\begin{equation}
 \cA_4[\Psi_i]
 =\int\prod_{i=1}^4\left(\dd\omega_i\dd^2z_i\,\Psi_i\right)
 \cM_4\,\delta^{(4)}\!\left(\sum_i\epsilon_i\omega_iq_i\right).
 \label{eq:wavepacketfour}
\end{equation}
Mellin packets take the form $\Psi_i=\omega_i^{\Delta_i-1}\chi_i$, where $\chi_i$ is smooth and has compact energy support. For every energy derivative, distributional integration by parts gives
\begin{equation}
 \begin{aligned}
&\int_0^\infty\dd\omega\,
 \omega^{\Delta-1}\chi(\omega)\,\omega\partial_\omega F(\omega)
 =-\int_0^\infty\dd\omega\,\omega^{\Delta-1}\\
 &\qquad\times\left[\Delta\chi(\omega)+\omega\chi'(\omega)\right]F(\omega).
 \end{aligned}
 \label{eq:mellinEnergyIntegrationByParts}
\end{equation}
Compact support removes the endpoint term, while the interior profile derivative in Eq.~\eqref{eq:mellinEnergyIntegrationByParts} remains for a general packet. Combining the $\Delta_i$ contribution with the spin part of the momentum-space Lorentz generator gives $h_i=\frac{\Delta_i+J_i}{2}$ and $\bar h_i=\frac{\Delta_i-J_i}{2}$ in Eq.~\eqref{eq:Pone} and Eq.~\eqref{eq:Mtwo}. To define the local test space, choose product neighborhoods $U=\prod_iU_i$ of the hard configuration, with compact closure, positive energies and separated angular points. Choose external energy cutoffs $f_i$ equal to unity on the energy projection of $\overline U_i$. The local tests are arbitrary sections of $\mathscr T(U)=C_c^\infty(U)$ with the hard-leg spin bundles understood. Every such section can be multiplied by $\prod_i f_i$ without change, and the derivatives of the external cutoffs have disjoint support. Eq.~\eqref{eq:mellinEnergyIntegrationByParts} retains derivatives of the local test section on this plateau. With $m_\Delta=\prod_i\omega_i^{\Delta_i-1}$ and $\chi\in\mathscr T(U)$, Eq.~\eqref{eq:fullAngularEulerCorrection} gives the compact-packet identity
\begin{equation}
 \begin{aligned}
&\left\langle(\Omega^{\mathrm{full}}_{+-}-\Omega^{\rm raw}_{+-})
 [\mathcal R_4\delta^{(4)}(P)],m_\Delta\chi\right\rangle\\
 &\quad=-\left\langle\mathcal R_4\delta^{(4)}(P),
 m_\Delta\sum_i k_iE_i\chi\right\rangle.
 \end{aligned}
 \label{eq:fullAngularPacketCorrection}
\end{equation}
The $\Delta_i$ terms cancel the derivatives of the Mellin density and integration measure. At $P=0$, an energy derivative of a test linear in $P^0$ contains $p_i^0$, so the normal witness below detects the correction. The fixed-weight Ward representative is recovered on Mellin sections with the required endpoint continuation; Eq.~\eqref{eq:fullAngularPacketCorrection} relates it to compact hard packets.

Let $\Psi\in\mathscr T(U)$ include any Mellin factors, with the normal-coordinate Jacobian retained as a separate density. Transferring the angular benchmark's hard derivatives to the test gives
\begin{equation}
 \left\langle
 \Omega^{\rm raw}_{+-}\!\left[\mathcal R_4\delta^{(4)}(P)\right],
 \Psi\right\rangle
 =\left\langle
 \mathcal R_4\delta^{(4)}(P),
 (\Omega^{\rm raw}_{+-})^{\mathsf t}\Psi\right\rangle.
 \label{eq:transposepairing}
\end{equation}
In the normal chart of Eq.~\eqref{eq:momentummapminor}, for either $\mathcal L=\Omega^{\rm raw}_{+-}$ or $\mathcal L=\Omega^{\mathrm{full}}_{+-}$, write $\mathcal L=V^a\partial_a+C$. The vector $V$ includes the energy directions when $\mathcal L=\Omega^{\mathrm{full}}_{+-}$. The distributional identity
\begin{equation}
 \begin{aligned}
&a(n,\vartheta)\,\partial_{n^\mu}\delta^{(4)}(n)
 =a(0,\vartheta)\,\partial_{n^\mu}\delta^{(4)}(n)\\
 &\qquad-\left.\partial_{n^\mu}a(n,\vartheta)\right|_{n=0}
 \delta^{(4)}(n)
 \end{aligned}
 \label{eq:normalproductidentity}
\end{equation}
puts the complete distribution into canonical normal form,
\begin{equation}
 \begin{gathered}
\begin{aligned}
 &\rho\,\mathcal L\!\left[\mathcal R_4\delta^{(4)}(P)\right]
 =A^\mu(\vartheta)\,\partial_{n^\mu}\delta^{(4)}(n)\\
 &\qquad+B(\vartheta)\delta^{(4)}(n),
 \end{aligned}\\
 A^\mu=\left.\rho\,\mathcal R_4\,V(P^\mu)\right|_{n=0}.
 \end{gathered}
 \label{eq:canonicalnormaljets}
\end{equation}
Coefficient and Jacobian derivatives enter only $B(\vartheta)$, leaving $A^\mu$ unchanged. Such a finite expansion in the conservation delta and its normal derivatives is the distributional normal jet.

Under an invertible change of normal coordinates $n'=M(\vartheta)n+O(n^2)$ at fixed tangential coordinates, $\rho'=\rho|\det M|^{-1}$ and $A'=MA$. Thus the nonvanishing of the normal coefficient is independent of the chart; nonlinear coordinate terms enter only $B$.

For the angular benchmark and kinematics used above, the vector $V_{\mathrm{ang}}$ gives
\begin{equation}
 \begin{aligned}
\left.V_{\mathrm{ang}}(P^0)\right|_*
 &=E\left(-\frac{177069492}{177555625}
 -\ii\frac{279020544}{177555625}\right)\\
 &\simeq(-0.99726-1.57145\ii)E.
 \end{aligned}
 \label{eq:normalcoefficientvalue}
\end{equation}
The energy Euler coefficients at this configuration are
\begin{equation}
 \begin{aligned}
 k_1&=-2,&
 k_2&=-\frac{250+600\ii}{169},\\
 k_3&=-\frac{352+264\ii}{125},&
 k_4&=-\frac{35378+41496\ii}{210125}.
 \end{aligned}
 \label{eq:realChannelEulerCoefficients}
\end{equation}
Since $E_iP^\mu=p_i^\mu$, their normal contribution $V_E=\sum_i k_iE_i$ and the full vector $V_{\mathrm{full}}=V_{\mathrm{ang}}+V_E$ satisfy
\begin{equation}
 \begin{aligned}
 \left.\frac{V_E(P^0)}{E}\right|_*
 &=\frac{3515058+8812536\ii}{7102225}\\
 &\simeq0.49492+1.24081\ii,\\
 \left.\frac{V_{\mathrm{full}}(P^0)}{E}\right|_*
 &=-\frac{89193042+58707144\ii}{177555625}\\
 &\simeq-0.50234-0.33064\ii,\\
 \frac{A^0_{\mathrm{full}}(\vartheta_*)}{\rho_*E^3}
 &=\frac{44596521}{19318052}
 +\ii\frac{7338393}{4829513}\ne0.
\end{aligned}
 \label{eq:fullNormalCoefficientValue}
\end{equation}
The comparable angular and energy Euler contributions have a nonzero sum. Multiplication by the hard coefficient in Eq.~\eqref{eq:kltEnergyNormalization} gives the canonical coefficient, invariant under tangential integration by parts and undifferentiated-delta terms. Its nonzero value gives a nonvanishing principal conormal symbol of the complete consecutive distribution at this real channel.

Choose $\eta\in C_c^\infty$ supported in a sufficiently small tangential neighborhood of $\vartheta_*$, with $\eta\geq0$, and choose $\chi\in C_c^\infty$ in the normal variables with $\chi=1$ near $n=0$. For a constant phase $\zeta$, set
\begin{equation}
 \Psi_{\zeta,\eta}(n,\vartheta)
 =\zeta\,n^0\chi(n)\eta(\vartheta).
 \label{eq:normaltestsection}
\end{equation}
Since $\Psi_{\zeta,\eta}(0,\vartheta)=0$, every undifferentiated-delta term, including the Jacobian contributions in $B$, has zero pairing. Applying Eq.~\eqref{eq:canonicalnormaljets} to the full consecutive operator gives
\begin{equation}
 \left\langle
 \Omega^{\mathrm{full}}_{+-}
 \!\left[\mathcal R_4\delta^{(4)}(P)\right],
 \Psi_{\zeta,\eta}
 \right\rangle
 =-\zeta\int\dd\vartheta\,
 A^0_{\mathrm{full}}(\vartheta)\eta(\vartheta)\ne0,
 \label{eq:compactnormalpairing}
\end{equation}
where the phase and support are chosen so that the real part of the integrand has fixed sign.

The hard energies satisfy $\omega_i>0$ on $\overline U_i$, so the Mellin factors are smooth and invertible there. Extend the normal witness by zero within $U$, choosing its support strictly inside the coordinate chart. The tensor-product density of $\bigotimes_i C_c^\infty(U_i)$ in $C_c^\infty(U)$ supplies finite sums of hard-leg packets converging to this witness, with supports in one compact subset of $U$. Multiplying every approximant by the fixed external cutoffs $f_i$ preserves both convergence and the plateau property. Continuity of the full distribution then implies a nonzero pairing with a finite sum, hence with at least one product packet in that sum. Together with Eq.~\eqref{eq:sixpointCurvatureResidue}, this establishes a compact Einstein-scattering witness for the strict consecutive tree-level residue on the defined local test space. The same construction applied to $V_E$ detects the nonzero correction to the angular benchmark. The BP-transformed response is determined by the additional angular continuation and metric contacts in Section~\ref{sec:mixedbeltrami}.

\subsection{A mixed Mellin divisor from the gravitational soft contact}
\label{subsec:contactMellinExample}

The simultaneous opposite-helicity soft expansion contains the gravitational contact term of Ref.~\cite{Klose:2015xoa}. In the hard channel above, this term contributes to the mixed-pole test in Eq.~\eqref{eq:BPPinchRemainder}. With independent outgoing soft momenta $\tau_s q_s$ and $\tau_t q_t$, define $A_i=-2p_i\cdot q_s$, $B_i=-2p_i\cdot q_t$ and $s_{st}=-2q_s\cdot q_t$. In the signature and reduced-bracket convention of Eq.~\eqref{eq:momentumparam}, this term is
\begin{equation}
 \begin{aligned}
 \mathcal C_{st}(\tau_s,\tau_t)
 &=\sum_i\frac{C_i}{A_i\tau_s+B_i\tau_t},\\
 C_i
 &=-\frac{16\omega_i^2
 (\bar w_1-\bar z_i)^3(w_2-z_i)^3}
 {s_{st}(w_1-z_i)(\bar w_2-\bar z_i)}.
 \end{aligned}
 \label{eq:physicalSoftContact}
\end{equation}
Generalized Wilson lines organize the double-soft expansion through external hard-particle emissions, three-graviton vertices and gauge-constrained internal emissions~\cite{FernandesLinWhite:2026DoubleSoft}. For scalar hard states, Fernandes et al. use momentum and angular-momentum conservation to relate forms of the double-soft factor. For graviton hard states, energy and helicity terms act on the full distribution. We relate the simultaneous contact~\cite{Klose:2015xoa} to both conservation-normal coefficients of the consecutive operators in the same normalization.

The common gravitational coupling and amplitude phase follow the convention for $\mathcal R_4$. The relative minus sign is fixed by the consecutive operator and the on-shell factorization calculation below. The factor $16$ converts the four net powers of reduced brackets in the contact numerator to canonical normalization. The factors of $\tau_s\tau_t$ from those brackets cancel the same factors in the soft-pair invariant, leaving the denominator displayed in Eq.~\eqref{eq:physicalSoftContact}.

Multiplication by the term linear in the soft momenta in the conservation delta function gives the degree-zero contribution $\mathcal R_4F^\mu(r)\partial_{P^\mu}\delta^{(4)}(P)$, where $\varrho=\tau_s+\tau_t$, $r=\frac{\tau_s}{\varrho}$ and
\begin{equation}
 F^\mu(r)=\sum_i C_i
 \frac{r q_s^\mu+(1-r)q_t^\mu}
 {A_i r+B_i(1-r)}.
 \label{eq:contactRatioProfile}
\end{equation}
The simultaneous soft theorem also contains products of the single-soft factors \cite{Klose:2015xoa}. At the normal order selected by a test linear in $P$, these terms and the analytic recoil of the leading amplitude have coordinate Laurent powers in the soft energies. Their two ordered finite residues commute. The stripped degree-zero amplitude multiplies an undifferentiated conservation delta and has zero pairing with this test. The denominators in Eq.~\eqref{eq:physicalSoftContact} consequently determine the antisymmetric normal derivative, while the other simultaneous terms remain part of the full degree-zero distribution.

This identification follows at arbitrary hard multiplicity from a local identity on the null cone. Write the vector part of the one-leg consecutive operator as $V_i=v_i\partial_i+\bar v_i\bar\partial_i+k_iE_i$, with
\begin{equation}
 v_i=a_i\bar\partial_i c_i-L_{12}c_i,
 \qquad
 \bar v_i=-c_i\partial_i a_i+\bar L_{21}a_i.
 \label{eq:contactNormalVectorComponents}
\end{equation}
Here $a_i,c_i$ and $k_i$ are the coefficients used in Eq.~\eqref{eq:fullAngularEulerCorrection}. Introduce the global conformal vector fields $f(z)=\frac{(z-t)^2}{s-t}$ and $\bar f(\bar z)=\frac{(\bar z-\bar s)^2}{\bar s-\bar t}$. Their orbital action is
\begin{equation}
 \begin{gathered}
\mathcal L_{s,t}=f\partial+\bar f\bar\partial
 -\frac12(f'+\bar f')E,\\
 \mathcal L_{s,t}p^\mu=\Lambda^\mu{}_{\nu}p^\nu,\qquad
 \eta\Lambda+\Lambda^{\mathsf T}\eta=0.
 \end{gathered}
 \label{eq:contactGlobalLorentz}
\end{equation}
The quadratic degree of $f$ and $\bar f$ makes $\Lambda$ independent of the hard momentum. Substitution of the soft coefficients into Eq.~\eqref{eq:contactNormalVectorComponents} gives
\begin{equation}
 V_i(p_i^\mu)
 -C_i\left(\frac{q_t^\mu}{B_i}-\frac{q_s^\mu}{A_i}\right)
 =\mathcal L_{s,t}p_i^\mu.
 \label{eq:contactLorentzRemainder}
\end{equation}
For example, dividing by $\epsilon_i\omega_i$ and using $q^0+q^3=2$, $q^1+\ii q^2=2z_i$ and $q^1-\ii q^2=2\bar z_i$ reduces the angular coefficients to $f$ and $\bar f$ and the energy coefficient to $-\frac12(f'+\bar f')$. This gives the component derivation of Eq.~\eqref{eq:contactLorentzRemainder}. The helicity multiplication terms commute with the conservation delta and therefore preserve this relation for its highest normal derivative.

Summing over the hard legs leaves the Lorentz variation $\Lambda P$, which vanishes on the momentum-conservation surface. The contact endpoints and the full consecutive distribution are thus related by
\begin{equation}
 A_{\mathrm{full}}^\mu(\vartheta)
 =\left.\rho\,\mathcal M_n\right|_{P=0}
 \left[F^\mu(0)-F^\mu(1)\right].
 \label{eq:contactConormalIdentity}
\end{equation}
For $n=4$, the hard coefficient is $\mathcal M_4=\mathcal R_4$. Eq.~\eqref{eq:contactConormalIdentity} holds for separated soft directions on any regular hard momentum chart whose support avoids the soft--hard collinear denominators, and is independent of the hard helicities. It identifies the coefficient of the highest normal derivative of the physical scattering distribution. The coefficient multiplying the undifferentiated delta also contains derivatives of the hard amplitude and of the coarea density, as in Eq.~\eqref{eq:canonicalnormaljets}.

The Lorentz remainder also controls the undifferentiated conservation distribution. Define the contact endpoint difference on the hard chart by $\mathcal J_{st}^{\mu}(P,\vartheta)=F^\mu(0)-F^\mu(1)$. Summing Eq.~\eqref{eq:contactLorentzRemainder} before imposing momentum conservation gives $V_{\mathrm{full}}(P^\mu)=\mathcal J_{st}^{\mu}+\Lambda^\mu{}_{\nu}P^\nu$.
The matrix $\Lambda$ has zero trace and is independent of the hard variables.
Consequently, $(\Lambda P)^\mu\partial_{P^\mu}\delta^{(4)}(P)=0$ as a
distribution. The full response therefore satisfies
\begin{equation}
 \begin{aligned}
&\Omega^{\mathrm{full}}_{+-}[\mathcal M_n\delta^{(4)}(P)]
 =(\Omega^{\mathrm{full}}_{+-}\mathcal M_n)\delta^{(4)}(P)\\
 &\qquad+\mathcal M_n\mathcal J_{st}^{\mu}\partial_{P^\mu}\delta^{(4)}(P).
 \end{aligned}
 \label{eq:contactFullDistribution}
\end{equation}
In the chart $n=P$, the complete lower normal coefficient is thus
\begin{equation}
 B(\vartheta)
 =\left[
 \rho\,\Omega^{\mathrm{full}}_{+-}\mathcal M_n
 -\partial_{n^\mu}(\rho\mathcal M_n\mathcal J_{st}^{\mu})
 \right]_{n=0}.
 \label{eq:contactCanonicalScalar}
\end{equation}
The derivative in Eq.~\eqref{eq:contactCanonicalScalar} is taken at
fixed tangential coordinates. It contains the hard-amplitude recoil and the
variation of the coarea density. The contact supplies the normal derivative;
the coefficient action supplies the other part of the same physical
distribution. A change of continuation
$\mathcal M_n\mapsto\mathcal M_n+P^\mu K_\mu$ changes the two terms in
Eq.~\eqref{eq:contactCanonicalScalar} by the same amount,
$[\rho \mathcal J_{st}^{\mu} K_\mu]_{P=0}$. Both canonical coefficients are therefore
independent of the continuation of the stripped amplitude.

Choose a smooth radial cutoff $\chi_{\rm s}(\varrho)$ that is unity near the origin and has compact support. The local Mellin transform of the coefficient in Eq.~\eqref{eq:contactRatioProfile} is
\begin{equation}
 \begin{gathered}
\begin{aligned}
 &\mathcal M_{\rm ct}^\mu(\alpha,\beta)
 =\int_0^\infty\dd\varrho\,\varrho^{\alpha+\beta-1}\chi_{\rm s}(\varrho)\\
 &\qquad\times\int_0^1\dd r\,r^{\alpha-1}(1-r)^{\beta-1}F^\mu(r),
 \end{aligned}\\
 (\alpha,\beta)=(\Delta_+,\Delta_-).
 \end{gathered}
 \label{eq:contactMellinIntegral}
\end{equation}
When $A_i B_i>0$, the interpolation denominators have no zero on the closed interval. Endpoint subtraction of the integral over $r$ then yields
\begin{equation}
 \mathcal M_{\rm ct}^\mu
 =\frac{F^\mu(0)}{\alpha(\alpha+\beta)}
 +\frac{F^\mu(1)}{\beta(\alpha+\beta)}
 +\frac{H^\mu(\alpha,\beta)}{\alpha+\beta}
 +\mathcal M_{\rm coord}^\mu(\alpha,\beta),
 \label{eq:contactMellinGerm}
\end{equation}
where $H^\mu$ is holomorphic near the origin and $\mathcal M_{\rm coord}^\mu$ has only coordinate poles there. The radial integral supplies the pole at $\alpha+\beta=0$. With ordinary operator composition, the order that extracts the $s$ residue and then the $t$ residue is $\Res_\beta\Res_\alpha$. Consequently,
\begin{equation}
 \left(\Res_\beta\Res_\alpha
 -\Res_\alpha\Res_\beta\right)
 \mathcal M_{\rm ct}^\mu
 =F^\mu(0)-F^\mu(1).
 \label{eq:contactMellinResidue}
\end{equation}
The holomorphic numerator over the mixed divisor and the coordinate-pole terms have vanishing ordered double-residue difference. The endpoint values therefore determine the contribution in Eq.~\eqref{eq:contactMellinResidue}.

The residue along the mixed Mellin divisor can be evaluated at generic conformal weight. Set
$\delta=\alpha+\beta$ and
$D_i^\mu=C_i(\frac{q_t^\mu}{B_i}-\frac{q_s^\mu}{A_i})$. For $A_iB_i>0$, the result is
\begin{equation}
 \underset{\delta=0}{\operatorname{Res}}
 \mathcal M_{\rm ct}^\mu(\alpha,\delta-\alpha)
 =\frac{\pi}{\sin(\pi\alpha)}
 \sum_iD_i^\mu\left(\frac{B_i}{A_i}\right)^\alpha.
 \label{eq:contactMixedDivisorResidue}
\end{equation}
To derive Eq.~\eqref{eq:contactMixedDivisorResidue}, put $v=\frac{r}{1-r}$ in
the ratio integral on $\beta=-\alpha$. Its measure becomes
$v^{\alpha-1}\dd v$, and the contribution of leg $i$ becomes
$\frac{C_iq_s^\mu}{A_i}+\frac{D_i^\mu}{1+\frac{A_i}{B_i}v}$. The constant part has zero
meromorphic restriction to this divisor, since
$\mathrm B(\alpha,-\alpha)=0$ away from integer $\alpha$.
For $0<\operatorname{Re}\alpha<1$, the remaining integral is
$\int_0^\infty\dd v\,\frac{v^{\alpha-1}}{1+cv}
=\frac{\pi c^{-\alpha}}{\sin(\pi\alpha)}$ with $c=\frac{A_i}{B_i}>0$.
Continuation gives Eq.~\eqref{eq:contactMixedDivisorResidue} on the branch
selected by the original ratio integral. The radial cutoff equals unity
near zero, so the coefficient of its simple pole is one.

Near the conformally soft point this residue has the expansion
\begin{equation}
 \begin{aligned}
&\underset{\delta=0}{\operatorname{Res}}
 \mathcal M_{\rm ct}^\mu(\alpha,\delta-\alpha)
 =\frac{F^\mu(0)-F^\mu(1)}{\alpha}\\
 &\qquad+\sum_iD_i^\mu\log\frac{B_i}{A_i}+O(\alpha).
 \end{aligned}
 \label{eq:contactMixedDivisorExpansion}
\end{equation}
The endpoint jump fixes the pole at the conformally soft intersection; the next coefficient retains the soft--hard invariant ratios. Eq.~\eqref{eq:contactMixedDivisorResidue} gives the generic-weight energy integral, whose four BP angular blocks and metric variations enter Eq.~\eqref{eq:shadowedmixedcurvature}.

\begin{figure}[t]
 \centering
 \includegraphics[width=\linewidth]{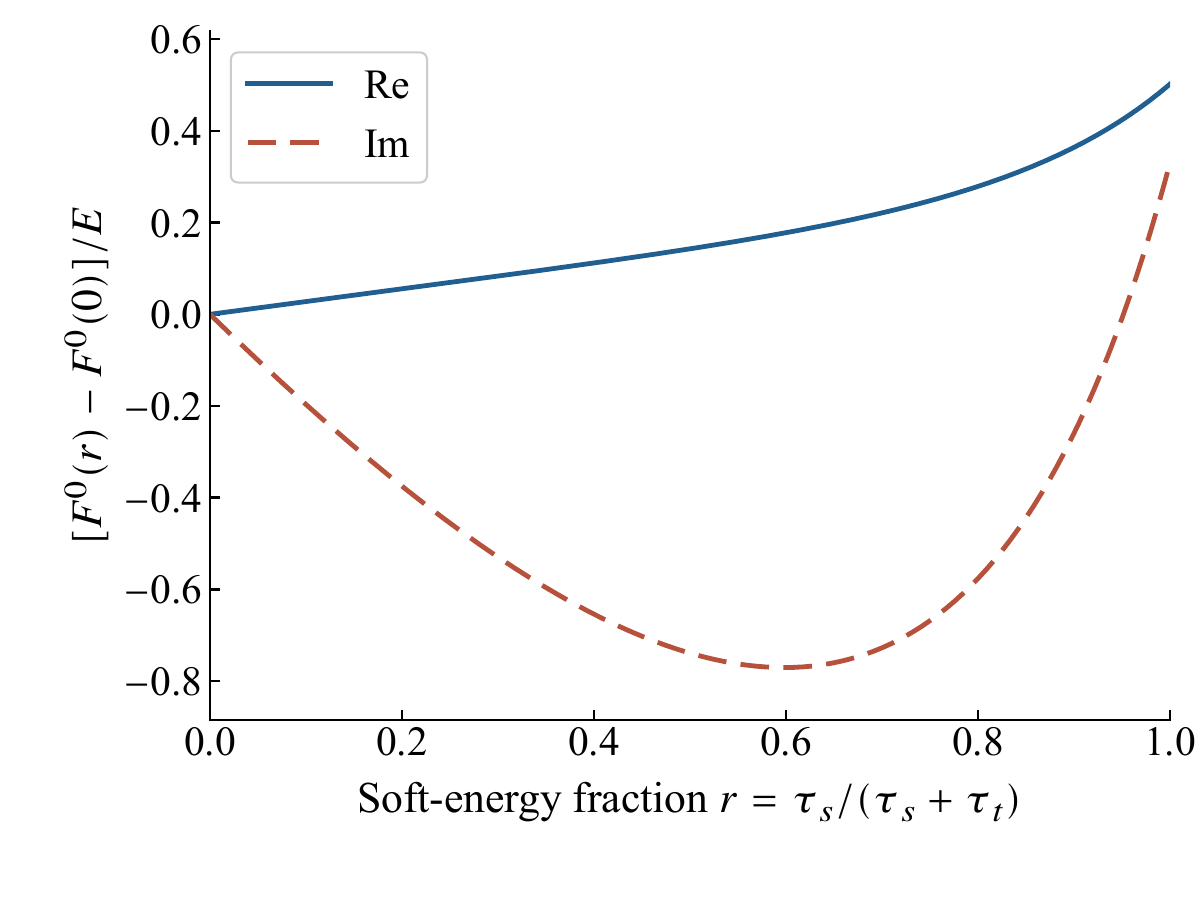}
 \caption{Real and imaginary parts of $\frac{F^0(r)-F^0(0)}{E}$ for the local gravitational contact, with hard kinematics from Table~\ref{tab:physicalKinematics} and soft directions from Eq.~\eqref{eq:physicalsoftpoints}. The parameter $r$ is the positive-helicity fraction of the sum of the two soft-energy parameters. At $r=1$, the curves give $-\frac{V_{\mathrm{full}}(P^0)}{E}$, which determines the highest normal-derivative coefficient through Eq.~\eqref{eq:contactConormalIdentity}. The curves describe the energy dependence of the scattering contact before the modified-shadow angular integration.}
 \label{fig:contactratio}
\end{figure}

For Table~\ref{tab:physicalKinematics} and Eq.~\eqref{eq:physicalsoftpoints}, $s_{st}=5$, $\frac{A_i}{E}=(-4,-8,4,8)$ and $\frac{B_i}{E}=(-8,-13,\frac{64}{5},\frac{41}{5})$. The signs obey $A_i B_i>0$ on each hard leg. Direct substitution gives
\begin{equation}
 \begin{aligned}
 \frac{F^0(0)}{E}
 &=\frac{55474902+225913464\ii}{35511125},\\
 \frac{F^0(1)}{E}
 &=\frac{687744+2229408\ii}{333125}.
\end{aligned}
 \label{eq:physicalContactEndpoints}
\end{equation}
$F^0(0)-F^0(1)$ equals $V_{\mathrm{full}}(P^0)$ in Eq.~\eqref{eq:fullNormalCoefficientValue}, and Eq.~\eqref{eq:contactConormalIdentity} recovers the coefficient measured by the compact normal test. Figure~\ref{fig:contactratio} displays the change of this contact from its $r=0$ endpoint as the soft-energy ratio varies. Its variation across the interval comes from the denominators $A_i r+B_i(1-r)$, and the value at $r=1$ equals $-\frac{V_{\mathrm{full}}(P^0)}{E}$, measuring the noncommuting nested residues. The imaginary part has a minimum near $r=0.598$, where $\operatorname{Im}[F^0(r)-F^0(0)]/E\simeq-0.771$, before reaching its positive endpoint difference. This nonmonotonic response retains the energy sharing between the two emissions, whereas the ordered residues select the endpoint values. The energy integral can also be expressed through hypergeometric functions, as given in Appendix~\ref{app:mixedsoft}.

Section~\ref{subsec:twoLegBPFamily} combines this physical mixed divisor with the four angular blocks, map variations and source contacts to determine $\mathcal R_{\rm pinch}$.

\section{Angular integration of the opposite-helicity gravitational contact}
\label{subsec:nmhvIntegratedContact}

\subsection{Contact kernel and source smearing}

Consider the homogeneous contact in Eq.~\eqref{eq:physicalSoftContact}
at conserved hard momentum, with $\tau_+=\lambda r$ and
$\tau_-=\lambda(1-r)$. The universal vector factor multiplying
$\mathcal R_4\partial_{P^\mu}\delta^{(4)}(P)$ is
$F^\mu(r)$ of Eq.~\eqref{eq:contactRatioProfile}. On every compact real
angular chart in a smooth tensor trivialization, the summed coefficient is
uniformly bounded for $0\le r\le1$, including the soft diagonal and hard
punctures. Momentum conservation cancels the apparent leading angular
singularity before this bound is taken.

Let $\varphi(u,v;\alpha,\beta)$ be a smooth compact angular
test, holomorphic in the weights with locally uniform smooth seminorms,
and include the compact tangential hard pairing on the conserved support in
$f^\mu(r;\alpha,\beta)=\langle F^\mu(r),\varphi\rangle$.
Then, uniformly for weights near zero,
\begin{equation}
 \begin{aligned}
 |f^\mu(r)-f^\mu(0)|&\le C r(1+|\log r|),\\
 |f^\mu(r)-f^\mu(1)|&\le C(1-r)(1+|\log(1-r)|).
 \end{aligned}
 \label{eq:realContactEndpointModulus}
\end{equation}
The bounds below include simultaneous approach to a hard puncture and an energy-ratio endpoint.

Set $A_j^\mu(\alpha,\beta)=f^\mu(j;\alpha,\beta)$ for $j=0,1$.
Subtracting $(1-r)A_0^\mu+rA_1^\mu$ makes the ratio remainder integrable
and holomorphic near the soft weights. The radial Mellin integral then gives
\begin{equation}
 \begin{aligned}
G_{\ct}^{\mu}(\alpha,\beta)
 &=\frac{\begin{gathered}
 A_0^\mu\mathrm B(\alpha,\beta+1)\\
 +A_1^\mu\mathrm B(\alpha+1,\beta)+H^\mu(\alpha,\beta)
 \end{gathered}}{\alpha+\beta}\\
 &\quad+G_{\mathrm{single}}^\mu,
 \end{aligned}
 \label{eq:integratedContactGerm}
\end{equation}
where $H^\mu$ is holomorphic and $G_{\mathrm{single}}^\mu$ has no
simultaneous double residue. In the ordering of
Eq.~\eqref{eq:contactMellinResidue},
\begin{equation}
 \begin{gathered}
\mathcal R_+G_{\ct}^{\mu}=A_0^\mu(0,0),\qquad
 \mathcal R_-G_{\ct}^{\mu}=A_1^\mu(0,0),\\
 \mathcal R_+=\Res_\beta\Res_\alpha,\quad
 \mathcal R_-=\Res_\alpha\Res_\beta.
 \end{gathered}
 \label{eq:integratedContactOrders}
\end{equation}
The endpoint result survives smooth angular integration. Transposing the standard shadows onto smooth source tests gives holomorphic families near zero, extending the result to the contact block with a standard shadow on each leg (the standard--standard block) at first normal-derivative order. Relative blocks, the lower normal coefficient and other homogeneous terms in $\mathcal M_6$ require separate joint estimates.

\subsection{Conservation cancellation and uniform angular bounds}
\label{app:nmhvContactBound}

Use the sign already fixed in Eq.~\eqref{eq:physicalSoftContact} and impose
hard momentum conservation. Write $u=v+d$,
$a_i=v-z_i$, $b_i=\bar v-\bar z_i$, and $w_i=\epsilon_i\omega_i$.
Direct cancellation in the contact gives
\begin{equation}
 \begin{gathered}
\lambda\mathcal C_{st}
 =-\frac1{d\bar d}\sum_iw_iH_r(a_i,b_i;d,\bar d),\\
 H_r(a,b;d,\bar d)=\frac{(b+\bar d)^3a^3}
 {(a+d)b[ab+r(a\bar d+bd+d\bar d)]}.
 \end{gathered}
 \label{eq:nmhvSummedContactKernel}
\end{equation}
The four conservation moments are
\begin{equation}
 \sum_iw_i=\sum_iw_iz_i=\sum_iw_i\bar z_i
 =\sum_iw_iz_i\bar z_i=0.
 \label{eq:nmhvContactConservationMoments}
\end{equation}
For $ab\ne0$, the Taylor polynomial through angular degree two is
\begin{equation}
 \begin{aligned}
 H_r={}&ab-(r+1)bd+(3-r)a\bar d
 +(r^2+r+1)\frac ba d^2\\
 &+(2r^2-3r-3)d\bar d
 +(r^2-3r+3)\frac ab\bar d^2+O(|d|^3).
 \end{aligned}
 \label{eq:nmhvContactTaylor}
\end{equation}
The constant, linear and displayed $d\bar d$ terms vanish after summing
over the hard legs. Away from hard punctures,
\begin{equation*}
 \begin{aligned}
 \lambda\mathcal C_{st}
 &=-\sum_iw_i\left[\begin{aligned}
 &(r^2+r+1)\frac{b_i}{a_i}\frac d{\bar d}\\
 &+(r^2-3r+3)\frac{a_i}{b_i}\frac{\bar d}d
 \end{aligned}\right]\\
 &\quad+O(|d|).
\end{aligned}
 \end{equation*}
On the real section these ratios have unit modulus. The diagonal limit
can depend on its direction; a smooth extension is not needed.

To control hard punctures as well, subtract only the universally vanishing
moments:
$E_r=H_r-ab+(r+1)bd-(3-r)a\bar d$.
For $|a|\ge2|d|$, the denominator
$r|a+d|^2+(1-r)|a|^2$ is at least $|a|^2/4$. Rescaling by $|a|$
makes the quadratic derivatives in $d$ uniformly bounded for $0\le r\le1$;
Taylor's theorem gives $|E_r|\le C|d|^2$.
For $|a|<2|d|$, use the modulus
\begin{equation}
 \begin{aligned}
 |H_r|&=\frac{|a+d|^2|a|^2}
 {r|a+d|^2+(1-r)|a|^2}\\
 &\le\max(|a+d|^2,|a|^2)\le9|d|^2.
\end{aligned}
 \label{eq:nmhvContactNearPuncture}
\end{equation}
The subtracted terms obey the same bound. Assigning arbitrary bounded
representatives at the measure-zero undefined ratios therefore yields
\begin{equation}
 \sup_{0\le r\le1}|\lambda\mathcal C_{st}|
 \le C\sum_i|w_i|.
 \label{eq:nmhvContactUniformBound}
\end{equation}
Multiplication by $rq_u^\mu+(1-r)q_v^\mu$ proves local boundedness of
$F^\mu(r)$, uniformly over a compact conserved hard support.

For the endpoint rate, in the region $|a|\geq2|d|$ the $r$ derivative of
$E_r/|d|^2$ is also uniformly bounded. In the region $|a|<2|d|$ set
$X=|a|^2/|d|^2$, $Y=|a+d|^2/|d|^2$. For $r\le1/2$,
\begin{equation*}
 \frac{|H_r-H_0|}{|d|^2}
 =\frac{rY|Y-X|}{rY+(1-r)X}
 \le C\min\left(1,\frac{r|d|^2}{|a|^2}\right).
\end{equation*}
The linear subtractions change by $O(r)$, and
\begin{equation}
 \int_{|a|\le2|d|}\dd^2a\,
 \min\left(1,\frac{r|d|^2}{|a|^2}\right)
 =\pi r|d|^2\left(1+\log\frac4r\right).
 \label{eq:nmhvContactEndpointIntegral}
\end{equation}
The real change $(u,v)\mapsto(d,a_i)$ has unit Jacobian.
Compact integration in $d$ gives the endpoint bound at $r=0$ in
Eq.~\eqref{eq:realContactEndpointModulus}; exchanging the endpoint roles
gives the bound at $r=1$. Smooth $r$ dependence in the tests adds an $O(r)$ or
$O(1-r)$ term. The argument is uniform for holomorphic smooth test families
and their weight derivatives. Physical-energy ratios are smooth,
angle-dependent reparametrizations comparable to the affine ratio at each
endpoint on a compact chart, so the same estimates apply.

For $g^\mu(r)=f^\mu(r)-(1-r)A_0^\mu-rA_1^\mu$, the endpoint integrands in
\begin{equation*}
 H^\mu(\alpha,\beta)
 =\int_0^1\dd r\,r^{\alpha-1}(1-r)^{\beta-1}g^\mu(r;\alpha,\beta)
\end{equation*}
are bounded by $r^{\operatorname{Re}\alpha}(1+|\log r|)$ and its
$r=1$ counterpart. This proves holomorphy near zero and the beta-function
decomposition in Eq.~\eqref{eq:integratedContactGerm}.
The compact radial cutoff equal to one near zero supplies
$(\alpha+\beta)^{-1}$ plus a holomorphic function. The latter multiplies
only single coordinate poles. The beta terms give the two endpoint values
in Eq.~\eqref{eq:integratedContactOrders}; $H^\mu/(\alpha+\beta)$ has
zero ordered double residue.

Finally, the positive-helicity standard-shadow kernel on a real chart has local
form $\rho^{2\alpha-4}e^{4\ii\theta}$. Against a Taylor monomial
$(x-u)^m(\bar x-\bar u)^n$, the phase selects $n=m+4$ and the radial
denominator is $2\alpha+2m+2$. There is no pole at $\alpha=0$.
Taylor subtraction also bounds each fixed output derivative in terms of
finitely many source derivatives. Thus the transpose of the standard map
takes smooth sources to a holomorphic smooth family near zero, on a common
finite chart cover with the bundle factors retained. This proves the
result stated above for the standard--standard block. A relative contour
is a different pullback operation and is not covered by this real estimate.

\section{Six-graviton scattering response}
\label{sec:independentSixPoint}
\label{subsec:independentNMHV}

The six-graviton next-to-MHV (NMHV) amplitude with helicities $(--+++-)$ independently determines the zeroth and first normal-derivative orders of the local mixed-soft ordering difference through on-shell factorization and the MHV determinant~\cite{Benincasa:2007qj,Hodges:2012ym}. It tests the contact sign in Eq.~\eqref{eq:physicalSoftContact} and the energy derivatives in Eq.~\eqref{eq:momentumSoftGenerators}.

Set $E=1$ in this section and keep $\omega_4=\frac15$, $z_2=-2$, $z_3=\frac{\ii}{2}$, $z_4=-2\ii$ and $\operatorname{Im}z_1=0$ fixed. Momentum conservation determines $x_1=\operatorname{Re}z_1$ and the remaining three hard energies in terms of the two outgoing soft energies:
\begin{equation}
 \begin{aligned}
 x_1&=\frac{2+7\tau_s+16\tau_t}{4-2\tau_s+4\tau_t},
 &\omega_1&=\frac{2(2-\tau_s+2\tau_t)^2}{10+3\tau_s+24\tau_t},\\
 \omega_2&=1-\tau_s-\omega_1,
 &\omega_3&=\frac45-2\tau_s-\tau_t.
 \end{aligned}
 \label{eq:realNMHVRecoil}
\end{equation}
The transverse component $P^2$ fixes $\omega_3$, the sum $P^0+P^3$ fixes $\omega_1+\omega_2$, and $P^1$ gives $\omega_1(2+x_1)=2-\tau_s+2\tau_t$. The remaining component then determines $x_1$ in Eq.~\eqref{eq:realNMHVRecoil}. Small positive soft energies give real momenta with positive hard energies and approach the channel in Table~\ref{tab:physicalKinematics}.

The momentum-map determinant along this slice and the corresponding density are
\begin{equation}
 \begin{gathered}
\det\frac{\partial(P^0,P^1,P^2,P^3)}
 {\partial(\omega_1,\omega_2,\omega_3,x_1)}=-4\omega_1(x_1+2)^2,\\
 \frac{\rho(\tau_s,\tau_t)}{\rho_*}
 =\frac{10}{10+3\tau_s+24\tau_t},\qquad\rho_*=\frac1{20}.
 \end{gathered}
 \label{eq:NMHVRecoilDensity}
\end{equation}
The remaining eight hard coordinates are tangential coordinates and carry the compact profile of Eq.~\eqref{eq:normaltestsection}. The density factor in Eq.~\eqref{eq:NMHVRecoilDensity} includes the recoil of the conservation surface in their fixed coordinate measure.

For the four-graviton channel, keep the tangential coordinates used in
Eq.~\eqref{eq:realNMHVRecoil} fixed. With
$y=(\omega_1,\omega_2,\omega_3,x_1)$ and
$\mathsf J^\mu{}_{a}=\frac{\partial P^\mu}{\partial y^a}$, the derivative in
Eq.~\eqref{eq:contactCanonicalScalar} is evaluated as
$(\mathsf J^{-1})^a{}_{\mu}\partial_{y^a}(\rho\mathcal R_4\mathcal J_{st}^{\mu})$.
The result is
\begin{equation}
 \begin{aligned}
\frac{B(\vartheta_*)}{\rho_*E^2}
 &=\frac{54812151321}{1642034420}
 +\ii\frac{14330612793}{410508605}\\
 &\simeq33.38063+34.90941\ii.
 \end{aligned}
 \label{eq:NMHVScalarCoefficient}
\end{equation}
This value and the four components of $A^\mu_{\mathrm{full}}$ determine the
complete local ordered difference at the reference tangential point. A
test with constant normal profile $\Psi_0=\chi(n)\eta(\vartheta)$, with $\chi=1$ near
$n=0$, has pairing $\int\dd\vartheta\,B(\vartheta)\eta(\vartheta)$.
Together with the tests linear in the normal coordinates, it resolves every normal coefficient
of this distribution. The value of $B$ refers to the specified normal and
tangential coordinates; the resulting paired distribution is independent
of that coordinate choice.

Use the reduced spinors $\lambda_i=\sqrt{\omega_i}(z_i,1)^{\mathsf T}$ and $\widetilde\lambda_i=\epsilon_i\sqrt{\omega_i}(\bar z_i,1)^{\mathsf T}$. For a negative-helicity leg $i$ and a positive-helicity leg $j$, the Britto--Cachazo--Feng--Witten (BCFW) deformation \cite{Britto:2005fw} is $\widetilde\lambda_i(\zeta)=\widetilde\lambda_i+\zeta\widetilde\lambda_j$, $\lambda_j(\zeta)=\lambda_j-\zeta\lambda_i$. Each partition with $i\in L$ and $j\notin L$ contributes the product of its two lower-point amplitudes divided by the propagator, with internal helicities summed over $\pm2$. The pole is fixed by $\det P_L(\zeta_L)=0$ for $P_L(\zeta)=P_L+\zeta\lambda_i\widetilde\lambda_j$. We use the two three-graviton amplitudes and the MHV determinant for the four- and five-graviton factors, with the common normalization fixed by Eq.~\eqref{eq:kltEnergyNormalization}. The four-graviton recursion and determinant agree with this normalization. The two five-graviton helicity sectors and three choices of the shifted pair give the same amplitudes in this convention. This construction uses on-shell factorization at every pole.

For the normal test profile $h(n)=n^0$, momentum conservation evaluates its normal argument at $n=-\tau_s q_s-\tau_tq_t$. Since $q_s^0=3$ and $q_t^0=\frac{21}{4}$, the local tangential density divided by $\rho_*$ is
\begin{equation}
 \mathfrak f(\tau_s,\tau_t)
 =-\left(3\tau_s+\frac{21}{4}\tau_t\right)
 \frac{\rho(\tau_s,\tau_t)}{\rho_*}
 \mathcal M_6(\tau_s,\tau_t).
 \label{eq:NMHVNormalDensity}
\end{equation}
The normal cutoff is constant on the recoil image near the origin. Tangential integration therefore tests the coefficient in Eq.~\eqref{eq:compactnormalpairing}.

The normal test profiles $h(n)=1$ and $h(n)=n^0$ separate the two coefficients of the full scattering distribution. The constant profile measures the coefficient of the undifferentiated delta function; the linear profile measures the normal recoil, with the sign fixed by distributional integration by parts. The amplitude calculation uses the positive-energy finite parts of Eq.~\eqref{eq:positiveEnergyFinitePart}, with the other soft energy held nonzero during each inner extraction. Table~\ref{tab:NMHVNormalCoefficients} gives the resulting local densities.

\begin{table}[t]
 \centering
 \setlength{\tabcolsep}{3pt}
 \renewcommand{\arraystretch}{1.3}
 \begin{tabular}{@{}c c c@{}}
 \toprule
 \shortstack{Normal test\\profile} & \shortstack{Distributional\\coefficient} & \shortstack{Six-graviton\\difference} \\
 \midrule
 $n^0$ & $-A^0_{\mathrm{full}}/\rho_*$ & $-2.308541-1.519489\,\ii$ \\
 $1$ & $B/\rho_*$ & $33.38063+34.90941\,\ii$ \\
 \bottomrule
 \end{tabular}
 \caption{Local mixed-helicity ordering difference at the hard configuration in Table~\ref{tab:physicalKinematics}, with $E=1$. The two profiles measure different normal orders. The values agree with Eq.~\eqref{eq:fullNormalCoefficientValue} and Eq.~\eqref{eq:NMHVScalarCoefficient}, respectively. The density is divided by $\rho_*=1/20$; integrating against the compact tangential profile gives the hard-packet response.}
 \label{tab:NMHVNormalCoefficients}
\end{table}

The extraction uses positive energies throughout. For a density $f$ with at most a simple pole in each single-soft limit, define
\begin{equation}
 \mathsf P_h f=-\frac52f(h)+8f(2h)-\frac92f(3h).
 \label{eq:positiveEnergyEstimator}
\end{equation}
For $f(\tau)=a_{-1}/\tau+a_0+a_1\tau+a_2\tau^2+\cdots$, this gives $\mathsf P_hf=a_0-11a_2h^2+O(h^3)$ at fixed nonzero remaining energy. We apply it in both variables and subtract the exchanged hierarchy, with $(h_s,h_t)=(\varepsilon^4,\varepsilon)$.

The constant-profile entry retains the action on the hard amplitude and the variation of the recoil density in Eq.~\eqref{eq:contactCanonicalScalar}. The linear-profile entry includes both the angular and energy parts of the soft generator. Their contributions to $V(P^0)/E$ are $-0.997262-1.571454\,\ii$ and $0.494923+1.240813\,\ii$, respectively, giving the nonzero sum in Eq.~\eqref{eq:fullNormalCoefficientValue}. Energy differentiation therefore changes the physical recoil coefficient substantially. The spatial components in Eq.~\eqref{eq:contactSpatialNormalCoefficients} determine the response to the other linear normal test profiles. Together, these coefficients specify the complete local ordering difference.

The simultaneous energy ratio separates the contact from products of single-soft factors. Put $\tau_s=\varrho r$, $\tau_t=\varrho(1-r)$, and denote the degree-zero coefficient of Eq.~\eqref{eq:NMHVNormalDensity} by $\mathfrak b(r)$. Its decomposition is
\begin{equation}
 \mathfrak b(r)=u_0+u_1\frac{r}{1-r}+u_2\frac{1-r}{r}
 -\left.\mathcal R_4\right|_* F^0(r).
 \label{eq:NMHVRatioDecomposition}
\end{equation}
The three coordinate Laurent functions arise from the product terms and analytic recoil. The factorization calculation fixes the remaining contribution to the gravitational contact in Eq.~\eqref{eq:physicalSoftContact}, including its relative sign. Its dependence on $r$ is displayed in Figure~\ref{fig:contactratio}. The nonzero endpoint difference is the normal component measured by the linear packet, while the invariant-ratio logarithms in Eq.~\eqref{eq:contactMixedDivisorExpansion} retain information along the mixed Mellin divisor.

\section{Quadratic metric-source responses}
\label{sec:bp}

We apply the Beltrami generating-functional construction of Refs.~\cite{AldrovandiTakhtajan:1996,Nguyen:2023VirasoroBlocks} to the physical hard distribution. The ordered Ward kernels are collected in Appendix~\ref{app:samechiral}. Smearing them with Beltrami sources determines the nonlinear coordinate response and the quadratic and linear normal-derivative coefficients evaluated below.

\subsection{Reference Ward normalization}

The selected transport is defined in Section~\ref{subsec:uniformization}. Its Hessian includes both the variation of $J_g=U_gJ$ and the nonlinear term $\langle W_{\hg,I}^{\rm norm},\delta_B\delta_AJ_g^I\rangle$, as in Eq.~\eqref{eq:fullhessian}.

The reference single-insertion action is
\begin{equation}
 \begin{aligned}
\left\langle\bar T_{\rm mod}^{\BP}(\bar z)X\right\rangle_{\sep}
 &=\bar D_{\bar z}A_X,\\
 \bar D_{\bar z}&=\sum_i\left[
 \frac{\bar h_i}{(\bar z-\bar z_i)^2}
 +\frac{\bar\partial_i}{\bar z-\bar z_i}\right].
 \end{aligned}
 \label{eq:bpsingleward}
\end{equation}
With the coupling in Eq.~\eqref{eq:firstvariation}, this agrees with the reference metric variation $\frac{\delta W}{\delta\bar\mu}=-\pi^{-1}\langle\bar T^{\met}\rangle_J$ in the tested hard-primary sector.

\subsection{Beltrami acceleration and the double Ward response}
\label{subsec:nonlinearBeltramiExample}

Along $\mu_\epsilon=\epsilon\mu_1$, $\bar\mu_\epsilon=0$, expand the normalized map as $F_\epsilon=z+\epsilon v+\epsilon^2w+O(\epsilon^3)$. The Beltrami equation gives $\bar\partial v=\mu_1$ and $\bar\partial w=\mu_1\partial v$. The primary-frame expansion gives
\begin{equation*}
 \begin{aligned}
A_{X,\trn}''(0)&=(D_v^2+D_{\zeta_v})A_X,\\
 D_v&=\sum_i(v\partial_i+h_i\partial_i v),\\
 \zeta_v&=2w-v\partial v.
 \end{aligned}
\end{equation*}
Distributional transport is defined by differentiating the transported test sections.

For two compact profiles, set $v_a=G_{\bar\partial}\mu_a$ and $v_b=G_{\bar\partial}\mu_b$, using the common normalized inverse
\begin{equation*}
 G_{\bar\partial}f(z)=\frac1\pi\int\dd^2x\,\frac{f(x)}{z-x}.
\end{equation*}
The mixed map coefficient and its acceleration are
\begin{equation}
 \begin{aligned}
 w_{ab}&=G_{\bar\partial}
 \bigl(\mu_a\partial v_b+\mu_b\partial v_a\bigr),\\
 \zeta_{ab}&=w_{ab}
 -\frac12\bigl(v_a\partial v_b+v_b\partial v_a\bigr),\\
 \bar\partial\zeta_{ab}
 &=\frac12\bigl(
 \mu_a\partial v_b+\mu_b\partial v_a
 -v_a\partial\mu_b-v_b\partial\mu_a\bigr).
 \end{aligned}
 \label{eq:polarizedBeltramiAcceleration}
\end{equation}
Smearing the product term of $G^X_{xy}=(\cV_{xy}D_y+D_xD_y)A_X$ gives $D_{v_a}D_{v_b}A_X$. Integration by parts in the stress-on-stress term gives $D_{u_{ab}}A_X$, with $u_{ab}=\zeta_{ab}-\frac12[v_a,v_b]$. The identity $[D_{v_a},D_{v_b}]=D_{[v_a,v_b]}$ then gives
\begin{equation}
 \begin{aligned}
&\frac1{\pi^2}\int\dd^2x\dd^2y\,
 \mu_a(x)\mu_b(y)G^X_{xy}\\
 &\quad=\left[\frac12\{D_{v_a},D_{v_b}\}+D_{\zeta_{ab}}\right]A_X\\
 &\quad=\left.\partial_\epsilon\partial_\eta
 A_{X,\trn}[\epsilon\mu_a+\eta\mu_b]\right|_0.
 \end{aligned}
 \label{eq:polarizedWardTransportEquality}
\end{equation}
Global conformal zero modes cancel by the hard Ward identity. The acceleration term is therefore already contained in the double Ward kernel. Its Cauchy inverses can couple separated source regions; local contacts remain the coincident or puncture-supported extensions defined in Appendix~\ref{app:conventions}.

Changing the source path to $\mu=\epsilon\mu_a+\eta\mu_b+\epsilon\eta\kappa$ preserves its linear profiles but gives
\begin{equation}
 \begin{aligned}
w_{ab}^{(\kappa)}&=w_{ab}+G_{\bar\partial}\kappa,\\
 \zeta_{ab}^{(\kappa)}&=\zeta_{ab}+G_{\bar\partial}\kappa,\\
 H_{ab}^{(\kappa)}-H_{ab}^{(0)}&=D_{G_{\bar\partial}\kappa}A_X,\\
 H_{ab}^{(0)}&=\left[\tfrac12\{D_{v_a},D_{v_b}\}+D_{\zeta_{ab}}\right]A_X.
 \end{aligned}
 \label{eq:beltramiSourcePathDependence}
\end{equation}
The choice $\kappa=-\bar\partial\zeta_{ab}$ removes the acceleration modulo the fixed global conformal zero modes. Eq.~\eqref{eq:polarizedBeltramiAcceleration} places this quadratic source on the union of the original source regions. Thus the metric path fixes the isolated nonlinear response. Its change affects the linear normal coefficient, while the highest quadratic normal coefficient remains fixed by the two linear generators.

\subsection{A quadratic same-chirality Einstein response}
\label{subsec:physicalChiralMetricExample}

Let $U$ be a compactly contained hard-coordinate neighborhood of Table~\ref{tab:physicalKinematics}, with positive energies and separated directions. Choose disjoint discs centered at $a,b$ away from its hard angular projections, and smooth unit-integral radial profiles $\rho_a,\rho_b$ supported there. For $\mu_a=\pi\rho_a$ and $\mu_b=\pi\rho_b$, the Cauchy moment identity gives
\begin{equation}
 \begin{aligned}
v_a(z)&=\frac1{z-a},\qquad v_b(z)=\frac1{z-b},\\
 w_{ab}(z)&=-\frac{v_a(z)+v_b(z)}{(a-b)^2}.
 \end{aligned}
 \label{eq:radialBeltramiProfiles}
\end{equation}
These values are independent of the radial shape and give nonzero $\zeta_{ab}$ at separated source supports.

Compact energy packets require the weight operators $\mathsf h_i=\frac{J_i-E_i}{2}$ and $\overline{\mathsf h}_i=-\frac{J_i+E_i}{2}$, where $E_i=\omega_i\partial_{\omega_i}$. Write $\mathsf D_v=\sum_i[v(z_i)\partial_i+\mathsf h_i\partial_i v(z_i)]$. The positive-helicity generator is
\begin{equation}
 \overline{\mathsf D}_v
 =\sum_i\left[
 v(\bar z_i)\bar\partial_i
 -\frac12\partial_{\bar z_i}v(\bar z_i)(E_i+J_i)
 \right].
 \label{eq:energyResolvedChiralWardGenerator}
\end{equation}
Eq.~\eqref{eq:momentumSoftGenerators} fixes its energy and spin terms. Mellin pairing differentiates both the Mellin factor and the compact energy profile, as in Eq.~\eqref{eq:mellinEnergyIntegrationByParts}.

The Einstein double-soft Ward kernel follows from the single-soft action and modified-shadow construction~\cite{Cachazo:2014fwa,Banerjee:2022wht}. On the disjoint supports above, its kernel and hard derivatives are smooth, so angular integration commutes with finite-order differentiation of the hard distribution. Let $\mathcal K^{\BP,\rm ref}_{\mu\mu}(x,y;X_4)$ denote this kernel for hard helicities $(--++)$. Its barred counterpart contains two positive-helicity soft gravitons. Eq.~\eqref{eq:polarizedWardTransportEquality}, with the energy operators retained, gives
\begin{equation}
 \begin{aligned}
&\mathfrak M^{\rm ref}_{a,b;X_4}[\Psi]:={}\frac1{\pi^2}\int\dd^2x\dd^2y\,
 \mu_a(x)\mu_b(y)\\
 &\qquad\left\langle\mathcal K^{\BP,\rm ref}_{\mu\mu}(x,y;X_4),\Psi\right\rangle\\
 &\quad-\left\langle
 \left[\frac12\{\mathsf D_{v_a},\mathsf D_{v_b}\}
 +\mathsf D_{\zeta_{ab}}\right]\mathcal F_4,\Psi
 \right\rangle=0,\\
 &\hfill\Psi\in C_c^\infty(U).
 \end{aligned}
 \label{eq:quadraticEinsteinMarkedTest}
\end{equation}
This evaluates the quadratic reference response of the selected source prescription for every $\Psi\in C_c^\infty(U)$.

Take $a=1+\ii$, $b=2+\frac{\ii}{2}$ and the barred profiles $v_a(\bar z)=(\bar z-\bar a)^{-1}$, $v_b(\bar z)=(\bar z-\bar b)^{-1}$. The vector parts of Eq.~\eqref{eq:energyResolvedChiralWardGenerator} have $\frac{V_a(P^0)}{E}=\frac{12}{25}$ and $\frac{V_b(P^0)}{E}=-\frac{1792767}{56817800}-\ii\frac{953418}{7102225}$ at the reference channel. The coefficient of $\partial_{P^0}^2\delta^{(4)}(P)$ is
\begin{equation}
 \begin{aligned}
 C^{00}_{ab}
 &=\left.\rho\mathcal R_4V_a(P^0)V_b(P^0)\right|_{P=0},\\
 \frac{C^{00}_{ab}|_*}{\rho_*E^4}
 &=\frac{5378301}{77272208}
 +\ii\frac{1430127}{4829513}
 \simeq0.069602+0.296122\ii.
 \end{aligned}
 \label{eq:chiralPhysicalQuadraticCoefficient}
\end{equation}
Choose $\chi=1$ near $P=0$ and nonnegative compact $\eta$ near the displayed tangential point. The packet $\Psi_2=\frac12\zeta(P^0)^2\chi(P)\eta(\vartheta)$ annihilates the lower normal derivatives and pairs to $\zeta\int\dd\vartheta\,C^{00}_{ab}\eta\ne0$ for a constant phase $\zeta$.

To isolate the coordinate response, subtract the complete anticommutator from the barred source smearing $\mathcal H^{\rm ref}_{ab}$:
\begin{equation}
 \mathcal R^{\rm ref}_{ab}
 :=\mathcal H^{\rm ref}_{ab}
 -\frac12\{\overline{\mathsf D}_{v_a},
 \overline{\mathsf D}_{v_b}\}\mathcal F_4
 =\overline{\mathsf D}_{\zeta_{ab}}\mathcal F_4.
 \label{eq:chiralNonlinearResponseResidual}
\end{equation}
Canonical reduction of its angular and energy action gives
\begin{equation}
 \begin{aligned}
A^0_\zeta
 &=\rho\mathcal R_4\sum_i\epsilon_i\omega_i
 \Bigl[z_i\zeta_{ab}(\bar z_i)\\
 &\qquad\left.
 -\frac{1+z_i\bar z_i}{2}\,
 \zeta'_{ab}(\bar z_i)\Bigr]\right|_{P=0},\\
 \frac{A^0_\zeta|_*}{\rho_*E^3}
 &=-\frac{147575102973}{82372173728}
 -\ii\frac{453706165357}{164744347456}\\
 &\simeq-1.791565-2.754001\ii.
 \end{aligned}
 \label{eq:chiralNonlinearNormalCoefficient}
\end{equation}
The helicity multiplication term and derivatives of the amplitude coefficient and coarea density enter the undifferentiated delta. With $\Psi_1=e^{\ii\theta_0}P^0\chi(P)\eta(\vartheta)$, they drop out and
\begin{equation}
 \left\langle\mathcal R^{\rm ref}_{ab},\Psi_1\right\rangle
 =-e^{\ii\theta_0}\int\dd\vartheta\,
 A^0_\zeta(\vartheta)\eta(\vartheta)\ne0.
 \label{eq:chiralNonlinearCompactPairing}
\end{equation}
A constant phase and sufficiently small tangential support make both displayed pairings nonzero. The tensor-product density argument of Section~\ref{subsec:physicalwavepacket} supplies product hard packets. Continuation independence follows from the full distributional reduction: changes $\Delta\mathcal R_4=P^\mu K_\mu$ cancel between coefficient and delta-derivative terms. The source-path variation in Eq.~\eqref{eq:beltramiSourcePathDependence} changes the linear normal response while preserving the highest quadratic coefficient.

\subsection{Relation to a metric stress tensor}
\label{subsec:relationbp}

For the normalized transport in Eq.~\eqref{eq:transportcompletion}, with stress tensor $T^{\trn}$, the marked reference data obey
\begin{equation}
 \begin{aligned}
&\left\langle
 \left(\bar T^{\trn}-\bar T_{\rm mod}^{\BP}\right)X
 \right\rangle_{\sep}=0,\\
 &\left\langle
 \left(\bar T^{\trn}\bar T^{\trn}
 -\bar T_{\rm mod}^{\BP}\bar T_{\rm mod}^{\BP}\right)X
 \right\rangle_{\sep}=0
 \end{aligned}
 \label{eq:restrictedequivalence}
\end{equation}
in the tested same-chirality sector. The acceleration is included in the Ward kernel and is counted once. Additional finite-metric map variations and local distributional extensions enter the matching below. A functional independent of the hard sources changes the vacuum stress Hessian and drops out of marked derivatives; Eq.~\eqref{eq:covariantcompletedstress} records the general completed stress.

\subsection{Mixed-helicity Ward response}
\label{subsec:evaluatedMixedWard}

To evaluate the reference Ward response and its contacts, use the natural Weyl--Beltrami coframe with reference conformal density absorbed into the primary trivialization. The finite coefficient transport is
\begin{equation}
 \begin{aligned}
A_{X,g}&=\prod_i e^{-\Delta_i\phi_i}
 (1-\mu_i\bar\mu_i)^{\Delta_i/2}\\
 &\qquad(\partial F_i)^{h_i}(\bar\partial\bar F_i)^{\bar h_i}
 A_X(F_i,\bar F_i).
 \end{aligned}
 \label{eq:explicitMixedPrimaryFrame}
\end{equation}
Here $\bar\partial F=\mu\partial F$ and $\partial\bar F=\bar\mu\bar\partial\bar F$ have the same fixed conformal normalization as Eq.~\eqref{eq:uniformization}. On the complexified metric-source space the two maps depend independently on $\mu$ and $\bar\mu$. This fixes the primary frame of the selected metric-source prescription.

Set $\mu=\epsilon\mu_a$, $\bar\mu=\eta\bar\mu_b$, $\bar\partial v=\mu_a$, $\partial\bar v=\bar\mu_b$, and
\begin{equation}
 \mathscr D_v=\sum_i(v_i\partial_i+h_i\partial v_i),\qquad
 \overline{\mathscr D}_{\bar v}=\sum_i(\bar v_i\bar\partial_i+\bar h_i\bar\partial\bar v_i).
 \label{eq:mixedSmearedWardGenerators}
\end{equation}
Direct differentiation gives the symmetric mixed Hessian
\begin{equation}
 \begin{aligned}
H^{\trn}_{a\bar b}[A_X]
 ={}&\mathscr D_v\overline{\mathscr D}_{\bar v}A_X\\
 &-\sum_i v_i(\bar\mu_{b,i}\bar\partial_i
       +\bar h_i\bar\partial\bar\mu_{b,i})A_X\\
 &-\frac12\sum_i\Delta_i\mu_{a,i}\bar\mu_{b,i}A_X\\
 ={}&\overline{\mathscr D}_{\bar v}\mathscr D_vA_X\\
 &-\sum_i\bar v_i(\mu_{a,i}\partial_i+h_i\partial\mu_{a,i})A_X\\
 &-\frac12\sum_i\Delta_i\mu_{a,i}\bar\mu_{b,i}A_X.
 \end{aligned}
 \label{eq:explicitMixedTransportHessian}
\end{equation}
For compact energy-resolved distributions, replace $h_i,\bar h_i$ by $(J_i-E_i)/2,(-J_i-E_i)/2$, with $E_i=\omega_i\partial_{\omega_i}$ acting on the full distribution. In particular the last term is $+\frac12\sum_i\mu_{a,i}\bar\mu_{b,i}E_i\mathcal F_n$. When both source profiles vanish near every hard puncture, only the product of Ward generators remains; there is no mixed-coordinate acceleration for these independent normalized maps.

Write $a_\pm=\langle S_\pm X\rangle$ and use $\bar D_{\bar x}$ for the antiholomorphic Ward operator at $x$. The two standard-shadow soft-leg contributions, with a common Cauchy extension, are
\begin{equation}
 \begin{aligned}
J_-(x,y)&=-\frac12\partial_y^3
       \left[\frac{a_-(y)}{\bar x-\bar y}\right],\\
 J_+(x,y)&=-\frac12\bar\partial_x^3
       \left[\frac{a_+(x)}{y-x}\right].
 \end{aligned}
 \label{eq:mixedSoftLegContactKernels}
\end{equation}
They vanish at separated soft and hard points but are nonzero distributions. Thus the reference ordered kernels are
\begin{equation}
 \begin{aligned}
 K^{\rm sh,sh}_{-\to+}&=D_y\bar D_{\bar x}A_X+J_+,\\
 K^{\rm sh,sh}_{+\to-}&=\bar D_{\bar x}D_yA_X+J_-.
 \end{aligned}
 \label{eq:evaluatedStandardMixedKernels}
\end{equation}
The upper line takes the negative-helicity residue before the positive-helicity residue, with the ordering in Eq.~\eqref{eq:regulatedDoubleBP}. Its puncture commutator is already canceled by the puncture part of $J_+-J_-$. The remaining soft-diagonal and coincident-hard-source contacts must still be retained.

Transporting both bundles in Eq.~\eqref{eq:transportedBPmap} gives, at the soft point,
\begin{equation}
 \begin{aligned}
 \delta_v\mathscr Q_+f
 &=v(x)\partial_x(\mathscr Q_+f)
   -\mathscr Q_+[(v\partial+\partial v)f],\\
 \delta_{\bar v}\mathscr Q_-f
 &=\bar v(y)\bar\partial_y(\mathscr Q_-f)
   -\mathscr Q_-[(\bar v\bar\partial+\bar\partial\bar v)f].
 \end{aligned}
 \label{eq:mixedMapIntertwinerVariation}
\end{equation}
For the relative correction this derivative includes its contours, endpoints, boundary conversion and projectors. For the standard-shadow block the Cauchy identity evaluates it locally:
\begin{equation}
 \begin{aligned}
 \delta_\mu\mathscr S_+
 &=\frac12\left[3\mu\bar\partial^2
    +3(\bar\partial\mu)\bar\partial+\bar\partial^2\mu\right],\\
 \delta_{\bar\mu}\mathscr S_-
 &=\frac12\left[3\bar\mu\partial^2
    +3(\partial\bar\mu)\partial+\partial^2\bar\mu\right].
 \end{aligned}
 \label{eq:explicitCrossShadowVariation}
\end{equation}
The input-bundle terms cancel the two soft-leg contributions. The hard commutator and output-bundle response then leave $\sum_iJ_i\mu_{a,i}\bar\mu_{b,i}A_X$. The finite primary frame in Eq.~\eqref{eq:explicitMixedPrimaryFrame} supplies the ordered local terms $-\frac12\sum_iJ_i\mu_{a,i}\bar\mu_{b,i}A_X$ and $+\frac12\sum_iJ_i\mu_{a,i}\bar\mu_{b,i}A_X$ in the negative-first and positive-first orders, respectively. Their difference cancels that remainder. Each completed reference response therefore equals Eq.~\eqref{eq:explicitMixedTransportHessian}, including its symmetric part. Appendix~\ref{app:mixedBPdomains} derives the contact terms. The source profiles below select distinct components of this reference Ward identity, and Section~\ref{subsec:physicalMixedResiduals} compares it with the generic BP family.

\subsection{Two orders on common physical packets}
\label{subsec:two-order-benchmarks}

We compare the transport prediction with each soft ordering on the real four-graviton channel of Table~\ref{tab:physicalKinematics}. All three calculations use $\mathcal F_4=\mathcal R_4\delta^{(4)}(P)$, the KLT normalization in Eq.~\eqref{eq:kltHardCoefficient}, and the same hard measure and coarea density. The angular operations use the common BP prescription: the $(UA)^\eta$ finite part is taken at generic weights before the ordered residues, with the same loaded paths and contact extension. The recoil comparison uses a local tangential density, and the overlap comparison uses an integrated packet ratio.

\paragraph{A nonzero recoil coefficient.}
Take disjoint radial sources centered at $a=1+\ii$ and $b=2+\ii/2$, away from the hard directions. Normalize their exterior vector fields to $\bar v=(\bar z-\bar a)^{-1}$ and $v=(z-b)^{-1}$. The compact normal test has germ $h(P)=(P^0)^2/2$ and a cutoff equal to one on the recoil image. At the hard conservation surface,
\begin{equation}
 \begin{aligned}
V_vP^0&=\sum_i\epsilon_i\omega_i
 \bigl[v(z_i)\bar z_i-\tfrac12v'(z_i)(1+|z_i|^2)\bigr],\\
 \frac{V_vP^0}{E}&=-\frac{1792767}{56817800}
             +\ii\frac{953418}{7102225},\\
 \frac{V_{\bar v}P^0}{E}&=\frac{12}{25},\qquad
 \frac{\mathcal R_4}{E^2}=-\frac{625}{136}.
 \end{aligned}
 \label{eq:mixed-recoil-factors}
\end{equation}
Differentiating the conservation distribution in the transport gives
$C^{00}_{b\bar a}=\rho_*\mathcal R_4(V_vP^0)(V_{\bar v}P^0)$.

The amplitude extraction gives the same coefficient independently at the highest normal order. On the recoiling surface $P=-s q_s-t q_t$, the leading Einstein soft term contributes
\begin{equation}
 \begin{aligned}
&\frac{W_s^{(0)}W_t^{(0)}\mathcal R_4}{st}\,
       \frac{(s q_s^0+t q_t^0)^2}{2}\\
 &\quad=W_s^{(0)}W_t^{(0)}\mathcal R_4
 \left[\frac{s(q_s^0)^2}{2t}+q_s^0q_t^0
       +\frac{t(q_t^0)^2}{2s}\right].
 \end{aligned}
 \label{eq:mixed-recoil-energy-extraction}
\end{equation}
Only the middle term contributes a constant in both soft energies, in either order. Terms of higher total soft degree contribute no such constant in this projection. Soft-leg and frame terms of normal order at most one annihilate the quadratic germ. The projected leading factors give $V_vP^0$ and $V_{\bar v}P^0$ in Eq.~\eqref{eq:mixed-recoil-factors}. Extending this extraction through angular integration uses the ordered remainder bounds summarized in Section~\ref{subsec:core-evaluated-matching} and proved in Supplementary Section~S2.12.9. Thus, with identical profiles and normalization,
\begin{equation}
 \frac{C^{00}_{\trn}}{\rho_*E^4}
 =\frac{C^{00}_{-\to+}}{\rho_*E^4}
 =\frac{C^{00}_{+\to-}}{\rho_*E^4}
 =\frac{5378301}{77272208}-\ii\frac{1430127}{4829513}.
 \label{eq:mixedQuadraticNormalBenchmark}
\end{equation}
These are local tangential densities; a compact tangential profile integrates them. This test resolves the highest conservation-normal derivative and leaves a term multiplying an undifferentiated delta invisible.

\paragraph{An overlapping-source contact.}
Keep the same hard channel, with $J_1=-2$ and $z_*=z_1=1/2$. Let a real smooth bump $\chi$ equal one for $|z-z_*|<1/8$ and vanish for $|z-z_*|>1/4$, with support disjoint from the other hard patches and the conformal fixing points. Choose
\begin{equation}
 v=(\bar z-\bar z_*)\chi,\qquad
 \bar v=(z-z_*)\chi.
 \label{eq:overlap-benchmark-profiles}
\end{equation}
On the leg-1 plateau, $\mu_a=\bar\partial v=1$, $\bar\mu_b=\partial\bar v=1$ and $\partial v=\bar\partial\bar v=0$. Use the $\Delta_i=1$ hard pairing $\prod_i\dd\omega_i\,\dd^2z_i$ and a compact packet $\Psi$ whose leg-1 factor is one near the conservation support. The other three legs can be localized so that this support lies inside the plateau. Conservation fixes the first momentum there, and the cutoff in $P$ can also have unit plateau. Define $A=\langle\mathcal F_4,\Psi\rangle\ne0$, including the conservation distribution and its coarea density once.

With $u=z_1-z_*$ and $E_1=\omega_1\partial_{\omega_1}$, direct differentiation of Eq.~\eqref{eq:explicitMixedPrimaryFrame} gives
\begin{equation}
 H^{\trn}[\mathcal F_4]
  =\left(|u|^2\partial_1\bar\partial_1+\tfrac12E_1\right)\mathcal F_4,
 \qquad \frac{\langle H^{\trn}[\mathcal F_4],\Psi\rangle}{A}=\frac12.
 \label{eq:overlap-transport-benchmark}
\end{equation}
Indeed, $\partial_1\bar\partial_1|u|^2=1$ and $E_1^t1=-1$ in the stated measure. For the finite maps $F=z+\epsilon v$ and $\bar F=\bar z+\eta\bar v$, on the plateau $r_g=1$, $\dd^2F=(1-\epsilon\eta)\dd^2z$ and $\omega'=(1-\epsilon\eta)^{-1/2}\omega$. Changing variables in the paired hard coefficient gives $A(1-\epsilon\eta)^{-1/2}$, whose mixed derivative is $A/2$.

The two ordered hard-generator products both pair to zero. For example,
$\mathscr D_v\overline{\mathscr D}_{\bar v}\mathcal F_4
=(|u|^2\partial_1\bar\partial_1+\bar u\bar\partial_1)\mathcal F_4$
has pairing $A-A$. The soft-leg and map terms give the overlap coefficients $-\bar h_1=(E_1+J_1)/2$ in the negative-first order and $-h_1=(E_1-J_1)/2$ in the positive-first order. The spin frame then contributes $-J_1/2$ and $+J_1/2$, respectively, as derived in Appendix~\ref{subsec:explicitMixedContacts}. Both completed operators reduce to Eq.~\eqref{eq:overlap-transport-benchmark}.

\begin{table}[t]
 \centering
 \begin{tabularx}{\linewidth}{@{}>{\raggedright\arraybackslash}Xcc@{}}
 \toprule
 Paired response divided by $A$ & $-\to+$ & $+\to-$\\
 \midrule
 Soft-leg and map terms included, before frame & $-1/2$ & $3/2$\\
 Spin-frame correction & $+1$ & $-1$\\
 Completed Ward response & $1/2$ & $1/2$\\
 Differentiated hard transport & $1/2$ & $1/2$\\
 \bottomrule
 \end{tabularx}
 \caption{Analytic overlapping-source benchmark with the same four-hard channel, packet, measure and BP prescription in both orders. The entries evaluate the reduced Ward operators and the independently differentiated transport. The completed scattering responses are identified with the Ward row through the physical residue identities of Supplementary Section~S2.12.9, which set both defects to zero.}
 \label{tab:overlap-two-orders}
\end{table}

This contact test complements the recoil projection: it retains the undifferentiated conservation distribution and resolves the opposite frame contributions before they produce the common symmetric response. Additional puncture and soft-diagonal tests appear in Appendix~\ref{subsec:additional-contact-tests}.

\section{Relative shadow corrections and metric-source matching}
\label{sec:mixedbeltrami}

\subsection{Source response and generic-weight maps}
\label{subsec:responseform}
\label{subsec:genericBPMaps}

The metric coupling defines the response one-form and its curvature by
\begin{equation}
 \begin{aligned}
\boldsymbol\omega
 &=\int_\Sigma\dd^2x\,\omega_A(x;s,J)\,\delta s^A(x),\\
 \omega_\phi&=\sqrt g\,\Theta,\qquad
 \omega_\mu=-\frac{1}{\pi}T^{\met},\\
 \omega_{\bar\mu}&=-\frac{1}{\pi}\bar T^{\met}.
 \end{aligned}
 \label{eq:responseoneform}
\end{equation}
\begin{equation}
 \cF_{AB}(x,y;s,J)
 =\frac{\delta\omega_B(y;s,J)}{\delta s^A(x)}
 -\frac{\delta\omega_A(x;s,J)}{\delta s^B(y)}.
 \label{eq:sourcecurvature}
\end{equation}
A metric Hessian requires vanishing curvature including local source contacts. We retain the soft-weight continuation before either residue.

The reference BP modified-shadow maps are
\begin{equation}
 \begin{aligned}
\mathscr Q_{+,\hg}^{\BP}\equiv\mathscr Q_+
 &=\mathscr S_++\frac12\overline{\mathcal B}_{3,\hg}G_\partial^{\BP},\\
 \mathscr Q_{-,\hg}^{\BP}\equiv\mathscr Q_-
 &=\mathscr S_-+\frac12\mathcal B_{3,\hg}G_{\bar\partial}^{\BP}.
 \end{aligned}
 \label{eq:modifiedshadowmaps}
\end{equation}
Here $\mathscr S_\pm$ are the standard shadows and $\overline{\mathcal B}_{3,\hg}$ is the cubic projective descendant, reducing to $\bar\partial^3$ in a projective chart and $\bar\eth^3$ on the round sphere. The relative inverse has a specified contour, endpoints, branches, domain and codomain, with
\begin{equation}
 \partial G_\partial^{\BP}=1-\widetilde\Pi_\partial^{\BP},
 \qquad
 G_\partial^{\BP}\partial=1-\Pi_\partial^{\BP},
 \label{eq:BPRelativeProjectors}
\end{equation}
and conjugate identities. For the continuation of Ref.~\cite{Banerjee:2022wht}, put $K(\Delta)=\frac{\Gamma(4-\Delta)}{\Gamma(1+\Delta)}$ and $r(\Delta)=\frac{\ii}{2}(1-e^{-2\pi\ii\Delta})$. Then
\begin{equation}
 \begin{aligned}
&[\mathscr Q_+^{\BP}(\Delta)\cO_{\Delta,+2}](x)\\
 &={}\frac{K(\Delta)}{2\pi}
 \left[\int_{\widehat{\mathbb C}}\dd^2u
 +r(\Delta)\int_{C_x}\dd u\int_{\bar C_{\bar x}}\dd\bar u\right]\\
 &\qquad\frac{(x-u)^\Delta}{(\bar x-\bar u)^{4-\Delta}}
 \cO_{\Delta,+2}(u,\bar u),\\[2pt]
 &[\mathscr Q_-^{\BP}(\Delta)\cO_{\Delta,-2}](y)\\
 &={}\frac{K(\Delta)}{2\pi}
 \left[\int_{\widehat{\mathbb C}}\dd^2v
 +r(\Delta)\int_{C_y}\dd v\int_{\bar C_{\bar y}}\dd\bar v\right]\\
 &\qquad\frac{(\bar y-\bar v)^\Delta}{(y-v)^{4-\Delta}}
 \cO_{\Delta,-2}(v,\bar v).
 \end{aligned}
 \label{eq:BPGenericIntertwiners}
\end{equation}
The plane term is the standard shadow, and the open-contour term is the relative correction; together they form the BP modified shadow. Contours run from the fixed base points to the transform points, with continuous logarithms, a common endpoint cutoff and exchange-covariant finite part. Winding retains the corresponding monodromy. The transform is continued from a joint convergence region with these data fixed. Near the soft point, $\frac{K(0)}{2\pi}=\frac{3}{\pi}$ and $\frac{K(\Delta)r(\Delta)}{2\pi}=-3\Delta+O(\Delta^2)$, so the relative residue depends on the continued endpoint singularity.

\subsection{Physical MHV continuation and endpoint residues}
\label{subsec:MHVEnergyCorner}
\label{subsec:BPEndpointResidues}

For hard helicities $(--++)$ and two positive-helicity soft legs, use the
six-point field-theory KLT representation~\cite{Kawai:1985xq}, equivalently
the MHV reduced determinant~\cite{Hodges:2012ym}, in the normalization
fixed by the four-point amplitude in Section~\ref{subsec:klt}.
These six-point representations give rational angular coefficients. A six-point Parke--Taylor factor carries $\frac{\omega_1\omega_2}{\omega_3\omega_4\tau_s\tau_t}$. With hard anchors $1,3,4$, the momentum kernel contains $\tau_s\tau_t$; after removing them, its remaining soft dependence has degree at most two. Thus
\begin{equation}
 \mathcal M_6^{\rm MHV}
 =\frac{1}{\tau_s\tau_t}
 \sum_{\substack{m,n\geq0\\m+n\leq2}}
 C_{mn}(X;u,v)\tau_s^m\tau_t^n.
 \label{eq:MHVSeparateEnergyDivisors}
\end{equation}
On the compact unpinched angular tubes, analytic recoil and the analytic normal test profile replace the polynomial by a convergent Taylor series. Independent energy cutoffs equal to one near zero give local Mellin terms
\begin{equation}
 \frac{\Lambda_s^{\Delta_s+m-1}
       \Lambda_t^{\Delta_t+n-1}}
 {(\Delta_s+m-1)(\Delta_t+n-1)}.
 \label{eq:MHVLocalMellinDivisors}
\end{equation}
Cutoff transitions add coordinate poles or holomorphic terms. The two residues therefore commute. For $h(n)=\frac12(n^0)^2$ in Eq.~\eqref{eq:analyticNormalPacketClass}, the mixed term of $h(-P_{\rm s})$ selects
\begin{equation}
 \begin{split}
 &\Res_{\Delta_s=0}\Res_{\Delta_t=0}
 \mathcal A_6^{++}[\chi_{a,b,h}]\\
 &\qquad=
 \int\dd\vartheta\,a(\vartheta)\rho(\vartheta,0)
 \mathcal R_4 S_s^{(0)}S_t^{(0)}q_s^0q_t^0.
 \end{split}
 \label{eq:MHVNormalSquareResidue}
\end{equation}
At the reference channel, the polarization $a^\mu(w)=2^{-\frac12}(\bar w,1,-\ii,-\bar w)$ gives $\mathcal R_4S_s^{(0)}S_t^{(0)}q_s^0q_t^0=\frac{32634-140238\ii}{9061}E^4$, so compact support and a phase yield a nonzero pairing. On these compact unpinched angular tubes, with compact energy cutoffs, $3<\Re\Delta_s,\Re\Delta_t<4$ is a sufficient common convergence strip for the four localized blocks. The relative endpoint requires $\Re\Delta>3$ and the soft energy pole requires $\Re\Delta>1$. This local strip does not control hard punctures, the soft diagonal, infinity or relative-chain pinches; those strata enter the joint continuation in Section~\ref{sec:nmhvContinuation}.

For one leg, write $\mathcal A(\Delta;u,\bar u)=\frac{a_{-1}(u,\bar u)}{\Delta}+O(1)$ on the analytic contour tube. Set $L=x-u_0$, $\bar L=\bar x-\bar u_0$. An endpoint monomial gives
\begin{equation}
 I_{mn}(\Delta)
 =\frac{L^{\Delta+m+1}\bar L^{\Delta+n-3}}
 {(\Delta+m+1)(\Delta+n-3)}.
 \label{eq:BPEndpointMonomial}
\end{equation}
Its pole at $\Delta=0$ has $n=3$. The Taylor coefficient $-\frac16\bar\partial_x^3a_{-1}(u,\bar x)$, combined with the relative prefactor $-3\Delta$, yields
\begin{equation}
 \Res_{\Delta=0}
 \bigl[\mathscr Q_+^{\rm rel}(\Delta)\mathcal A(\Delta)\bigr](x)
 =\frac12\bar\partial_x^3
 \int_{u_0}^{x}\dd u\,a_{-1}(u,\bar x).
 \label{eq:BPAnalyticEndpointResidue}
\end{equation}
Subtracting the cubic Taylor polynomial makes the remaining endpoint operator holomorphic near zero, uniformly on compact tubes. In the standard-shadow block, the angular phase $e^{4\ii\theta}$ selects $n=m+4$ and the radial denominator is $2\Delta+2m+2$; this block has no local pole at zero. The two descendant maps therefore act coefficientwise on the MHV double residue. A Ward polynomial of conjugate degree at most two gives zero in Eq.~\eqref{eq:BPAnalyticEndpointResidue}. The global transform additionally retains punctures, infinity, endpoint overlaps and monodromy~\cite{Banerjee:2022wht}.

\subsection{Relative boundary normalization}
\label{subsec:relativeBoundaryNormalization}

The endpoint formula Eq.~\eqref{eq:BPAnalyticEndpointResidue} differentiates on an independently complexified cut chamber. Ordinary weak differentiation of a discontinuous branch representative includes a boundary current. For example, with $z=x+\ii y$ and $a\in\mathbb R$, the principal logarithm has cut current $C_a=\Theta(a-x)\delta(y)$ and $\bar\partial\operatorname{Log}(z-a)=-\pi C_a$. A quadratic Ward polynomial therefore gives
\begin{equation}
 \begin{aligned}
&\frac12\bar\partial^3[P(\bar z)\operatorname{Log}(z-a)]\\
 &\quad=-\frac\pi2\left[P\bar\partial^2C_a
              +3P'\bar\partial C_a+3P''C_a\right].
 \end{aligned}
 \label{eq:BPExplicitCutCurrent}
\end{equation}
The right side is nonzero across the cut, although the descendant vanishes within each chamber. A physical packet with a linear normal test profile at the real four-graviton channel detects it with metric-source coefficient $875/272$ in units of the local density $\rho_*E^3$.

The generic relative kernel selects a different boundary operation. For a based normalized fractional integral,
\begin{equation}
 (I_C^\lambda f)(x)=\frac1{\Gamma(\lambda)}
 \int_{x_0}^{x}\dd u\,(x-u)^{\lambda-1}f(u),
 \label{eq:BPBasedFractionalIntegral}
\end{equation}
Euler reflection and the branch orientation in Eq.~\eqref{eq:BPGenericIntertwiners} give
\begin{equation}
 \begin{aligned}
\mathscr Q_+^{\rm rel}(\Delta)
 &=\frac12e^{-\ii\pi\Delta}I_x^{\Delta+1}I_{\bar x}^{\Delta-3},\\
 \mathscr Q_-^{\rm rel}(\Delta)
 &=\frac12e^{-\ii\pi\Delta}I_y^{\Delta-3}I_{\bar y}^{\Delta+1}.
 \end{aligned}
 \label{eq:BPFractionalRelativeMap}
\end{equation}
For the isolated single-soft Ward pole $\Delta^{-1}P(\bar u)/(u-a)$, its relative residue vanishes, including the locally integrable logarithmic boundary value. It consequently differs from Eq.~\eqref{eq:BPExplicitCutCurrent}. The conversion follows from the jump derivatives: if $j_k$ is the jump of the regular $k$th antiholomorphic derivative of a primitive $\epsilon_C$, and $\mathcal N_Cj$ is the current supplied by one weak derivative, set
\begin{equation}
 \begin{aligned}
 E_C[\epsilon_C]&=\frac12\left[
 \mathcal N_Cj_2+\bar\partial(\mathcal N_Cj_1)
 +\bar\partial^2(\mathcal N_Cj_0)\right],\\
 T_{{\rm rel},C}&=T_{\rm sh}
       +\frac12\bar\partial^3\epsilon_C-E_C[\epsilon_C].
 \end{aligned}
 \label{eq:BPBoundaryCompletedDescendant}
\end{equation}
For quadratic Ward monodromies this is cut independent and reproduces the global single Ward tensor. Its boundary subtraction has a nonzero shape derivative and must move with the contour. Appendix~\ref{app:BPboundary} derives these statements and the physical coefficients. This fixes the single-Ward boundary normalization on the specified module. The full two-weight family additionally uses the joint boundary calculation in Section~\ref{sec:core-constructed-matching}.

\subsection{The two-leg family and the mixed response}
\label{subsec:twoLegBPFamily}

For a compact hard packet, define the generic two-soft family and its four angular blocks by
\begin{equation}
 \begin{aligned}
&\mathcal A_6(\Delta_+,\Delta_-;u,v;X)\\
 &\quad=\int_0^\infty\dd\tau_+\,\tau_+^{\Delta_+-1}
 \int_0^\infty\dd\tau_-\,\tau_-^{\Delta_--1}\\
 &\qquad\left\langle\mathcal F_6(\tau_+,u;\tau_-,v;X),\chi_X\right\rangle.
 \end{aligned}
 \label{eq:genericTwoSoftAmplitude}
\end{equation}
\begin{equation}
 \begin{aligned}
&\mathcal G^{ab}(\Delta_+,\Delta_-;x,y;X)\\
 &\quad=\int \mathscr Q_+^a(\Delta_+;x,u)
 \mathscr Q_-^b(\Delta_-;y,v)\\
 &\qquad\mathcal A_6(\Delta_+,\Delta_-;u,v;X).
 \end{aligned}
 \label{eq:BPFourGenericBlocks}
\end{equation}
Here $a,b\in\{\mathrm{sh},\mathrm{rel}\}$ select the measures in Eq.~\eqref{eq:BPGenericIntertwiners}. The ordered boundary values on a common branch give
\begin{equation}
 \begin{aligned}
 \mathcal K_{\bar\mu|\mu}^{\BP}(x|y;X)
 &:={}
 \Res_{\Delta_+=0}\Res_{\Delta_-=0}
 \sum_{a,b}\mathcal G^{ab}_{\bar\mu|\mu}
 (\Delta_+,\Delta_-;x,y;X),\\
 \mathcal K_{\mu|\bar\mu}^{\BP}(y|x;X)
 &:={}
 \Res_{\Delta_-=0}\Res_{\Delta_+=0}
 \sum_{a,b}\mathcal G^{ab}_{\mu|\bar\mu}
 (\Delta_+,\Delta_-;x,y;X).
 \end{aligned}
 \label{eq:regulatedDoubleBP}
\end{equation}
Composition acts from right to left: the upper line takes the negative-helicity residue earlier. Its small contour lies inside the positive-helicity contour; the lower line exchanges them, avoiding the polar set. Thus the consecutive orientation of Section~\ref{sec:amplitude} is $\mathcal K_{\mu|\bar\mu}^{\BP}-\mathcal K_{\bar\mu|\mu}^{\BP}$, opposite to the source curvature below.

The evaluated contact residue in Eq.~\eqref{eq:contactMixedDivisorResidue} determines which coefficients an angular pole selects. Suppose the angular extension commutes with the residue transverse to $\delta=\alpha+\beta=0$ and is holomorphic in $\delta$ at its generic points. Denote the resulting simple-pole contribution by $\mathcal G_{\rm ct,\delta}^\mu$. Write $\mathscr T(\alpha)=\mathscr Q_+^{\BP}(\alpha)\mathscr Q_-^{\BP}(-\alpha)=\sum_{k=-p}^{\infty}\mathscr T_k\alpha^k$, with finite $p$, and the contact residue as $\sum_{j=-1}^{\infty}a_j^\mu\alpha^j$. Deformation of the nested cycles gives
\begin{equation}
 \begin{aligned}
&\left(\Res_\beta\Res_\alpha-\Res_\alpha\Res_\beta\right)
 \mathcal G_{\rm ct,\delta}^\mu\\
 &\quad=\Res_{\alpha=0}\!\left[
 \mathscr T(\alpha)\sum_{j=-1}^{\infty}a_j^\mu\alpha^j\right]\\
 &\quad=\sum_{k=-p}^{0}\mathscr T_k a_{-1-k}^\mu.
 \end{aligned}
 \label{eq:contactAngularLaurentSelection}
\end{equation}
Eq.~\eqref{eq:contactMixedDivisorExpansion} gives $a_{-1}^\mu$ and $a_0^\mu$, while $a_1^\mu=\sum_iD_i^\mu[\frac12(\log\frac{B_i}{A_i})^2+\frac{\pi^2}{6}]$. Holomorphic angular continuation selects the endpoint jump. An angular pole also selects logarithmic coefficients. Other mixed divisors have separate residues, and higher transverse poles require transverse derivatives.

Define the intertwining defects relative to the product Ward action:
\begin{equation}
 \begin{aligned}
 \mathfrak I_{\bar\mu|\mu}(x|y;X)
 &:={}
 \mathcal K_{\bar\mu|\mu}^{\BP}(x|y;X)
 -D_y\bar D_{\bar x}A_X,\\
 \mathfrak I_{\mu|\bar\mu}(y|x;X)
 &:={}
 \mathcal K_{\mu|\bar\mu}^{\BP}(y|x;X)
 -\bar D_{\bar x}D_yA_X.
 \end{aligned}
 \label{eq:BPIntertwiningDefects}
\end{equation}
Their standard-shadow and relative-block contributions sum to these expressions, with the Ward subtraction assigned to the standard--standard block. In the reference Ward calculation these intertwining defects include the nonzero soft-leg distributions $J_+$ and $J_-$ evaluated in Eq.~\eqref{eq:mixedSoftLegContactKernels}; they are not by themselves physical matching failures. The difference between the two operator orderings is
\begin{equation}
 \Delta^{\BP,\rm op}_{\bar\mu\mu}(x,y;X)
 =\mathcal K^{\BP}_{\bar\mu|\mu}(x|y;X)
 -\mathcal K^{\BP}_{\mu|\bar\mu}(y|x;X).
 \label{eq:shadoworderingdifference}
\end{equation}
The full metric-source curvature obeys $\cF^{\rm seed}_{\mu\bar\mu}=\pi^{-2}\Delta^{\BP}_{\bar\mu\mu}$, where
\begin{equation}
 \begin{aligned}
&\Delta^{\BP}_{\bar\mu\mu}(x,y;X)={}\mathcal K^{\BP}_{\bar\mu|\mu}(x|y;X)
 -\mathcal K^{\BP}_{\mu|\bar\mu}(y|x;X)\\
 &-\pi\int\dd^2u\,
 \bigl(\delta^{\rm full}_{\mu(y)}\mathscr Q_+\bigr)(x,u)
 \langle S_+(u)X\rangle\\
 &+\pi\int\dd^2v\,
 \bigl(\delta^{\rm full}_{\bar\mu(x)}\mathscr Q_-\bigr)(y,v)
 \langle S_-(v)X\rangle\\
 &+C^{\rm src}_{\bar\mu\mu}(x,y;X).
 \end{aligned}
 \label{eq:shadowedmixedcurvature}
\end{equation}
The factors $-\pi$ and $+\pi$ follow from $\omega_{\bar\mu}=-\pi^{-1}\mathscr Q_+S_+$. Each full map variation includes its kernel, descendant, relative inverse, contour, endpoints and projectors. The term $C^{\rm src}$ contains the nonlinear source map and covariant-delta or coincident-source contacts, counting each contribution once.

For compact Beltrami profiles, the integrated curvature is
\begin{equation}
 \mathfrak C_{ab}[\chi_X]
 =\frac{1}{\pi^2}\int\dd^2x\,\dd^2y\,
 \bar\mu_a(x)\mu_b(y)
 \Delta^{\BP}_{\bar\mu\mu}(x,y;X).
 \label{eq:pairedMixedMetricCurvature}
\end{equation}
It is the coefficient of $\epsilon\eta$ in the source-loop response with the $\mu_b$ direction traversed before $\bar\mu_a$. This tests the antisymmetric response; the symmetric susceptibility is additional matching data.

\subsection{Contour transport and finite-metric matching}
\label{subsec:transportedmixed}

Let $\mathbb U_g^{S_\pm}$ and $\mathbb U_g^{T_\pm}$ carry the soft and stress bundles, contours and endpoints to the reference sphere. Define
\begin{equation}
 \mathscr Q_{\pm,g}^{\trn}
 =\bigl(\mathbb U_g^{T_\pm}\bigr)^{-1}
 \mathscr Q_{\pm,\hg}^{\BP}\mathbb U_g^{S_\pm}.
 \label{eq:transportedBPmap}
\end{equation}
For $\mathcal A_A^E=(\mathbb U_g^E)^{-1}\delta_A\mathbb U_g^E$, this gives
\begin{equation}
 \nabla_A\mathscr Q_{\pm,g}^{\trn}
 :=\delta_A\mathscr Q_{\pm,g}^{\trn}
 +\mathcal A_A^{T_\pm}\mathscr Q_{\pm,g}^{\trn}
 -\mathscr Q_{\pm,g}^{\trn}\mathcal A_A^{S_\pm}=0.
 \label{eq:BPmapParallelTransport}
\end{equation}
The operator transport fixes the input and output bundles and their relative domains. The full metric derivative also contains the spin-frame contact fixed by Eq.~\eqref{eq:explicitMixedPrimaryFrame}. We denote the completed critical insertion by
\begin{equation}
 \mathcal I_{\pm,g}^{\mathfrak P}
 =\operatorname{Crit}\!\left(\mathscr Q_{\pm,g}^{\trn}S_\pm[g]\right)
 +C_{\pm,g},
 \label{eq:completedPhysicalInsertion}
\end{equation}
where $\operatorname{Crit}$ is the prescribed conformally soft extraction, including the relative boundary operation. Section~\ref{sec:core-constructed-matching} specifies the common prescription $\mathfrak P$ and the finite frame terms $C_{\pm,g}$. The marked comparison is
\begin{equation}
 \left\langle\mathcal I_{+,g}^{\mathfrak P}\right\rangle_{g,J,c}
 =-\pi\frac{\delta W_{\trn}}{\delta\bar\mu},
 \qquad
 \left\langle\mathcal I_{-,g}^{\mathfrak P}\right\rangle_{g,J,c}
 =-\pi\frac{\delta W_{\trn}}{\delta\mu}.
 \label{eq:transportedSoftMetricMatching}
\end{equation}
Section~\ref{subsec:finite-external-states} derives this identity for the chosen finite family by transporting the state and its complete current pairing. It is a change-of-variables identity for that family. The independent scattering comparison is performed at the reference metric in Section~\ref{subsec:physicalMixedResiduals}. The displayed Beltrami variation holds at fixed Weyl source; a general variation also contains the Weyl insertion.

Define the matching defect relative to the transported metric response by
\begin{equation}
 \begin{aligned}
 \mathfrak E_{\bar\mu}(x;g,J)
 &:=-\frac{1}{\pi}
 \left\langle\mathcal I_{+,g}^{\mathfrak P}\right\rangle_{g,J,c}
 -\frac{\delta W_{\trn}}{\delta\bar\mu(x)},\\
 \mathfrak E_{\mu}(y;g,J)
 &:=-\frac{1}{\pi}
 \left\langle\mathcal I_{-,g}^{\mathfrak P}\right\rangle_{g,J,c}
 -\frac{\delta W_{\trn}}{\delta\mu(y)}.
 \end{aligned}
 \label{eq:softMetricMatchingDefect}
\end{equation}
Symmetry of the transport Hessian gives
\begin{equation}
 \frac{\delta\mathfrak E_{\bar\mu}(x)}{\delta\mu(y)}
 -\frac{\delta\mathfrak E_{\mu}(y)}{\delta\bar\mu(x)}
 =\frac{1}{\pi^2}\Delta^{\trn}_{\bar\mu\mu}(x,y).
 \label{eq:softMetricDefectCurl}
\end{equation}
Closure constrains this curl; equality in Eq.~\eqref{eq:transportedSoftMetricMatching} also fixes the symmetric marked functional. The curvature coupling in Appendix~\ref{subsec:symmetricMarkedCouplings} illustrates that freedom, and Appendix~\ref{app:homotopy} gives the separate matching condition for the radial prescription.
\subsection{Two physical matching defects}
\label{subsec:physicalMixedResiduals}

The comparison with the generic family uses the global distributional single-insertion normalization
\begin{equation}
 \begin{aligned}
e_+(x)&:=\mathscr Q_+a_+(x)-\bar D_{\bar x}A_X=0,\\
 e_-(y)&:=\mathscr Q_-a_-(y)-D_yA_X=0.
 \end{aligned}
 \label{eq:globalOneLegBPNormalization}
\end{equation}
Eq.~\eqref{eq:BPBoundaryCompletedDescendant} supplies this normalization for the single-Ward module. Its application to a larger continued family retains the same boundary and transition data; otherwise the output-bundle variations of $e_\pm$ enter the comparison in the linear response.

Let $k^{\rm phys}_{-\to+}$ and $k^{\rm phys}_{+\to-}$ be the two kernels in Eq.~\eqref{eq:regulatedDoubleBP} smeared with $\pi^{-2}\bar\mu_b(x)\mu_a(y)$ and the compact hard packet $\Psi$. Define the soft-leg terms using the same completed relative boundary prescription as the maps:
\begin{equation}
 \begin{aligned}
 j_+^{\BP}&=-\frac1\pi\int\dd^2x\,\bar\mu_b(x)
 \mathscr Q_+[(v\partial+\partial v)a_+](x),\\
 j_-^{\BP}&=-\frac1\pi\int\dd^2y\,\mu_a(y)
 \mathscr Q_-[(\bar v\bar\partial+\bar\partial\bar v)a_-](y).
 \end{aligned}
 \label{eq:BPConsistentSoftLegTerms}
\end{equation}
The hard pairing is understood on the right. On the energy-resolved distribution
$\mathcal F_n$, the Ward generators use
$h_i=(J_i-E_i)/2$ and $\bar h_i=-(J_i+E_i)/2$, as in
Eq.~\eqref{eq:energyResolvedChiralWardGenerator}; Mellin pairing recovers
the numerical celestial weights. The two physical matching defects are
\begin{equation}
 \begin{aligned}
 \mathcal B_{-\to+}
 &=k^{\rm phys}_{-\to+}
 -\langle\mathscr D_v\overline{\mathscr D}_{\bar v}\mathcal F_n,\Psi\rangle
 -j_+^{\BP},\\
 \mathcal B_{+\to-}
 &=k^{\rm phys}_{+\to-}
 -\langle\overline{\mathscr D}_{\bar v}\mathscr D_v\mathcal F_n,\Psi\rangle
 -j_-^{\BP}.
 \end{aligned}
 \label{eq:physicalMixedResidueDefects}
\end{equation}
They compare the generic two-weight continuation with the iterated single-soft Ward distribution on the chosen global domain. With Eq.~\eqref{eq:globalOneLegBPNormalization}, transported endpoint data and the primary frame Eq.~\eqref{eq:explicitMixedPrimaryFrame}, the evaluated cancellation gives
\begin{equation}
 \begin{aligned}
 H^{\rm phys}_{-\to+}&=\langle H^{\trn}[\mathcal F_n],\Psi\rangle
                         +\mathcal B_{-\to+},\\
 H^{\rm phys}_{+\to-}&=\langle H^{\trn}[\mathcal F_n],\Psi\rangle
                         +\mathcal B_{+\to-},\\
 \mathfrak C^{\rm phys}_{a\bar b}
 &=\mathcal B_{-\to+}-\mathcal B_{+\to-}.
 \end{aligned}
 \label{eq:transportedMixedCurvatureZero}
\end{equation}
Thus closure tests their difference, whereas matching to the specified transport requires both defects to vanish. A common nonzero defect changes the symmetric marked Hessian at the reference point; integrability throughout a source neighborhood requires higher-order matching conditions.

Section~\ref{subsec:core-evaluated-matching} evaluates these defects with the common prescription in Section~\ref{subsec:early-matching-statement}.

\section{Joint soft limits and relative-contour factorization}
\label{sec:nmhvContinuation}

The remaining NMHV terms require joint energy and angular continuation beyond the real-angular contact estimate of Section~\ref{subsec:nmhvIntegratedContact}. With a common boundary prescription for both residue orders and source variations, factorization supplies the pole coefficients and the resolved expansion determines which coefficient derivatives the Mellin integrals select.

\subsection{One prescribed family and its physical singularities}
\label{subsec:nmhvCommonFamily}

Put $\alpha=\Delta_+$, $\beta=\Delta_-$ and $Q=\tau_+q(u,\bar u)+\tau_-q(v,\bar v)$. The physical pairing is Eq.~\eqref{eq:analyticCoareaPairing}, with $b(-Q)=1$ on the soft support. Its normal factors are continued while the compact tangential profile remains a smooth real test function. Appendix~\ref{app:nmhvRecoilResolution} gives the recoil solution for a general $Q$ and the corresponding Jacobian.

A prescription $\mathcal P$ fixes the four standard and relative chains, their
base points, branches, physical boundary values, chart transitions, regulator
removal and source transport. With compact soft cutoffs understood, the common
object is
\begin{equation}
 \begin{aligned}
&G^{\mathcal P}(\alpha,\beta)=\sum_{a,b\in\{\mathrm{sh},\mathrm{rel}\}}
 \int_{\mathcal C^{ab}_{\mathcal P}}
 \mathscr Q_+^a(\alpha)\mathscr Q_-^b(\beta)\\
 &\qquad\tau_+^{\alpha-1}\tau_-^{\beta-1}
 a(\vartheta)\rho\mathcal M_6h(-Q).
 \end{aligned}
 \label{eq:nmhvCommonPrescribedFamily}
\end{equation}
The measure includes $\dd\tau_+\dd\tau_-\dd\vartheta$ and the corresponding
oriented angular measures. All integration variables are retained through
the joint singularity analysis. Pairing a rational angular kernel with an
arbitrary smooth hard profile need not produce a meromorphic angular
function; the appropriate continuation is a family of distributions. If
auxiliary analytic regulators are necessary, their finite parts are taken
at the same stage in both soft orders.

For six external legs there are $15$ two-particle and $10$ three-particle
factorization partitions. On the regular compact hard support, thirteen candidate physical denominators can vanish at the soft corner:
\begin{equation}
 \begin{gathered}
 d_{i+}=A_i\tau_+,\qquad d_{i-}=B_i\tau_-,\qquad
 d_{+-}=C\tau_+\tau_-,\\
 d_{i+-}=A_i\tau_++B_i\tau_-+C\tau_+\tau_-,\qquad
 d_I=-\Bigl(\sum_{j\in I}p_j\Bigr)^2.
 \end{gathered}
 \label{eq:nmhvPhysicalDivisors}
\end{equation}
There are eight hard--soft, one soft--soft and four mixed three-particle
channels. Here $A_i=-2p_i\cdot q_+$, $B_i=-2p_i\cdot q_-$ and
$C=-2q_+\cdot q_-$. The coefficients retain the hard recoil. The remaining
twelve channels stay away from zero on a sufficiently small neighborhood.
In the $(--++,+,-)$ sector, each mixed three-particle residue has a single
allowed internal helicity and is a product of two four-point MHV amplitudes.
Appendix~\ref{app:nmhvRecoilResolution} gives their normalization and an
independent factorization check. This inventory concerns the summed
amplitude; spurious denominators of individual recursion terms do not
receive independent boundary prescriptions.

Physical soft energies $e_+=\tau_+(1+|u|^2)$ and
$e_-=\tau_-(1+|v|^2)$ control recoil uniformly on the real sphere, since
$\|Q\|_{\mathbb R^4}\le\sqrt2(e_++e_-)$. Their Mellin measures contain the
factors $(1+|u|^2)^{-\alpha}(1+|v|^2)^{-\beta}$. A real kinematic estimate
also excludes noncollinear zeros of $d_{i+-}$ when
$\max(e_+,e_-)<\frac12\min_i\mathcal E_i$, where
$p_i=\epsilon_i\mathcal E_i(1,\mathbf n_i)$.
The soft and collinear boundary faces remain, as do the independent complex
zeros on relative chains.

\subsection{Residues from the joint boundary expansion}
\label{subsec:nmhvFiniteResidue}

Fix $\mathcal P$ and suppose that every relevant block, its boundary terms
and its required source derivatives admit a finite resolution on compact
support into integrals
\begin{equation}
 \begin{aligned}
&I_c(\alpha,\beta)=\int_{K_c}\dd y\int_{[0,1]^{N_c}}\dd t\,\\
 &\quad\prod_j t_j^{a_{cj}\alpha+b_{cj}\beta+c_{cj}-1}
 (\log t_j)^{k_{cj}}F_c(t,y;\alpha,\beta).
 \end{aligned}
 \label{eq:nmhvResolvedClass}
\end{equation}
The coefficients are smooth in the integration variables, holomorphic in
the weights and locally uniform in the necessary test seminorms. Singular
pullbacks and boundary values are part of these hypotheses. We require a
nonempty joint convergence domain, or a fixed prior extension in
$\mathcal P$ of every weight-independent divergent face. Every surviving
resonant linear form is nonzero: $(a_j,b_j)\ne(0,0)$.
Real analytic
continuation and sector decomposition provide the established framework
\cite{Dang:2015ComplexPowers,Bogner:2007cr}; they do not automatically verify
the hypotheses on independently complex relative cycles.

Under these hypotheses, each paired family has a meromorphic germ with
affine linear candidate poles. Both ordered double residues depend on
finitely many Taylor coefficients normal to the resolved boundary faces. After
nonresonant factors have been absorbed into a holomorphic numerator, a term
\begin{equation}
 \frac{H(\alpha,\beta)}{\prod_j(a_j\alpha+b_j\beta)^{p_j}},
 \qquad P_{\mathrm{pol}}=\sum_jp_j,
 \label{eq:nmhvPolarTerm}
\end{equation}
contributes only through the homogeneous Taylor term
$H_{P_{\mathrm{pol}}-2}(\alpha,\beta)$.
If $P_{\mathrm{pol}}<2$, both residues vanish. With
$f(z)=H_{P_{\mathrm{pol}}-2}(z,1)/\prod_j(a_jz+b_j)^{p_j}$, the two values
and their difference are
\begin{equation}
 \begin{aligned}
\mathcal R_+I&=\Res_{z=0}f(z)\dd z,\\
 \mathcal R_-I&=-\Res_{z=\infty}f(z)\dd z,\\
 (\mathcal R_+-\mathcal R_-)I
 &=-\sum_{z_*\ne0,\infty}\Res_{z=z_*}f(z)\dd z.
 \end{aligned}
 \label{eq:nmhvProjectiveResidues}
\end{equation}
The degree principle for hyperplane-arrangement residues gives this termwise selection~\cite{SzenesVergne:2003Residues}. Appendix~\ref{app:nmhvFiniteResidues} derives the overlap subtractions and uniform integrable bound for the holomorphic remainder.

At total pole degree two only the constant numerator is needed. Higher
degree selects derivatives of the gamma normalization, phases, angular
kernels, recoil and coefficient integrals. In particular, a relative map
that annihilates a Ward polynomial at zero weight need not annihilate the
weight derivatives selected by a higher pole. The BP weight derivatives are
given in Appendix~\ref{app:nmhvWeightJets}. Finite Taylor order transverse to these faces leaves
coefficient functions along them; it does not make arbitrary source
dependence finite dimensional.
For the physical prescription in Section~\ref{sec:core-constructed-matching}, the required estimates include the finite polar division, the common cutoff and overlap subtraction, and the joint source and path coordinates at endpoint folds. Source derivatives act on the same continued current. The general reduction above retains its hypotheses for other choices of relative cycles or extensions.

\subsection{Symmetric response and contour-dependent terms}

The joint hypothesis cannot be replaced by a soft expansion at separated
angles. For example,
\begin{equation}
 \frac1\pi\int_{|w|<1}\dd^2w\,
 \frac{st}{(|w|^2+st)^2}=\frac1{1+st},\qquad s,t>0.
 \label{eq:nmhvBoundaryLayerDiagnostic}
\end{equation}
At every fixed $w\ne0$ the finite coefficient in either soft energy
vanishes, whereas the integrated double Mellin germ has coefficient
$1/(\alpha\beta)$. Both ordered residues are one. This diagnostic is not
an Einstein contact; it shows why a common symmetric contribution can be
missed by fixed-angle reasoning and by a curvature test alone.

The four two-leg blocks determine both matching defects in Eq.~\eqref{eq:physicalMixedResidueDefects}; the eight three-leg blocks specify the global cubic comparison that remains to be completed. Their BP assembly includes the input, map, contour and frame variations with the same regulator and global relative transitions. Table~\ref{tab:matching-proof-dependencies} identifies the decisive comparisons.

\section{Matching to Einstein tree amplitudes}
\label{sec:core-constructed-matching}

We separate the reference scattering comparison from the finite construction. The latter realizes Eq.~\eqref{eq:transportcompletion} by a common transport of external states and current data. The seven-graviton calculation then tests a local ingredient of the cubic comparison.

\subsection{Physical domain and relative-current prescription}

On the hard chamber and source neighborhood of Section~\ref{subsec:early-matching-statement}, let $\mathcal D_0$ be the reference test space of analytic conservation-normal germs with compact smooth tangential profiles. Transport of the hard external states at finite metric sources defines
\begin{equation}
 \mathcal D_g=
 \{\psi:(U_g^{\rm hard})^*\psi\in\mathcal D_0\}.
 \label{eq:core-transported-normal-domain}
\end{equation}
Generic-weight continuation uses this domain; source derivatives of each critical residue act on its smooth-packet extension. The metric sources require only smooth real jets.

We use one reference prescription $\mathfrak P$ for all standard-shadow and relative
blocks. The high-order relative paths are straight. The negative barred
low-order path carries the detour $e^{\ii\theta}\sin(\pi q)/4$, while the
other low-order path is straight. On each compact chart the base lies outside
the output support and hard neighborhoods, and the detour angle is transverse
to the finite set of nonzero hard offsets. Open connectors, closed periods,
cuts and projective transitions, including the sphere endpoint, are carried
on a finite adapted cover. Their evaluated periods remain part of the
current.

Pole overlaps and the scalar finite part follow the prescription in
Section~\ref{subsec:early-matching-statement}. Supplementary
Section~S1 specifies the path loadings and extensions.

\subsection{External states at finite metric sources and the completed insertion}
\label{subsec:finite-external-states}

The finite primary frame in Eq.~\eqref{eq:explicitMixedPrimaryFrame} has an
on-shell realization. Put
\begin{equation}
 D_g=1-\mu\bar\mu,\qquad
 r_g=\frac{F_z}{\bar F_{\bar z}},\qquad
 \nu_g=\frac{e^\phi}{\sqrt{D_gF_z\bar F_{\bar z}}}.
 \label{eq:core-finite-state-scales}
\end{equation}
On the real source slice $\nu_g>0$ and $|r_g|=1$. With spin-frame roots
continued from the reference source, the spinors representing the massless momenta are
\begin{equation}
 \lambda_{i,g}=\sqrt{\nu_i\omega_i}\,r_i^{-1/4}
       \begin{pmatrix}F_i\\1\end{pmatrix},\qquad
 \widetilde\lambda_{i,g}=\epsilon_i\sqrt{\nu_i\omega_i}\,r_i^{1/4}
       \begin{pmatrix}\bar F_i\\1\end{pmatrix}.
 \label{eq:core-finite-null-spinors}
\end{equation}
They give $p_{i,g}=\epsilon_i\nu_i\omega_i q(F_i,\bar F_i)$ with the same
incoming or outgoing sign. Little-group homogeneity fixes
\begin{equation}
 \mathcal F^{\rm kin}_{n,g}
 =U_g^{\rm hard}\mathcal F_n
 =\prod_i r_i^{J_i/2}\,
     \mathcal F_n(\nu_i\omega_i,F_i,\bar F_i).
 \label{eq:core-original-hard-target}
\end{equation}
The transformed conservation surface is $\sum_i p_{i,g}=0$. Equation~
\eqref{eq:core-original-hard-target} is the hard coefficient of
Eq.~\eqref{eq:transportcompletion}, since $\Delta_i=-E_i$ and
$E_i=\omega_i\partial_{\omega_i}$. The change of variables also fixes the
test transport and measure:
\begin{equation}
 \omega'\,\dd\omega'\,\dd^2F
   =\nu_g^2D_gF_z\bar F_{\bar z}\,
       \omega\,\dd\omega\,\dd^2z
   =e^{2\phi}\omega\,\dd\omega\,\dd^2z.
 \label{eq:core-finite-state-measure}
\end{equation}
The adjoint transformation of the hard packet consequently retains the
coarea Jacobian and conservation-normal derivatives.

Let $\Phi_g=(F_g,\bar F_g)$, with uniformized coordinates
$(\zeta,\bar\zeta)$, and write
$v=\delta F\circ\Phi_g^{-1}$ and
$u=\delta\bar F\circ\Phi_g^{-1}$. Differentiating the Beltrami equations
gives
\begin{equation}
 \begin{aligned}
\delta\log\nu_g&=\delta\phi-\tfrac12
       (\partial_\zeta v+\partial_{\bar\zeta}u),\\
 \delta\log r_g&=\partial_\zeta v-\partial_{\bar\zeta}u
       +\frac{\bar\mu\delta\mu-\mu\delta\bar\mu}{D_g}.
 \end{aligned}
 \label{eq:core-finite-state-log-variation}
\end{equation}
All source coefficients on the right are expressed in the same uniformized
chart. The full derivative of the Einstein distribution is therefore
\begin{equation}
 \begin{aligned}
\delta\mathcal F^{\rm kin}_{n,g}
 ={}&U_g^{\rm hard}\sum_i\Bigl[
  v_i\partial_{\zeta_i}+u_i\partial_{\bar\zeta_i}\\
 &+\frac{J_i-E_i}{2}\partial_\zeta v_i
  +\frac{-J_i-E_i}{2}\partial_{\bar\zeta}u_i\\
 &+E_i\delta\phi_i+\frac{J_i}{2D_{g,i}}
       (\bar\mu_i\delta\mu_i-\mu_i\delta\bar\mu_i)
 \Bigr]\mathcal F_n.
 \end{aligned}
 \label{eq:core-finite-state-generator}
\end{equation}
The last term is the finite helicity-frame connection; its nonzero curvature combines with the coordinate and Jacobian terms in the full derivative.

Away from the real angular section, we transport the on-shell massless external states along the radial source path of Supplementary Section~S4. On the real angular section this transport is $U_g^{\rm hard}$. The projected state connection may have off-real curvature; the radial transgression in Supplementary Eq.~\textup{(S4.10)} fixes the correction to the finite flow.

Let $\Psi_g$ denote the resulting real flow on the ambient angular space and
$a_{\beta,J,g}$ its Mellin and helicity factor. The generic soft input,
relative form and inverse chain are
\begin{align}
 S_{\pm,g}(\beta)
 &=a_{\beta,\pm2,g}\,U_g^{\rm hard}
                S_{\pm,0}(\beta)\circ\Psi_g,\notag\\
 \Omega_{\pm,\beta,g}
 &=B_{\pm,g}(\beta)\,a_{\beta,\pm2,g}^{-1}
                \Psi_g^*\Omega_{\pm,\beta,0},\qquad
 C_g=\Psi_g^{-1}C_0,
 \label{eq:core-finite-current}
\end{align}
where the output argument is carried by $\Phi_g$ and
$B_{-,g}=F_z^2\nu_g^\beta$,
$B_{+,g}=\bar F_{\bar z}^{\,2}\nu_g^\beta$.
The form is the full real pullback, including all six exterior minors of Supplementary Eq.~\textup{(S4.12)}. The inverse chain carries base edges, cuts, endpoints and resolved joint
faces. The standard shadow transform uses the same real change of variables, and
the energy contour and regulator follow the state energy scale. The input
and reciprocal kernel factors cancel before pairing, giving
\begin{equation}
 Q_{\pm,g}(\beta)S_{\pm,g}(\beta)
 =B_{\pm,g}(\beta)U_g^{\rm hard}
       [Q_{\pm,0}(\beta)S_{\pm,0}(\beta)]\circ\Phi_g.
 \label{eq:core-generic-transport}
\end{equation}
Equation~\eqref{eq:core-generic-transport} follows by change of variables because the input, reciprocal form and inverse chain in Eq.~\eqref{eq:core-finite-current} were chosen together. It establishes covariance of this finite family; an independently defined finite-metric scattering prescription would require a further comparison.

Here $Q_{\pm,g}$ denotes the transported BP operation
$\mathscr Q_{\pm,g}^{\trn}$. The identity is first established in a
convergence domain and then continued on Eq.~\eqref{eq:core-transported-normal-domain}.

The completed insertion in Eq.~\eqref{eq:completedPhysicalInsertion}
contains the frame currents $C_{-,g}=C_{\mu,g}$ and
$C_{+,g}=C_{\bar\mu,g}$ fixed by
Eq.~\eqref{eq:core-finite-state-generator}:
\begin{align}
 C_{\mu,g}(z)
 &=-\pi U_g^{\rm hard}\sum_i
       \frac{J_i\bar\mu_i}{2D_{g,i}}
           \delta^{(2)}(z-z_i)\mathcal F_n,\notag\\
 C_{\bar\mu,g}(z)
 &=+\pi U_g^{\rm hard}\sum_i
       \frac{J_i\mu_i}{2D_{g,i}}
           \delta^{(2)}(z-z_i)\mathcal F_n.
 \label{eq:core-spin-contacts}
\end{align}
The coefficients and source deltas use the original source chart, with the
pullback understood as in Eq.~\eqref{eq:core-finite-state-generator}.
The critical single-Ward coefficient of Eq.~\eqref{eq:core-generic-transport},
including these contacts, gives the finite marked identity
\begin{equation}
 -\frac1\pi\left[
  \langle\delta\bar\mu,\mathcal I_{+,g}^{\mathfrak P}\rangle
 +\langle\delta\mu,\mathcal I_{-,g}^{\mathfrak P}\rangle\right]
       =\delta_g\mathcal F^{\rm kin}_{n,g},\qquad \delta\phi=0.
 \label{eq:core-finite-marked-Ward}
\end{equation}
For a general variation, its left side also contains
\begin{equation}
 \mathcal W_{\phi,g}(\delta\phi)
     =U_g^{\rm hard}\sum_i E_i\delta\phi_i\mathcal F_n.
 \label{eq:core-weyl-insertion}
\end{equation}
Every source derivative of the relative pairing retains both boundary and
bulk terms:
\begin{equation}
 \delta\langle\Omega,C\rangle
 =\langle\delta\Omega,C\rangle
  +\langle\iota_V\Omega,\partial C\rangle
  +\langle\iota_V\dd\Omega,C\rangle.
 \label{eq:core-current-variation}
\end{equation}
Coherent changes of ambient lift transform the input, full form and inverse chain together. Change of variables equates their periods before continuation, and the physical endpoint generator includes the spin frame. Supplementary Section~S4 proves the lift and path equivalence using this same current variation.

\subsection{Reference defects and ordered third responses}
\label{subsec:core-evaluated-matching}
\label{thm:core-constructed-matching}

For the prescribed current, the reference calculation gives
\begin{equation}
 \mathcal B_{-\to+}^{\mathfrak P}=0,
 \qquad \mathcal B_{+\to-}^{\mathfrak P}=0.
 \label{eq:core-two-defects}
\end{equation}
The calculation extends the complete mixed polar coefficient together with
its physical pair poles, then integrates the cap and annular pieces with one
partition. It retains the moving-root Cauchy currents and their supported
real-translation contacts. For a negative-helicity hard leg, the disappearing
energy has a double pole, which selects the second-order recoil and conservation-normal
jet. The block with a relative positive-helicity leg and a standard-shadow negative-helicity
leg uses its own zero-offset bound. After the first residue, each remainder
has a positive power of that soft energy, at most logarithms, and finitely
many inverse powers of the retained energy. Keeping the weight of the retained energy in a
common convergence half-plane justifies the extraction before continuation.
The remaining single-Ward numerator is quadratic in the critical high-path
variable, so its relative hard term is annihilated by the cubic operation.
The soft-leg density action and generic map variation supply the retained BP
current. Supplementary Section~S2 evaluates both orders,
their overlap terms and the two hard-helicity cases.

For the $++-$ helicity assignment, the three-soft calculation treats the pair channels, the three channels
containing a hard leg and two soft legs, the all-soft channel, and the channel
containing a hard leg and three soft legs. Other hard channels remain regular
on the chosen chamber. Section~\ref{subsec:seven-graviton-current} displays the complete inner five-point factor and its nonzero three-root current, including terms regular on a selected nested pole.
Real derivatives of the roots include both Cauchy derivatives and supported
delta terms. Coalescence is controlled by
\begin{equation}
 \begin{aligned}
&\frac{C_1(z-a)-C_1(z-b)}{a-b}\\
 &\quad=\int_0^1\Bigl[
 C_2(z-b-u(a-b))\\
 &\qquad-\pi\frac{\bar a-\bar b}{a-b}
       \delta^{(2)}(z-b-u(a-b))\Bigr]\,\dd u.
 \end{aligned}
 \label{eq:core-real-divided-difference}
\end{equation}
Here $C_1(z)=1/z$ is the locally integrable Cauchy current and
$C_2=-\partial C_1$. The phase has unit modulus, giving a bound in fixed
smooth-test seminorms. For general hard helicities the additional anti-MHV
inner five-point helicity assignment contributes a distinct branch. When the positive-helicity soft limit is taken first, its cap
has amplitude order $s^{-3}$ and a measure in two complex normal coordinates of order $s^4$, leaving
$O_{\mathcal D'}(s\log^Ns)$ after the protected numerator and compatible
overlaps are retained. This branch is absent in the four-hard MHV benchmark.

A simultaneous lower-path endpoint requires a joint source and path normal.
The source-only rank argument valid in the interior fails at a noncollinear
fold. A remaining interior path ratio supplies the missing normal; an active
sine detour supplies two source normals through common dilation and phase
rotation. The corresponding correlated energy face also requires its
complete coefficient. Write the positive, negative and positive inputs as $(U,A),(V,B),(Z,C)$ with energies
$s,t,e$, and scale $U=tu$, $Z=tz$, $C=sc_0$, $B=sb_0$ at fixed
$e>0$ with $A,V$ bounded away from zero. With $d=u-z$ and $f=b_0-c_0$,
the coefficient including the lower-path measures is
\begin{equation}
 K_{\rm pol}=
 \frac{AVe^3(Ad+Vf)}{df\{A(ed-V)+eVf\}}.
 \label{eq:core-crossed-complete-polar}
\end{equation}
The physical pair correction is included in this expression. Recombining the
negative high path gives a current of the form $p^2F(tp;\beta)$, with its
cutoffs and tests retained. With a sharp cutoff, its transform is
\begin{equation}
 \begin{aligned}
&\mathcal T_\beta^\epsilon[F]=\frac{1}{2\Gamma(\beta-3)}
   \int_0^\epsilon t^{\beta-1}\,\dd t\\
 &\qquad\int_0^1(1-p)^{\beta-4}p^2F(tp;\beta)\,\dd p\\
 &\quad=\frac{1}{2\Gamma(\beta-2)}
   \int_0^\epsilon v^2(\epsilon-v)^{\beta-3}F(v;\beta)\,\dd v.
 \end{aligned}
 \label{eq:core-direct-finite-cutoff}
\end{equation}
The substitutions $v=tp$ and $r=(1-p)/p$ give the second line in the
convergence region. For the original smooth cutoff, this becomes
\begin{equation}
 \begin{aligned}
&\mathcal T_\beta^\chi[F]
  =\frac12\int_0^\infty v^2F(v;\beta)\mathcal K_\beta^\chi(v)\,\dd v,\\
 &\mathcal K_\beta^\chi(v)
  =\frac{1}{\Gamma(\beta-2)}
       \int_v^\infty[-\chi'(\epsilon)]
                    (\epsilon-v)^{\beta-3}\,\dd\epsilon.
 \end{aligned}
 \label{eq:core-smooth-finite-cutoff}
\end{equation}
The normalized endpoint distribution makes this a smooth entire kernel in
$\beta$. In particular,
\begin{equation}
 \mathcal T_0^\chi[F]
     =-\frac12\int_0^\infty v^2F(v;0)\chi'''(v)\,\dd v,
 \qquad \mathcal T_0^\chi[1]=1.
 \label{eq:core-finite-shell-value}
\end{equation}
The weighted two-Cauchy estimate gives
$v^2\operatorname{Mellin}_s F(\alpha,v)
=O_{\mathcal D'}(v^{\Re\alpha}\log^Nv)$ after the prescribed scalar
continuation. A finite spectator polar multiplier can be chosen independently
of $\beta$. The continued pairing is therefore holomorphic in $\beta$ with
meromorphic $\alpha$ coefficients, and this face has zero critical residue in
both orders. The finite polar division and source/regulator normals in
Supplementary Section~S3.9.5 control the remainder through the
closed energy-ratio interval and its adjacent faces.

A global cubic comparison requires connectors between base points and projective changes to act on the physical and Ward currents through the same completed loaded primitive. With three relative maps their
tensor product has seven nonempty connector terms, including the terms with
two or three changed maps. Their boundary, cut, weight and source derivatives
are retained before extraction. Their coherent-transport identities determine the geometric variation. Extending the physical comparison over the adapted cover also requires equality with the Ward variation of the amplitude on each singular overlap. The additional anti-MHV branch, all-soft
faces and transition currents are treated in Supplementary
Section~S3.

With the third energy held nonzero, its coefficient paired with the third metric-source profile is smooth
at the hard punctures; the required seminorms cost finitely many inverse
powers of that energy. Its weight can therefore be kept in a convergence
half-plane while the first two residues are taken, and then continued in the
same family. The candidate Ward recursion for each of the six residue orders and each of the eight combinations of standard-shadow and relative maps is
\begin{equation}
 K^{\rm W}_{abc}
  =(D_a+T_{ab}+T_{ac})(D_b+T_{bc})D_c\mathcal F_n,
 \qquad (a,b,c)\in S_3,
 \label{eq:core-ordered-Ward-recursion}
\end{equation}
where $D_a$ is the hard Ward action and $T_{ab}$ acts on the remaining soft
leg, including its energy Euler term. Assemble the prescribed BP combination
and differentiate its input, kernel, anchor, inverse chain and source data
using Eq.~\eqref{eq:core-current-variation}. Together with the frame terms,
the proposed completed scattering comparison is
\begin{equation}
 \mathcal H_{ABC}^{\mathrm{BP},\pi}
     =\langle\mathcal T_{ABC}\mathcal F_n,\psi\rangle,
 \qquad \pi\in S_3.
 \label{eq:core-completed-third}
\end{equation}
Equation~\eqref{eq:core-completed-third} is conditional on the remaining physical overlap comparisons; coherent transport alone proves the corresponding identity for the constructed family. Here $\mathcal T_{ABC}\mathcal F_n$ is the full third metric-source derivative of the finite family in Eq.~\eqref{eq:core-original-hard-target}. It retains nonlinear source-map acceleration, the cubic terms generated by the quadratic frame, and all conservation-normal derivatives. Supplementary Eq.~\textup{(S4.17)} gives its expanded coefficients.

\subsection{Finite reconstruction of marked coefficients}
\label{sec:marked-finite-functional-reconstruction}

For an admitted polarized hard packet, let $C_{n,I}(\boldsymbol s)$ be the pairing of
$\mathcal F^{\rm kin}_{n,g}$, with its prescribed test and coarea transport.
The index $I$ specifies the hard insertions and their helicities. Factorized sources may be
used with a fixed sector cutoff that keeps their support in the regular hard
chamber. Pairing the completed insertion and Weyl term in the same way
defines $\omega_{n,I}^{\rm BP}$. Equation~\eqref{eq:core-finite-marked-Ward}
gives
\begin{equation}
 \omega_{n,I}^{\rm BP}(\boldsymbol s;\delta\boldsymbol s)
      =\dd_{\boldsymbol s} C_{n,I}(\boldsymbol s)[\delta\boldsymbol s].
 \label{eq:marked-coefficient-exact-one-form}
\end{equation}
Integration along the radial path gives
\begin{equation}
 C_{n,I}(\boldsymbol s)=C_{n,I}(0)
      +\int_0^1\omega_{n,I}^{\rm BP}(t\boldsymbol s;\boldsymbol s)\,\dd t,
 \qquad \oint_\gamma\omega_{n,I}^{\rm BP}=0.
 \label{eq:marked-radial-reconstruction}
\end{equation}
Two smooth coefficients with the same finite Ward response and reference
value have a difference with zero derivative throughout the connected chart,
and hence agree there. This neighborhood identity also excludes marked
corrections whose Taylor coefficients vanish only at the reference metric.

The reconstructed moments determine the normalized generating series and its
connected logarithm through the convention in Appendix~\ref{app:distributional}.
This is a coefficientwise statement: every fixed finite set of
hard sectors has a common source neighborhood, while convergence of the full
hard-source series and a radius uniform in multiplicity are not assumed. The result fixes the marked part of the selected transport on its test domain, with the vacuum freedom treated in Appendix~\ref{app:vacuumresponse}. It supplies no extension on the excluded hard--hard coincidence loci.

\subsection{Seven-graviton factorization and a nonzero relative current}
\label{subsec:seven-graviton-current}

The three-soft comparison uses the seven-graviton tree with soft helicities
$++-$. Its four-particle channel contains a full inner five-point amplitude.
The terms regular on a selected nested pole contribute when the third soft
leg approaches the same hard puncture, so the nested residue alone does not
fix this current.

Write the selected hard direction as $(c,d)$, with $d=\bar c$ on the real hard section. Measure the positive soft directions from this point by
$(U,A)$ and $(Z,C)$, and the negative direction by $(V,B)$. Their energies
are $s,e,t>0$, and the signed hard energy is $w$. Put
\begin{equation}
 \begin{gathered}
 K=w+s+e+t,\qquad H=sU+eZ+tV,\\
 G_3=KV-H,\qquad Z_s=KU-H,\qquad Z_e=KZ-H,\\
 \mathcal T=sAZ_s+eCZ_e+tBG_3.
 \end{gathered}
 \label{eq:seven-cluster-normal}
\end{equation}
For a positive-helicity hard leg, tree factorization gives
$\operatorname{Res}_{\mathcal T=0}\mathcal M_7
=\mathcal H_5\mathcal M_{4,\mathcal T}$, where the complete inner factor is
\begin{align}
 \mathcal H_5={}&
 \frac{t^2G_3^6AC}{seK^2UZ(V-U)(V-Z)Z_sZ_e}\notag\\
 &-\frac{t^2G_3^6A(A-C)}{weK^2U(U-Z)(V-Z)Z_eHV}\notag\\
 &-\frac{t^2G_3^6C(A-C)}{wsK^2Z(U-Z)(V-U)Z_sHV}.
 \label{eq:seven-full-inner-five}
\end{align}
This is the full two-by-two Hodges minor in the reduced-spinor tree
normalization. It is symmetric under $(s,U,A)\leftrightarrow(e,Z,C)$.
Supplementary Section~S3.3 derives it and
retains the corresponding conservation distribution. For a general hard
sector the additional anti-MHV inner branch is also allowed; its compatible
division and energy gain are calculated in Supplementary
Sections~S3.7
and~S3.8. That branch vanishes in
the four-hard-graviton MHV reference because its outer four-point helicity sector
vanishes.

For the third-direction limit, keep $e>0$ fixed and choose the soft cutoff so that
$e+s+t<|w|/2$. Set $s=\varepsilon S$, $t=\varepsilon T$, $Z=\varepsilon\zeta$,
$k=w+e$ and $D=SU+TV$, with $S,T>0$ and
$U,V,A,C,U-V,D$ bounded away from zero on a compact chart. The complete pole has the leading current
\begin{equation}
 \begin{aligned}
\mathcal F_{7,\mathcal T}
 ={}&\varepsilon^{-2}\mathcal Q R(\zeta)
 C_1(B-B_0-B_1\zeta)\mathcal F_{4,0}\\
 &+\text{higher scaled orders},
 \end{aligned}
 \label{eq:seven-concentrating-current}
\end{equation}
where
\begin{equation}
 \begin{aligned}
&B_0=-\frac{SAkU-eCD}{TkV},\qquad B_1=-\frac{ewC}{TkV},\\
 &\mathcal Q=\frac{Tk^2V^4C}{SewU^2(V-U)},\\
 &R(\zeta)=\sum_{j=0}^{2}\frac{h_j}{\zeta-r_j},\\
 &(r_0,r_1,r_2)=\left(0,\frac Dw,-\frac De\right),\\
 &(h_0,h_1,h_2)=\frac1D\bigl(-(kA-eC),wA,e(A-C)\bigr).
 \end{aligned}
 \label{eq:seven-three-root-data}
\end{equation}
Here $\mathcal F_{4,0}$ includes the outer conservation distribution; the
coarea density enters its pairing once. The identities
$\sum_jh_j=0$ and $\sum_jh_j\bar r_j=C\bar D/D$ remove the noncompact
Cauchy tail while retaining a finite concentration. Keeping the moving
barred root inside the integral, with $b=d+B$, gives
\begin{equation}
 \begin{aligned}
&\int_{\mathbb C}\frac{R(\zeta)}{b-d-B_0-B_1\zeta}\,\dd^2\zeta\\
 &\quad=\frac\pi{B_1}\left(\bar q_b-\frac{\bar D}{D}q_b\right)R(q_b),\\
 &q_b=\frac{b-d-B_0}{B_1}.
 \end{aligned}
 \label{eq:seven-measured-three-root-current}
\end{equation}
The integral uses the common locally integrable Cauchy product and is
paired in the remaining complex source normal before $q_b$ meets a root
$r_j$. This prescription fixes the extension at that collision without
an additional supported term. The actual source cutoff gives a remainder of order
$\varepsilon\log(1/\varepsilon)$ on that chart. Its adjacent
endpoint and energy-ratio charts require the additional estimates in
Supplementary Sections~S3.5
and~S3.9.

For the barred relative inverse, let $b_j=d+B_0+B_1r_j$ and let $Y$ be the negative barred output. Retain the
prescribed path $b(q)=qY+e^{\ii\theta}\sin(\pi q)/4$. At zero barred
weight define
\begin{equation}
 \mathcal H_0(Y;b_j)=\int_0^1 b'(q)
       \frac{\overline{b(q)}-\bar b_j}{b(q)-b_j}\,\dd q.
 \label{eq:seven-relative-primitive-example}
\end{equation}
The coefficient multiplying $\mathcal F_{4,0}\delta^{(2)}(Z)$ is
$\mathcal Q\pi\sum_jh_j\mathcal H_0(Y;b_j)/\bar B_1$, as in
Supplementary Eq.~\textup{(S3.58)}.
For the following evaluation within this chart,
\begin{equation}
 \begin{aligned}
&w=1,\quad e=\tfrac14,\quad S=T=U=A=1,\\
 &V=3,\quad C=2,\quad c=d=2,\quad Y=5,\quad\theta=\tfrac\pi4.
 \end{aligned}
 \label{eq:seven-relative-example-data}
\end{equation}
we have $B_0=1/5$, $B_1=-2/15$, $\mathcal Q=2025/4$,
$(h_0,h_1,h_2)=(-3/16,1/4,-1/16)$ and
$(b_0,b_1,b_2)=(11/5,5/3,13/3)$.
On the prescribed real source graph, the two positive
external points are $3$ and $4$, and the negative external point is $5$.
The first positive unbarred path at its upper endpoint gives $U=A=1$;
the third positive straight path $a_e(p)=4p$ reaches the hard point at
$p=1/2$. Near that crossing its source coordinate is $4+2Z$, so the
actual barred graph is $C=2+2\bar Z$. The negative barred path is
$b(q)=5q+e^{\ii\pi/4}\sin(\pi q)/4$, with its actual tangent retained.

Take
$0<\varepsilon\leq1/100$ and a smooth source cutoff equal to one on
$|Z|\leq1/5$ and supported in $|Z|<1/4$. Then
$e+s+t\leq27/100<|w|/2$, and the three finite unbarred poles and their
barred roots are
\begin{equation}
 \begin{aligned}
 (Z_0,Z_1,Z_2)
 &=\left(0,\frac{4\varepsilon}{1+2\varepsilon},-16\varepsilon\right),\\
 (b_0^{(\varepsilon)},b_1^{(\varepsilon)},b_2^{(\varepsilon)})
 &=2+\left(\frac{3/4+2\varepsilon}{15/4+2\varepsilon},
       -\frac{1-2\varepsilon}{3+2\varepsilon},
       \frac{7-128\varepsilon}{3}\right).
 \end{aligned}
 \label{eq:seven-finite-graph-roots}
\end{equation}
All barred roots are real and lie in $[5/3,13/3]$. On
$1/4\leq q\leq9/10$, the path has imaginary part at least
$\sin(\pi/10)/(4\sqrt2)$; the complementary endpoint intervals have
larger real separation from the roots. This lower bound is uniform
in the stated finite cutoff, and the remaining angular denominators
are units on the source support. The nonzero coefficient persists
on nearby interior charts of the first positive path.
Direct integration at the limiting concentration gives
\begin{equation}
 \begin{aligned}
&\sum_jh_j\mathcal H_0(5;b_j)\\
 &\quad=0.00204374099982-0.123762586687\,\ii,\\
 &\frac{\pi\mathcal Q}{\bar B_1}
       \sum_jh_j\mathcal H_0(5;b_j)\\
 &\quad=-24.3782221210+1476.26916952\,\ii.
 \end{aligned}
 \label{eq:seven-nonzero-relative-current}
\end{equation}
This is a local relative-current coefficient before the remaining maps, ordered Mellin residues and physical packet pairing. Its variation requires the moving roots and the conjugate bulk term derived below. The value by itself establishes neither the six ordered cubic residues nor their equality with the third metric derivative.

\subsection{Connector variation of the nonholomorphic current}
\label{subsec:connector-bulk-repair}

The current in Eq.~\eqref{eq:seven-measured-three-root-current} depends on both source coordinates. This dependence changes integration by parts along an open connector. Orient $C$ from $a$ to $b$ on one loaded sheet and put $K=(Y-z)^{\gamma-1}$, $h=(\gamma+1)/2$. The Wirtinger derivative $\partial_z$ keeps $\bar z$ fixed, whereas the differential along the real path contains both components. The product rule gives
\begin{equation}
 \begin{aligned}
&\int_C K(\xi\partial_z f+h\xi'f)\,\dd z={}[K\xi f]_a^b\\
 &\qquad+(\gamma-1)\int_C K\left(\frac{\xi'}2+\frac{\xi}{Y-z}\right)f\,\dd z\\
 &\qquad-\int_C\partial_{\bar z}(K\xi f)\,\dd\bar z.
 \end{aligned}
 \label{eq:connector-complete-product-rule}
\end{equation}
For holomorphic $K$ and $\xi$, the last integrand is $K\xi\partial_{\bar z}f$. A smooth projected source also contributes its conjugate derivative. For $K=\xi=1$, $f=\bar z/z$, $C=[1,2]$, both sides equal $-\log2$: the endpoint difference is zero and the conjugate bulk supplies the result.

The seven-graviton pole determines this bulk directly. With $d_0=d+B_0$, $q_b=(b-d_0)/B_1$ and $b_j=d_0+B_1r_j$, suppress the common outer distribution and write
\begin{equation}
 \begin{aligned}
I(b)&=\int\dd^2\zeta\,R(\zeta)C_1(b-d_0-B_1\zeta)\\
 &=\frac\pi{\bar B_1}\sum_jh_j(\bar b-\bar b_j)C_1(b-b_j),\\
 \partial_{\bar b}I(b)
 &=\pi\int\dd^2\zeta\,R(\zeta)
                   \delta^{(2)}(b-d_0-B_1\zeta)\\
 &=\frac\pi{|B_1|^2}R(q_b)
       =\frac\pi{\bar B_1}\sum_jh_jC_1(b-b_j).
 \end{aligned}
 \label{eq:seven-parent-conjugate-current}
\end{equation}
The Jacobian is $|B_1|^2$. At a collision the parent variables are independent complex normals, so the tensor Cauchy product is paired jointly before restriction. The factor $(\bar b-\bar b_j)\delta^{(2)}(b-b_j)$ vanishes when the evaluated expression is differentiated. A cutoff that varies on this chart remains inside the derivative and supplies its product-rule terms. Thus a moving-pole contact before angular integration becomes a Cauchy current in the remaining source coordinate.

For $\omega=Kf\,\dd z$ and $X=\xi\partial_z$ with holomorphic $\xi$ on the sheet, the combined input-density and reciprocal-kernel variation is $\mathcal L_X\omega$. The inverse connector moves with velocity $-X$ and contributes
\begin{equation}
 -\int_C\mathcal L_X\omega
 =-[K\xi f]_a^b
   +\int_C\xi\partial_{\bar z}(Kf)\,\dd\bar z.
 \label{eq:connector-inverse-chain-cancellation}
\end{equation}
The endpoint and conjugate bulk cancel the corresponding input terms; the kernel variation cancels the remaining $(\gamma-1)$ integral. For a smooth real lift the full one-form variation also contains the $\dd\bar z$ component of $\dd\xi$. This is the infinitesimal form of $\int_{\Phi^{-1}C}\Phi^*\omega=\int_C\omega$.

On the actual sine detour of Eq.~\eqref{eq:seven-relative-example-data}, strip $\pi/\bar B_1$ and let $f=\sum_jh_j(\bar b-\bar b_j)/(b-b_j)$. Both endpoints have $f=0$. At $\gamma=1$ and $\xi=1$, the translation terms are
\begin{equation}
 \begin{aligned}
&W:=\int_C\partial_bf\,\dd b\\
 &\quad=-0.15558286339444+0.00298631864484\,\ii,\\
 &B:=\int_C\partial_{\bar b}f\,\dd\bar b\\
 &\quad=+0.15558286339444-0.00298631864484\,\ii,\\
 &W=-B.
 \end{aligned}
 \label{eq:connector-translation-check}
\end{equation}
The primitive in Eq.~\eqref{eq:seven-nonzero-relative-current} and this translation derivative are distinct quantities; both retain the complex path tangent. On the auxiliary segment $[5,6]$, the same roots give the elementary integral
\begin{equation}
 \begin{aligned}
&\int_5^6\partial_bf\,\dd b=\frac3{16}\log\frac{19}{14}-\frac14\log\frac{13}{10}
   +\frac1{16}\log\frac52\\
 &\quad=0.0489361639161\ldots.
 \end{aligned}
 \label{eq:connector-real-segment-check}
\end{equation}

This establishes the physical conjugate derivative on the three-root chart and its cancellation under coherent transport. Global cubic matching additionally requires the independently obtained amplitude Ward variation to induce this completed variation on coalescing roots, adjacent caps and annuli, lower endpoints, cut crossings, the sphere transition, and terms with two or three connectors. The common regulator and the weight derivatives selected by higher poles must be retained in each comparison. Supplementary Section~S3.9.11 states this remaining condition for Eq.~\eqref{eq:core-completed-third}.

\section{Discussion}
\label{sec:discussion}

The two reference matching defects compare each opposite-helicity soft order with the same metric-source Hessian. Their separate vanishing fixes the symmetric response that a charge-commutator calculation leaves open. The quadratic normal packet displays a common nonzero recoil coefficient, while the overlapping packet detects the spin-frame contact multiplying the undifferentiated conservation distribution.

The contact--recoil relation connects these responses to displacement of the hard conservation surface. The contact endpoints fix the normal recoil; the undifferentiated coefficient also contains the hard-amplitude action and coarea variation. The six-graviton calculation tests the normal-derivative and undifferentiated coefficients of the ordering difference, retaining the energy part of the soft generator.

Angular integration can convert a moving-pole contact into a nonholomorphic relative current. The seven-graviton example exhibits this mechanism and the resulting conjugate bulk term in a connector variation. Input, reciprocal kernel and inverse-chain transport cancel that bulk term when they move coherently. The global cubic scattering comparison additionally requires equality with the physical Ward action on coalescing-root, endpoint, cut and multiple-connector overlaps. The local calculation supplies one ingredient of that global comparison.

For the selected finite family, simultaneous transport of external states, kernels and inverse chains gives a Ward identity by change of variables. Its reference value and finite Ward response reconstruct each admitted marked coefficient. This construction is meaningful on the regular-packet domain; an alternative finite-metric ensemble and extensions supported solely on excluded hard--hard coincidences require additional input.

Every marked response contains a hard-source derivative, so adding $P[g]-P[\hg]$ leaves the scattering comparisons unchanged. The vacuum functional, which is independent of the hard sources, therefore remains separate data, including its Euler anomaly coefficient. Appendix~\ref{app:vacuumresponse} gives the scalar-determinant example and its spectral estimator under a specified derivative-order bound. A gravitational vacuum coefficient also requires a boundary action, soft operator and quantum measure.

\section*{Acknowledgments}
During preparation of this manuscript, ChatGPT-5.6-Sol and ChatGPT-6-Astra was used for language editing, literature organization and algebraic cross-checks.

\section*{Conflict of interest}
The authors declare that they have no conflict of interest.

\section*{Data Availability Statement}
This study is theoretical. No data were generated or analyzed.

\appendix
\renewcommand{\thesection}{\Alph{section}}
\renewcommand{\thesubsection}{\thesection.\arabic{subsection}}
\renewcommand{\thesubsubsection}{\thesubsection.\arabic{subsubsection}}
\renewcommand{\theequation}{\thesection.\arabic{equation}}

\section{Source normalization and completion freedom}
\label{app:sourcecompletion}

\subsection{Conventions}
\label{app:conventions}

Distributional derivatives, boundary values and extensions use the conventions of Refs.~\cite{GelfandShilov:1964,Hormander:2003}. We use Euclidean complex coordinates on the celestial cut and the positive scalar Laplacian $\Delta_g=-\nabla_g^2$.
For $g=e^{2\phi}\hg$,
\begin{equation}
 \Delta_g=e^{-2\phi}\hD,
 \qquad
 R_g=e^{-2\phi}(\hR+2\hD\phi).
 \label{eq:weylgeometry}
\end{equation}
The connected functional is $W=\log Z$. Its metric variation is normalized by Eq.~\eqref{eq:firstvariation}; thus $\Theta=(\sqrt g)^{-1}\frac{\delta W}{\delta\phi}$, $T^{\met}=-\pi\frac{\delta W}{\delta\mu}$ and $\bar T^{\met}=-\pi\frac{\delta W}{\delta\bar\mu}$.
The anomaly convention in Eq.~\eqref{eq:anomalydef} assigns $\cW=1$ to one ordinary real scalar determinant. Local contact terms are distributions supported when metric-source points or marked insertions coincide. Separated-point equalities are denoted by a subscript $\sep$.

\subsection{Connected source normalization and Ward kernels}
\label{app:distributional}
\label{app:samechiral}

Interpret $W=\log Z$ coefficientwise in Eq.~\eqref{eq:distributionalZ}. Cumulant products pair disjoint source blocks, whose proper-block conservation loci lie outside the compact four-graviton support. Disconnected terms and their selected local source derivatives therefore pair to zero, leaving the KLT amplitude times the total conservation distribution as the connected tree coefficient.

Metric differentiation acts on the geometric sources, transformed operator sources and additive measure terms in Eq.~\eqref{eq:completefunctional}. Writing $Q^M=(\cG^\alpha,F^I)$ and $W_{\rm add}=W_{\meas}+W_{\ct}+W_{\bare}$, the full response is
\begin{equation}
 W_{,AB}=W_{,MN}Q^M{}_{,A}Q^N{}_{,B}
 +W_{,M}Q^M{}_{,AB}+W_{{\rm add},AB}.
 \label{eq:fullhessian}
\end{equation}
DeWitt indices include the source integrations. The term containing $Q^M{}_{,AB}$ retains the nonlinear source map: local metric dependence produces coincident contacts, while inverse operators can contribute at separated points. In the physical frame $q_{AB}=g_{AB}$, the density ratio is unity and its compensating geometric variations remain in this formula.

The source normalization of Eq.~\eqref{eq:firstvariation} gives a factor $\pi^{-2}$ for a double stress insertion. We collect the Ward operators used in the physical quadratic calculation:

\begin{equation}
 D_w=\sum_i\left[
 \frac{h_i}{(w-z_i)^2}
 +\frac{1}{w-z_i}\partial_i\right].
 \label{eq:Dhol}
\end{equation}

\begin{equation}
 \cV_{zw}=\frac{2}{(z-w)^2}+\frac{1}{z-w}\partial_w,
 \label{eq:Vhol}
\end{equation}

\begin{equation}
 G^X_{\bar z\bar w}
 =\left\langle
 \bar T_{\rm mod}^{\BP}(\bar z)
 \bar T_{\rm mod}^{\BP}(\bar w)X
 \right\rangle_{\sep}
 =\bar\cV_{\bar z\bar w}\bar D_{\bar w}A_X
 +\bar D_{\bar z}\bar D_{\bar w}A_X.
 \label{eq:bpdoublesame}
\end{equation}

The antiholomorphic operators conjugate $D_w$ and $\cV_{zw}$. The centerless tree kernel in Eq.~\eqref{eq:bpdoublesame} includes the nonlinear coordinate response derived in Section~\ref{subsec:nonlinearBeltramiExample}. Local contacts and further metric dependence enter $C^W=C^{W,\rm loc}+C^{W,\rm add}$ once. Derivatives of $W$ give connected coefficients; normalization by a nonzero hard matrix element instead gives a sector-dependent susceptibility and cumulant subtraction.

\subsection{Carrollian source pairing}
\label{app:sourceframes}

Equality of the Carrollian and celestial source couplings fixes the map in Eq.~\eqref{eq:completefunctional}. On $\mathscr I^+\simeq\mathbb R_u\times S^2$ with cut metric $q_{AB}$, a regulated operator transform with retarded-time weight $\mu_B(u)$ has density adjoint
\begin{equation}
 \begin{aligned}
 \widetilde\cO_A(z,\bar z)
 &=\int_{\mathbb R}\dd u\,\mu_B(u)
 K_{\epsilon,L,A}{}^I(u)\cO_I(u,z,\bar z),\\
 \lambda_{\Carr,\epsilon,L}^I(u,z)
 &=\frac{\sqrt g}{\sqrt q}\,\mu_B(u)
 K_{\epsilon,L,A}{}^I(u)J^A(z,\bar z).
 \end{aligned}
 \label{eq:adjointmap}
\end{equation}
The density ratio equals one in the physical frame $q=g$ and varies with $g$ in a fixed reference frame. The full chain rule, Eq.~\eqref{eq:fullhessian}, includes the compensating variations of geometric sources, operator bundles and measure. Nonlinear operator redefinitions use the complete pullback $F^I_{\epsilon,L}[g,J]$.

The cut metric, clock form and null generator form Carrollian geometric sources~\cite{Duval:2014uoa,Hartong:2015xda,Ciambelli:2019lap,Herfray:2021qmp}. A dynamical metric also requires the radiative symplectic form and charge prescription~\cite{Ashtekar:1981bq,Wald:1999wa}, including boundary and corner terms under frame changes~\cite{Compere:2018ylh,Odak:2022vfp,RignonBret:2024}. Generalized superrotations combine the cut metric and the distinct radiative shear in a common phase space~\cite{Sudhakar:2023uan,Campiglia:2024pbn}.

Carrollian current and amplitude constructions represent massless scattering~\cite{Ciambelli:2018wre,Donnay:2022aba,Donnay:2022wvx,Bagchi:2023cen,Mason:2023mti,Liu:2024llk,Ruzziconi:2024kzo,Kulkarni:2025qcx}, with stress tensors supplying pre-transform operator data~\cite{Dutta:2022vkg,Bagchi:2024gnn}. Shadow OPEs~\cite{HimwichPate:2025ShadowOPE,LiuLiuMa:2026ShadowCompletion} and mixed-helicity brackets~\cite{PranzettiSalluce:2025,PranzettiSalluce:2026Mixed} constrain that data; the metric response also differentiates the source pairing. Compact hard packets fix the local distributional pairing within a celestial dictionary whose broader ultraviolet and infrared dependence is discussed in Ref.~\cite{Arkani-Hamed:2020gyp}.

The vacuum functional additionally specifies a variational problem and quantum measure~\cite{Polyakov:1981rd,Henningson:1998gx,deHaro:2000xn,Papadimitriou:2004ap}. A dynamical boundary metric changes the boundary conditions~\cite{Compere:2008us}. The relative determinant in Eq.~\eqref{eq:polyakovnonlocal} has a susceptibility governed by the constrained inverse in Appendix~\ref{sec:induced}. Carrollian Weyl anomalies and logarithmic celestial theories supply further anomaly sectors~\cite{Bagchi:2021gai,Fiorucci:2023lpb}. Loop-corrected Einstein soft Ward identities require a common infrared subtraction and stress prescription~\cite{He:2017fsb,Pasterski:2022djr,Donnay:2020lur}, while the classical memory kernel~\cite{Nguyen:2021dpa} leaves its curved operator, measure and zero-mode domain to be specified. These choices determine vacuum data independently of marked matching.

\subsection{Geometric sources and boundary ensembles}
\label{app:geometric-source-ensembles}

The gravitational variational principle supplies a covariant symplectic potential with boundary ambiguities~\cite{Iyer:1994ys}. Null-boundary phase-space constructions distinguish local charge variations, fluxes and global generators~\cite{Chandrasekaran:2018aop}. For the null Brown--York tensor, fixing a Carroll structure defines the Dirichlet sources; its charges agree with canonical charges for covariant boundary transformations~\cite{ChandrasekaranEtAl:2022NullBrownYork}.

The Carroll contraction of the Poincar\'e group~\cite{Levy-Leblond:1965dsc,SenGupta:1966qer} and its conformal extension~\cite{Duval:2014uva} organize null-boundary geometry. Non-Lorentzian classifications distinguish its metric, preferred directions and connection data~\cite{Bergshoeff:2022eog,Figueroa-OFarrill:2022nui}; Newton--Cartan gravity provides the Galilean framework~\cite{Hartong:2022lsy}. Carroll expansions of general relativity and magnetic Carrollian gravity furnish bulk realizations~\cite{Hansen:2021fxi,Campoleoni:2022ebj}; recent geometric and holographic treatments develop boundary structures~\cite{Ciambelli:2025unn,Bagchi:2025kaleidoscope,Ruzziconi:2026bix,Nguyen:2025zhg}. A source prescription must specify these additional geometric data along with the varying cut metric.

Carrollian partition functions, boundary conformal field theories and universal two-dimensional sectors provide settings for boundary generating functionals~\cite{Poulias:2025eck,Bagchi:2024qsb,Aggarwal:2025hji}.\footnote{The related deformed light-cone preprint of Ref.~\cite{Zeng:2026dlc} was withdrawn by its author on 15 June 2026 because of oversights in its derivations.} Comparing their metric susceptibilities requires field content and source couplings. Here the regulated transform and its density adjoint fix the marked coupling; the vacuum functional is separate data.

At null horizons, Carrollian fluids carry a response and symplectic structure fixed by the gravitational boundary problem~\cite{Donnay:2019jiz,Freidel:2022bai,Freidel:2022vjq}, including nonlinear evolution and external sources~\cite{RedondoYuste:2022czg,Husnugil:2025carroll}. Carroll black holes, higher-curvature limits and restrictions on rotating solutions depend on the bulk theory and contraction~\cite{Ecker:2023uwm,Tadros:2023teq,Tadros:2024,Kolar:2025rotating}. Tantum gravity and phase-space contractions of black-hole thermodynamics give further realizations~\cite{Ecker:2024czh,Xu:2026vpj}. The phase space and boundary ensemble fix the dynamical meaning of the cut-metric response.

Nonrelativistic brane limits transform the bulk geometry and boundary dictionary together~\cite{Lambert:2024uue,Lambert:2024yjk}. String Newton--Cartan holography and its D-brane constructions provide examples~\cite{Fontanella:2024kyl,Fontanella:2024rvn,Fontanella:2026gaq,Guijosa:2023qym,Guijosa:2025mwh}; a proposed Carroll contraction of the anti-de Sitter/conformal field theory (AdS/CFT) correspondence relates this question to flat-space holography~\cite{Fontanella:2025tbs}. Such limits act on both the geometric multiplet and marked sources in Eq.~\eqref{eq:completefunctional}, requiring data beyond the local matching of Einstein tree amplitudes with the BP prescription.

The induced-gravity hypothesis relates metric dynamics to the curvature dependence of quantum fluctuations~\cite{Sakharov:1967pk}. In AdS/CFT~\cite{Maldacena:1997re}, the bulk partition function or on-shell action with prescribed boundary data specifies a boundary generating functional~\cite{Gubser:1998bc,Witten:1998qj}; the Hamilton--Jacobi formulation organizes local counterterms~\cite{deBoer:1999tgo}. In five-dimensional asymptotically AdS gravity, counterterm and Hamiltonian charges can differ by an offset determined solely by the fixed boundary metric~\cite{Hollands:2005wt}, exemplifying geometric normalization beyond the symmetry action. Localized gravity~\cite{Randall:1999vf} and its proposed string realization~\cite{Karch:2001cw}, AdS/BCFT constructions with a boundary conformal field theory~\cite{Takayanagi:2011zk,Fujita:2011fp}, and Jackiw--Teitelboim (JT) gravity coupled to a non-gravitating bath~\cite{Almheiri:2019yqk} illustrate distinct choices of boundary actions and couplings. Here the relative determinant and fixed-area constraint determine the vacuum susceptibility independently of marked scattering.

\subsection{Metric-source completion and marked matching}
\label{subsec:classification}

With fixed operator bundles, field content and anomaly class, we use the locally covariant metric-source completion
\begin{equation}
 \begin{aligned}
W_{\rm class}[g,J]
 ={}&W_{\hg}^{\rm norm}[U_gJ]+\Gamma_{\anom}[g,J]\\
 &+I_{\inv}[g,J]+C_{\loc}[g,J].
 \end{aligned}
 \label{eq:generalcompletion}
\end{equation}

The three additive terms vanish at $g=\hg$: $\Gamma_{\anom}$ specifies the anomaly, $I_{\inv}$ is anomaly-free, and $C_{\loc}$ contains local counterterms. The marked first metric derivatives and matched same-chirality second derivatives of their sum also vanish at the reference metric. This preserves the tested Ward data while leaving symmetric mixed susceptibilities to be fixed by Eq.~\eqref{eq:radialMarkedMatchingCondition}. The associated stress response is
\begin{equation}
 T^{\met}=T^{\trn}+T_{\anom}+T_{\inv}+T_{\loc}.
 \label{eq:covariantcompletedstress}
\end{equation}

Here $T^{\trn}=-\pi\frac{\delta W_{\trn}}{\delta\mu}$, and $T_{\anom}$, $T_{\inv}$ and $T_{\loc}$ are $-\pi\frac{\delta}{\delta\mu}$ applied to the corresponding additive functionals. The radial correction in Appendix~\ref{app:homotopy} belongs to its separately chosen seed prescription. Eq.~\eqref{eq:transportedSoftMetricMatching} identifies the completed insertion with the marked metric response of the selected transported family. Independently, a specified gravitational Dirichlet functional supplies a relative vacuum term $-\Gamma_{\rm D}^{E}[g]+\Gamma_{\rm D}^{E}[\hg]$; its evaluation requires the bulk solution and boundary conditions. It leaves every marked coefficient unchanged.

\subsection{Source transport and symmetric marked response}

\subsubsection{Source connection and Ward normalization}

With density and spin factors included in the source sections, the coordinate pairing is $\langle a,J\rangle=\int\dd^2z\,aJ$ and $\bar\partial(z^{-1})=\pi\delta^{(2)}(z)$. For a compact vector field $v$, the path $\mu_\epsilon=\epsilon\bar\partial v$ has $F_\epsilon=z+\epsilon v+O(\epsilon^2)$ in the common global normalization. Its chiral response is

\begin{equation}
 \begin{aligned}
\left.\frac{\dd A_X}{\dd\epsilon}\right|_0
 &=-\frac{1}{\pi}\int\dd^2w\,\bar\partial v(w)D_wA_X\\
 &=\sum_i\bigl[v(z_i)\partial_i+h_i\partial_i v(z_i)\bigr]A_X.
 \end{aligned}
 \label{eq:uniformizationnormalization}
\end{equation}

The connection $\cA=U_g^{-1}\delta_sU_g$ acts on sources; its transpose $\mathcal B_A=\cA_A^{\mathsf t}$ acts on amplitude coefficients. At the reference metric, $\mathcal B_\mu=-\pi^{-1}D$, $\mathcal B_{\bar\mu}=-\pi^{-1}\bar D$ and $\mathcal B_\phi[\sigma]=-M_\sigma$. Transposing the Maurer--Cartan relation gives

\begin{equation}
 \pi^2\bigl(\delta_{\mu(y)}\mathcal B_{\bar\mu(x)}
 -\delta_{\bar\mu(x)}\mathcal B_{\mu(y)}\bigr)A_X
 =[D_y,\bar D_{\bar x}]A_X.
 \label{eq:dualconnectioncontact}
\end{equation}

The shadow soft-leg terms already cancel the hard Ward commutator, so Eq.~\eqref{eq:dualconnectioncontact} checks the source connection without adding another puncture contact. Input and output map variations leave the spin-frame difference in Eq.~\eqref{eq:transportedContactCancellation} and the symmetric response in Eq.~\eqref{eq:explicitMixedTransportHessian}.

\subsubsection{A covariant marked coupling}
\label{subsec:symmetricMarkedCouplings}

A symmetric source term $H_{\rm mark}=\frac12s^As^B\mathscr H_{AB}[J]$, with $\mathscr H_{AB}[0]=0$, preserves the reference marked coefficients, linear metric responses and vacuum functional while shifting the marked Hessian. The sphere admits a covariant realization. For a scalar hard source $j_\star$, take

\begin{equation}
 H_{\rm curv}[g,j_\star]
 =\kappa_{\rm m}\int_{S^2}\dd^2x\sqrt g\,
 j_\star(x)^2\bigl(R_g-2\bigr)^2.
 \label{eq:markedCurvatureCoupling}
\end{equation}

The coefficient $\kappa_{\rm m}$ has the units fixed by the source convention. A scalar bilinear formed with the bundle metric replaces $j_\star^2$ for spinning source sections. At the unit round reference, $H_{\rm curv}$ and its linear metric variation vanish. For metric tangents $h,k$ its Hessian is

\begin{equation}
 \left.\delta_h\delta_kH_{\rm curv}\right|_{\hg}
 =2\kappa_{\rm m}\int_{S^2}\dd^2x\sqrt{\hg}\,
 j_\star^2\,\delta_hR\,\delta_kR.
 \label{eq:markedCurvatureHessian}
\end{equation}

Variations of the measure and bundle norm multiply the vanishing background curvature factor. For a unit-norm real harmonic $Y$ with $\Delta_{\hg}Y=\lambda Y$, $\lambda=\ell(\ell+1)$ and $\ell\geq2$, Eq.~\eqref{eq:weylgeometry} gives $\delta_YR=2(\lambda-2)Y$. Therefore

\begin{equation}
 \begin{aligned}
H_{\rm curv}[g_\epsilon,j_\star]
 ={}&4\kappa_{\rm m}(\lambda-2)^2\epsilon^2\\
 &\times\int_{S^2}\dd^2x\sqrt{\hg}\,j_\star^2Y^2+O(\epsilon^3),\\
 \left.\frac{\dd^2H_{\rm curv}[g_\epsilon,j_0]}{\dd\epsilon^2}\right|_0
 ={}&8\kappa_{\rm m}j_0^2(\lambda-2)^2\\
 & (j_\star=j_0\ \hbox{constant}).
 \end{aligned}
 \label{eq:markedCurvatureHarmonicExample}
\end{equation}

The fixed-area path gives the same coefficient because the reference linear variation vanishes. At $\ell=2,3$, the Hessians are $128\kappa_{\rm m}j_0^2$ and $800\kappa_{\rm m}j_0^2$.

The traceless tangent

\begin{equation}
 \begin{gathered}
 \begin{aligned}
 h_{ab}^{Y}={}&\sqrt{\frac{2}{\lambda(\lambda-2)}}\\
 &\times\left(\widehat\nabla_a\widehat\nabla_b
 +\frac12\hg_{ab}\Delta_{\hg}\right)Y,
 \end{aligned}\\
 \int_{S^2}\dd^2x\sqrt{\hg}\,h_{ab}^{Y}h_Y^{ab}=1.
 \end{gathered}
 \label{eq:markedCurvatureBeltramiTangent}
\end{equation}

has $\delta_{h^Y}R=\sqrt{\frac{\lambda(\lambda-2)}{2}}\,Y$. Eq.~\eqref{eq:markedCurvatureHessian} then gives the marked Beltrami Hessian $\kappa_{\rm m}j_0^2\lambda(\lambda-2)$, equal to $24\kappa_{\rm m}j_0^2$ and $120\kappa_{\rm m}j_0^2$ at $\ell=2,3$. This coupling changes the symmetric metric susceptibility and marked Weyl data while preserving reference amplitudes and linear matching. The normalized transport sets its coefficient to zero as part of the chosen source prescription. A nonzero term of this form changes the finite Ward response and therefore defines a different marked completion.

For a closed matching defect $\mathfrak E$, the relative functional is

\begin{equation}
 W_{\intc}[s,J]-W_{\trn}[s,J]
 =\int_0^1\dd t\int_{S^2}\dd^2x\,
 s^A(x)\mathfrak E_A(x;ts,J),
 \label{eq:radialMarkedMatchingCondition}
\end{equation}

The seed is $\delta_sW_{\trn}+\mathfrak E$, and the two functionals have the same reference value. Their equality is equivalent to vanishing of the radial pairing throughout the source neighborhood. For the shifted seed $\delta_sW_{\trn}+\delta_sH_{\rm mark}$, the defect is $\mathfrak E_A=s^B\mathscr H_{AB}[J]$ and the integral returns $H_{\rm mark}$. Thus the closed response still carries symmetric marked information to be fixed by finite-metric matching.

\subsection{Radial completion and its physical matching}
\label{app:homotopy}

On a star-shaped source neighborhood, the Poincar\'e homotopy defines
\begin{equation}
 \begin{aligned}
W_{\intc}[s,J]={}&W_0[J]+\int_0^1\dd t\int_\Sigma\dd^2x\,\\
 &\times s^A(x)\omega_{{\rm seed},A}(x;ts,J),\\
 \delta_sW_{\intc}={}&\boldsymbol\omega_{\intc}.
 \end{aligned}
 \label{eq:radialfunctional}
\end{equation}
Writing $\cF=\delta_s\boldsymbol\omega_{\rm seed}$ gives $\omega_{\intc,A}=\omega_{{\rm seed},A}-\int_0^1\dd t\,t\int\dd^2y\,s^B(y)\cF_{BA}(y,x;ts)$. This prescription retains the reference linear response and symmetrizes the ordered double response. For $\omega_\mu=-T/\pi$, the stress corrections satisfy
\begin{equation}
 \begin{aligned}
\left.\frac{\delta N_\mu(1)}{\delta\bar\mu(2)}\right|_0
 &=-\frac{\pi}{2}\cF^{\rm seed}_{\mu\bar\mu}(1,2),\\
 \left.\frac{\delta N_{\bar\mu}(2)}{\delta\mu(1)}\right|_0
 &=+\frac{\pi}{2}\cF^{\rm seed}_{\mu\bar\mu}(1,2).
 \end{aligned}
 \label{eq:mixedcompletionterms}
\end{equation}
The source origin, path and coordinates select the primitive, whose symmetric susceptibility includes the covariant source-map response. Eq.~\eqref{eq:transportedMixedCurvatureZero} makes the correction derivatives vanish when the physical matching defects agree; matching requires each to vanish, with symmetric marked condition Eq.~\eqref{eq:radialMarkedMatchingCondition}.

Preserving a matched hard factorization pole requires
\begin{equation}
 \int_0^1\dd t\,t\int_\Sigma\dd^2y\,
 s^B(y)\operatorname{Res}_{P_I^2=0}
 \cF^{\rm seed}_{BA}(y,x;ts,J)=0,
 \label{eq:radialFactorizationCondition}
\end{equation}
where residue extraction commutes with radial integration on the selected unpinched neighborhood. Crossing and the real structure require an equivariant seed, path and distributional extension; a pure-chirality hierarchy is preserved when its seed curvature vanishes on that source subspace.

The mixed shadow OPEs~\cite{PranzettiSalluce:2025,PranzettiSalluce:2026Mixed} supply singular and contact data. Metric matching also varies both modified-shadow maps and their relative domains, while Eq.~\eqref{eq:standardShadowHomogeneousWard} controls the regular remainder. For the selected transported family in Section~\ref{sec:core-constructed-matching}, the radial integral reconstructs the marked coefficients from the finite Ward response in Eq.~\eqref{eq:transportedSoftMetricMatching}. Other seeds require the matching and factorization tests above.

\section{Contact, recoil and Mellin continuation}
\label{app:contactrecoil}

\subsection{Contact energy integral and spatial normal coefficients}
\label{app:mixedsoft}

For the physical contact in Eq.~\eqref{eq:contactRatioProfile}, define $\xi_i=1-\frac{A_i}{B_i}$. The Mellin integral over the energy ratio is determined by

\begin{equation}
 \begin{aligned}
 I_0^{(i)}(\alpha,\beta)
 &=\frac{\mathrm B(\alpha,\beta)}{B_i}
 {}_2F_1(1,\alpha;\alpha+\beta;\xi_i),\\
 I_1^{(i)}(\alpha,\beta)
 &=\frac{\mathrm B(\alpha+1,\beta)}{B_i}
 {}_2F_1(1,\alpha+1;\alpha+\beta+1;\xi_i).
 \end{aligned}
 \label{eq:contactHypergeometricIntegrals}
\end{equation}

\begin{equation}
 \begin{aligned}
&\int_0^1\dd r\,r^{\alpha-1}(1-r)^{\beta-1}F^\mu(r)\\
 &\qquad=\sum_i C_i\left[q_t^\mu I_0^{(i)}
 +(q_s^\mu-q_t^\mu)I_1^{(i)}\right].
 \end{aligned}
 \label{eq:contactHypergeometricProfile}
\end{equation}

Euler's representation applies for $\operatorname{Re}\alpha,\operatorname{Re}\beta>0$ with an unpinched denominator. The decomposition $r q_s+(1-r)q_t=q_t+r(q_s-q_t)$ gives the two terms, and $A_ir+B_i(1-r)=B_i(1-\xi_i r)$ fixes their branch. Meromorphic continuation and the radial integral in Eq.~\eqref{eq:contactMellinIntegral} give the mixed divisor and endpoint residues. The angular variables remain arguments of the subsequent modified-shadow integrals.

The Lorentz identity in Eq.~\eqref{eq:contactLorentzRemainder} applies to every momentum component. At the real channel in Table~\ref{tab:physicalKinematics}, the spatial components of Eq.~\eqref{eq:contactConormalIdentity} are

\begin{equation}
 \begin{aligned}
 \frac{A^1_{\mathrm{full}}}{\rho_*E^3}
 &=\frac{4130676}{4829513}-\ii\frac{8554968}{4829513},\\
 \frac{A^2_{\mathrm{full}}}{\rho_*E^3}
 &=\frac{23943141}{4829513}+\ii\frac{72128412}{4829513},\\
 \frac{A^3_{\mathrm{full}}}{\rho_*E^3}
 &=\frac{1904571}{1486004}+\ii\frac{2915043}{371501}.
 \end{aligned}
 \label{eq:contactSpatialNormalCoefficients}
\end{equation}

Together with the time component in Eq.~\eqref{eq:fullNormalCoefficientValue}, these coefficients determine the pairing with $\nu_\mu n^\mu\chi(n)\eta(\vartheta)$. Any real covector with nonzero contraction detects the normal response after fixing the phase and restricting the tangential support. The change of normal coordinates described after Eq.~\eqref{eq:canonicalnormaljets} preserves its nonvanishing. The physical amplitude tests in the main text evaluate this response together with the undifferentiated-delta coefficient.

\subsection{General recoil and joint factorization coordinates}
\label{app:nmhvRecoilResolution}

\subsubsection{Recoil for a general total soft momentum}

Keep $z_2=-2$, $z_3=\ii/2$, $z_4=-2\ii$,
$\operatorname{Im}z_1=0$ and $\omega_4=E/5$ fixed, and write $z_1=x$.
For the incoming signs $(-,-,+,+)$, conservation is
\begin{equation}
 \begin{aligned}
&-\omega_1q(x,x)-\omega_2q(-2,-2)\\
 &\quad+\omega_3q(\ii/2,-\ii/2)+\frac E5q(-2\ii,2\ii)=-Q.
 \end{aligned}
 \label{eq:nmhvGeneralConservation}
\end{equation}
The $P^2$, $P^0+P^3$, $P^1$ and $P^0-P^3$ equations successively give
\begin{equation}
 \begin{gathered}
\omega_3=\frac{4E}{5}-Q^2,\\
 \omega_1+\omega_2=T=E-Q^2+\frac{Q^0+Q^3}{2},\\
 \omega_1(x+2)=U=2T+\frac{Q^1}{2},\\
 \begin{aligned}
 \omega_1(4-x^2)=V
 ={}&4T-\frac14\left(\frac{4E}{5}-Q^2\right)\\
 &-\frac{4E}{5}-\frac{Q^0-Q^3}{2}.
 \end{aligned}
 \end{gathered}
 \label{eq:nmhvRecoilLinearEquations}
\end{equation}
Dividing the last two equations gives $2-x=V/U$. Define
\begin{equation}
 \begin{aligned}
 U&=2E+Q^0+\frac{Q^1}{2}-2Q^2+Q^3,\\
 W&=4U-V=5E+\frac52Q^0+2Q^1-\frac{17}{4}Q^2+\frac32Q^3.
 \end{aligned}
 \label{eq:nmhvRecoilUW}
\end{equation}
The solution and the pure coordinate coarea factor are
\begin{equation}
 \begin{gathered}
x=\frac{2U-V}{U},\qquad\omega_1=\frac{U^2}{W},\\
 \omega_2=T-\frac{U^2}{W},\qquad\omega_3=\frac{4E}{5}-Q^2,\\
 \begin{aligned}
 \det\frac{\partial P}{\partial(\omega_1,\omega_2,\omega_3,x)}
 &=-4\omega_1(x+2)^2\\
 &=-4W,
 \end{aligned}\\
 \rho(\vartheta_*,-Q)=\frac1{4W},\\
 \frac{\rho(\vartheta_*,-Q)}{\rho_*}=\frac{5E}{W},
 \quad\rho_*=\frac1{20E}.
 \end{gathered}
 \label{eq:nmhvGeneralRecoil}
\end{equation}
Here $Q^2$ denotes the Cartesian component with index $2$, while the Lorentz norm is written $Q_\mu Q^\mu$.
For $\max_\mu|Q^\mu|<E/100$,
$|U-2E|<9E/200$ and $|W-5E|<41E/400$. The denominators therefore do not
vanish in this complex polydisc. On its real part, the hard energies remain
positive; for example with $\varepsilon=1/100$,
\begin{equation*}
 \omega_2\ge E\left[1-2\varepsilon
 -\frac{(2+\frac92\varepsilon)^2}{5-\frac{41}{4}\varepsilon}\right]>0.
\end{equation*}
Square-root spinor branches are fixed from the physical point. Substituting
\begin{equation*}
 Q=\left(3\tau_s+\frac{21}{4}\tau_t,\,
 2\tau_s+4\tau_t,\,2\tau_s+\tau_t,\,
 -\tau_s-\frac{13}{4}\tau_t\right)
\end{equation*}
and $E=1$ reproduces Eq.~\eqref{eq:realNMHVRecoil} and
Eq.~\eqref{eq:NMHVRecoilDensity}. Zeros of $U$ or $W$ outside this domain are
failures of this chart and must not be assigned physical propagator residues.

At nearby tangential positions, the analytic implicit function theorem
applies uniformly on a smaller compact neighborhood. If $X^a(\vartheta,Q)$
denotes the eliminated coordinates,
$J^\mu{}_a=\partial_aP^\mu$ and $A=J^{-1}$, then
\begin{equation}
 X^a{}_{,\mu}=-A^a{}_\mu,\qquad
 X^a{}_{,\mu\nu}
 =-A^a{}_\rho P^\rho{}_{,bc}A^b{}_\mu A^c{}_\nu.
 \label{eq:nmhvImplicitRecoilJets}
\end{equation}
Higher recoil derivatives follow by differentiating $P(X,\vartheta)+Q=0$ and isolating
$JX_{,\mu_1\cdots\mu_k}$; the other terms are the partitions
of the derivative indices into at least two nonempty blocks, contracted
with the corresponding derivatives of $P$. For the coarea density,
$\partial_{Q^\mu}\log\rho=-\operatorname{tr}(J^{-1}\partial_{Q^\mu}J)$.
On the selected slice, writing $W=5E+c_\mu Q^\mu$ gives
\begin{equation*}
 \partial_Q^\nu\rho
 =\frac{(-1)^{|\nu|}|\nu|!c^\nu}{4W^{|\nu|+1}}.
\end{equation*}
Any hard Mellin factors or other measures in the packet convention supply
additional, separately differentiated factors.

\subsubsection{Physical residues, real bounds and joint coordinates}

With the reduced spinors of the main text,
$\sigma_I=\det\sum_{i\in I}\lambda_i\widetilde\lambda_i=d_I/4$.
In this normalization factorization reads
\begin{equation}
 \begin{gathered}
 \lim_{\sigma_I\to0}\sigma_I\mathcal M_6
 =\sum_{h=\pm2}\mathcal M_L(\ldots,-K^h)
                       \mathcal M_R(K^{-h},\ldots),\\
 \mathcal M_4(a^-,b^-,c^+,d^+)
 =\frac{\langle ab\rangle^7[ab]}
 {\langle ac\rangle\langle ad\rangle
  \langle bc\rangle\langle bd\rangle\langle cd\rangle^2}.
 \end{gathered}
 \label{eq:nmhvFactorizationNormalization}
\end{equation}
Thus the coefficient of $1/d_I$ is four times the displayed product.
Writing $s^+,t^-$ for the two soft legs, for $I=\{i,s,t\}$ the products are
\begin{equation*}
 \begin{gathered}
\begin{cases}
 \begin{gathered}
 \mathcal M_4(i^-,t^-,s^+,(-K)^+)\\
 \times\mathcal M_4(K^-,j^-,k^+,l^+),\quad i\text{ negative},
 \end{gathered}\\[4pt]
 \begin{gathered}
 \mathcal M_4(t^-,(-K)^-,i^+,s^+)\\
 \times\mathcal M_4(j^-,k^-,K^+,l^+),\quad i\text{ positive}.
 \end{gathered}
 \end{cases}
 \end{gathered}
\end{equation*}
In the upper line $j$ is the other negative hard leg.
Both statements are the usual tree factorization rule
\cite{Benincasa:2007qj,Hodges:2012ym} with the present normalization.

The four mixed channel residues are fixed by these products in the normalization of Eq.~\eqref{eq:kltHardCoefficient}. Regular transverse coefficients and intersections with another divisor are retained in the joint expansion.

The real-angular bound and its uniform recoil control are proved in
Supplementary Section~S2.1. With
$d_{i+-}=4\sigma_{i+-}$, the definite quadratic form in Supplementary
Eqs.~\textup{(S2.3)} and
\textup{(S2.4)} excludes noncollinear real roots throughout
the chosen soft neighborhood. The complex relative chains also meet
mixed roots. Their compatible pair and mixed coefficients are retained
together in Supplementary
Eq.~\textup{(S2.30)}; the cap and annular
calculations following it supply the joint energy and angular bounds.

The lower conservation-normal coefficients require derivatives of the recoil map, coarea density and
packet as well as the contact at highest normal-derivative order. Set
$s=\lambda r$, $t=\lambda(1-r)$ and $Q=\lambda q(r)$. Writing the
pulled-back amplitude and density as
$\mathcal M_6=\lambda^{-2}m_{-2}+\lambda^{-1}m_{-1}+m_0+O(\lambda)$ and
$\rho=\rho_0+\lambda\rho_1+\lambda^2\rho_2+O(\lambda^3)$, with all
$m_j$ including their required recoil, their degree-zero coefficients are
\begin{equation}
 \begin{gathered}
[\rho\mathcal M_6]_{\lambda^0}
 =\rho_0m_0+\rho_1m_{-1}+\rho_2m_{-2},\\
 \begin{aligned}
 &[\rho\mathcal M_6h(-Q)]_{\lambda^0}\big|_{h(n)=n^\mu}\\
 &\qquad=-q^\mu(r)(\rho_0m_{-1}+\rho_1m_{-2}),
 \end{aligned}\\
 q(r)=rq_++(1-r)q_-.
 \end{gathered}
 \label{eq:nmhvRadialCoefficientBookkeeping}
\end{equation}
These expressions preserve all ratio-dependent terms at the selected
normal order, including the contributions through
coordinate Mellin poles.

\subsection{Boundary expansions and ordered Mellin residues}
\label{app:nmhvFiniteResidues}

\subsubsection{Taylor subtraction and residue extraction}

For one boundary coordinate put $L=a\alpha+b\beta+c$ and subtract the
Taylor polynomial of a smooth holomorphic family $F$ through degree $N$.
Elementary integration and $k$ derivatives with respect to $L$ give
\begin{equation}
 \begin{aligned}
 \int_0^1\dd t\,t^{L-1}(\log t)^kF(t)
 ={}&\sum_{n=0}^{N}
 \frac{(-1)^kk!}{(L+n)^{k+1}}\frac{\partial_t^nF(0)}{n!}\\
 &+\int_0^1\dd t\,t^{L-1}(\log t)^kR_N(t).
 \end{aligned}
 \label{eq:nmhvOneFaceSubtraction}
\end{equation}
Choose $N$ so that $\operatorname{Re}L+N+1>0$ uniformly in the weight
neighborhood. The remainder is holomorphic there. A face can be resonant
at the soft point only if $c+n=0$; angular selection rules or numerator
zeros can remove the candidate pole.

In product coordinates let $T_j$ denote these Taylor operations and
$R_j=1-T_j$. Operators in distinct coordinates commute, and
\begin{equation}
 F=\sum_{S\subset\{1,\ldots,N_c\}}
 \left(\prod_{j\in S}T_j\right)
 \left(\prod_{j\notin S}R_j\right)F.
 \label{eq:nmhvFaceOverlapSubtraction}
\end{equation}
For two coordinates this is
$F=T_1F+T_2F-T_1T_2F+R_1R_2F$; the overlap is subtracted once.
Integrating every Taylor term leaves a finite sum of
holomorphic numerators divided by products of affine linear factors.
Partial remainders retain poles from the other faces; only the fully
subtracted bulk has all such poles removed. Intersecting original strata
must be resolved into the product charts used here.

Choose the subtraction orders such that
$\operatorname{Re}L_j+N_j+1\ge\delta_j>0$ on a closed weight bidisc.
The integral Taylor remainder gives, for every finite weight multi-index
$\nu$,
\begin{equation}
 \begin{gathered}
|\partial_{\alpha,\beta}^{\nu}\mathcal I_{\mathrm{rem}}|
 \le C_\nu\prod_j t_j^{-1+\delta_j}(1+|\log t_j|)^{M_\nu},\\
 \int_0^1 t^{-1+\delta}|\log t|^m\dd t
 =\frac{m!}{\delta^{m+1}}.
 \end{gathered}
 \label{eq:nmhvHolomorphicRemainderBound}
\end{equation}
Dominated complex differentiation gives holomorphy, also in compact-test seminorms for distribution-valued spectators when the singular pullbacks are defined and satisfy the stated uniform bounds.

After absorbing factors nonzero at the origin, expand the numerator in
Eq.~\eqref{eq:nmhvPolarTerm} as $H=\sum_{d\ge0}H_d$. A term
$H_d/\prod_j\ell_j^{p_j}$ has homogeneous degree $d-P_{\mathrm{pol}}$.
The double residue selects degree $-2$, hence only
$d=P_{\mathrm{pol}}-2$. Equivalently the corresponding two-form scales
as $t^{d-P_{\mathrm{pol}}+2}$ under simultaneous rescaling of both nested
residue circles, while its residue is unchanged. Nested circles can be
chosen to avoid the finite set of polar slopes. The convergent Taylor
series on a smaller bidisc justifies termwise extraction.

The selected rational term is $\beta^{-2}f(\alpha/\beta)$.
The chamber $|\alpha|\ll|\beta|$ selects $\Res_0f(z)\dd z$;
the opposite chamber selects $-\Res_\infty f(z)\dd z$.
The ordinary residue theorem on the projective line gives
Eq.~\eqref{eq:nmhvProjectiveResidues}. Coincident slopes must be grouped
after numerator cancellations. A pole of order $p$ at $z_*$ contributes
\begin{equation*}
 \Res_{z=z_*}f(z)\dd z
 =\left.\frac1{(p-1)!}\frac{\dd^{p-1}}{\dd z^{p-1}}
 [(z-z_*)^pf(z)]\right|_{z=z_*}.
\end{equation*}
Higher total pole degree selects higher weight derivatives of the
numerator and the angular maps. For $k$ soft weights the same
homogeneity argument selects degree $P_{\mathrm{pol}}-k$; the ordered
extractions are then multivariable flag residues. The local three-soft currents and the remaining global matching condition are treated in Section~\ref{sec:core-constructed-matching}.

\subsubsection{Boundary values and extension freedom}

Regulator finite parts and soft residues can differ:
\begin{equation*}
 \Res_{\alpha=0}\FP_{\eta=0}\frac1{\alpha+\eta}=1,
 \qquad
 \FP_{\eta=0}\Res_{\alpha=0}\frac1{\alpha+\eta}=0
\end{equation*}
when the finite part in the left expression is evaluated at generic nonzero $\alpha$.
Direct continuation from a common convergence domain avoids this extra
choice where that domain exists. Auxiliary powers otherwise define a
specified extension whose agreement with the desired physical boundary
value must be checked.

For a fixed off-stratum distribution and a smooth collision stratum $S$
of codimension $c$, a finite normal scaling-degree bound $s$ restricts
the difference of two extensions to
\begin{equation}
 \Delta T=\sum_{|\nu|\le\lfloor s-c\rfloor}
 c_\nu(y)\partial_n^\nu\delta(n).
 \label{eq:nmhvExtensionFreedom}
\end{equation}
This is the standard extension result on submanifolds
\cite{Brunetti:1999ps}. Below codimension the sum is empty and the extension
with that bound is unique. Intersections require compatible successive
extensions. The bound must be established for the actual continued
remainder; agreement away from $S$ does not exclude an undetermined delta
term. Moreover a term supported on an internal integration stratum can
push forward to a nonlocal kernel in external source coordinates.

Changing a relative path around a simple physical root is a different
operation from changing an extension of the same off-stratum distribution.
If $\sigma_I(u_I)=0$, $\partial_u\sigma_I(u_I)\ne0$, and all other
factors are regular there, a positive loop changes the integral by
\begin{equation}
 \Delta_C\int_C\dd u\,\mathcal K(u)\mathcal M_6(u)
 =2\pi\ii\left.
 \frac{\mathcal K(u)\sum_h\mathcal M_L\mathcal M_R}
 {\partial_u\sigma_I(u)}\right|_{u=u_I}.
 \label{eq:nmhvFactorizationPeriod}
\end{equation}
The derivative is the total derivative on the recoil chart. Generic BP
powers require the corresponding twisted-cycle monodromy factors. If the
root reaches an endpoint or another divisor, a joint local calculation
replaces this formula. Landau conditions can help locate candidate pinches
only when their real boundary-value and contour-deformation hypotheses
hold \cite{Collins:2020euz}; they do not classify arbitrary complex relative
chains without those checks.

\section{Relative-current geometry and source variations}
\label{app:relativegeometry}

\subsection{Relative inverses and completion data}
\label{app:contourinverse}
\label{sec:completiondata}
\label{subsec:covariantshadow}

The general theory of generalized matrix and Hilbert-space inverses is developed in Ref.~\cite{BenIsraelGreville:2003}.

The BP relative inverse and the smooth closed-sphere inverse act on different domains. With $\eth\,{}_sY_{\ell m}=\sqrt{(\ell-s)(\ell+s+1)}\,{}_{s+1}Y_{\ell m}$ and $\bar\eth\,{}_sY_{\ell m}=-\sqrt{(\ell+s)(\ell-s+1)}\,{}_{s-1}Y_{\ell m}$, set $\alpha_\ell=\sqrt{(\ell-1)(\ell+2)}$. The Moore--Penrose inverse $G_+$ of $\eth:{}_1\mathcal H\to{}_2\mathcal H$ obeys
\begin{equation}
 G_+\,{}_2Y_{\ell m}=\frac{1}{\alpha_\ell}\,{}_1Y_{\ell m},
 \qquad
 \bar\eth^3G_+\,{}_2Y_{\ell m}
 =-\ell(\ell+1)\,{}_{-2}Y_{\ell m}.
 \label{eq:smoothInverseModes}
\end{equation}

The descendant-shadow convention $\eth\mathscr S_+=-\frac12\bar\eth^3$ then gives $\mathscr S_+\,{}_2Y_{\ell m}=\frac12\ell(\ell+1)\,{}_{-2}Y_{\ell m}$, and hence
\begin{equation}
 \left(\mathscr S_++\frac12\bar\eth^3G_+\right)
 {}_2Y_{\ell m}=0,
 \qquad \ell\geq2,
 \label{eq:smoothModifiedCancellation}
\end{equation}

The BP Ward representative is therefore carried by the relative endpoint, puncture and monodromy data.

For comparison, the smooth Moore--Penrose inverse $D^+$ satisfies $DD^+=1-\widetilde\Pi$ and $D^+D=1-\Pi$. Differentiating the Moore--Penrose identities in a common isometric trivialization, at constant kernel and cokernel ranks, gives the three-term formula familiar from the constant-rank matrix case \cite{GolubPereyra:1973}
\begin{equation}
 \begin{split}
 \delta D^+={}&-D^+(\delta D)D^+
 +D^+D^{+\dagger}(\delta D^\dagger)(1-DD^+)\\
 &+(1-D^+D)(\delta D^\dagger)D^{+\dagger}D^+.
 \end{split}
 \label{eq:moorePenroseVariation}
\end{equation}

The projector terms distinguish a moving inverse domain from a fixed inverse. For the BP prescription, the corresponding contour, endpoint and relative-projector data are transported together by Eq.~\eqref{eq:transportedBPmap}.

\subsection{Mixed Ward contacts and global matching conditions}
\label{app:mixedBPdomains}

\subsubsection{Finite primary frame and the symmetric Hessian}

The natural coframe of the area-normalized source metric is
\begin{equation}
 e^+=\frac{e^\phi\sqrt{\rho_{\hg}}}{\sqrt{1-\mu\bar\mu}}
          (\dd z+\mu\dd\bar z),\qquad
 e^-=\frac{e^\phi\sqrt{\rho_{\hg}}}{\sqrt{1-\mu\bar\mu}}
          (\dd\bar z+\bar\mu\dd z).
 \label{eq:mixedNaturalCoframe}
\end{equation}
The uniformization Weyl factor satisfies
\begin{equation*}
 e^{2\sigma_g}=
 \frac{e^{2\phi}\rho_{\hg}(z)}
 {(1-\mu\bar\mu)(\partial F)(\bar\partial\bar F)\rho_{\hg}(F,\bar F)}.
\end{equation*}
Combining the Weyl weight and spin-frame ratio and absorbing $\rho_{\hg}^{\Delta/2}$ gives Eq.~\eqref{eq:explicitMixedPrimaryFrame}. In particular the exponent of $1-\mu\bar\mu$ is positive. On independent source directions $F=z+\epsilon v+O(\epsilon^2)$ and $\bar F=\bar z+\eta\bar v+O(\eta^2)$ have no mixed coefficient. Direct differentiation therefore contains the argument shifts and Jacobians, but no derivative of $\bar v$ by $v$. The latter is present in $\mathscr D_v\overline{\mathscr D}_{\bar v}$ and must be subtracted, giving Eq.~\eqref{eq:explicitMixedTransportHessian}. The equality of its two lines is the operator identity
\begin{equation}
 \begin{aligned}
&[\mathscr D_v,\overline{\mathscr D}_{\bar v}] =\sum_i\Bigl\{
 v_i(\bar\mu_{b,i}\bar\partial_i+\bar h_i\bar\partial\bar\mu_{b,i})\\
 &\hspace{30mm}-\bar v_i(\mu_{a,i}\partial_i+h_i\partial\mu_{a,i})\Bigr\}.
 \end{aligned}
 \label{eq:mixedSmearedCommutator}
\end{equation}
The energy Euler operators commute with source profiles, so this derivation also holds before hard Mellin pairing.

\subsubsection{Distributional soft-leg terms and map variation}
\label{subsec:explicitMixedContacts}

After the positive-helicity residue and its standard shadow, the remaining negative soft leg has antiholomorphic weight one. Its contribution is
\begin{equation}
 J_-(x,y)=\frac3\pi\int\dd^2t\,\frac1{(y-t)^4}
 \left[\frac1{(\bar x-\bar t)^2}
       +\frac1{\bar x-\bar t}\bar\partial_t\right]a_-(t).
 \label{eq:mixedSoftLegIntegral}
\end{equation}
The bracket is $\bar\partial_t[a_-(t)/(\bar x-\bar t)]$. Integrating by parts and using $\bar\partial_t(y-t)^{-4}=-(\pi/6)\partial_t^3\delta^{(2)}(t-y)$ gives Eq.~\eqref{eq:mixedSoftLegContactKernels}; the other helicity is its conjugate. The integration covers the whole soft plane in the selected sphere chart, with Cauchy finite parts. The rational Ward coefficients and kernels have vanishing affine boundary terms, and chart changes use the global Ward normalization.

To display the puncture contacts, write
\begin{equation*}
 a_-(y)=\sum_i\frac{-(y-z_i)^2\partial_i+2h_i(y-z_i)}
                         {\bar y-\bar z_i}A_X.
\end{equation*}
Away from the intersection $x=y=z_i$, partial fractions and the Cauchy identity yield
\begin{equation}
 \begin{aligned}
J_-^{\rm punct}
 ={}&\pi\sum_i\Bigl[-\frac{h_i}{\bar x-\bar z_i}
                 \partial_y\delta^{(2)}(y-z_i)\\
 &\hspace{8mm}+\frac{\delta^{(2)}(y-z_i)}{\bar x-\bar z_i}\partial_i\Bigr]A_X,\\
 J_+^{\rm punct}
 ={}&\pi\sum_i\Bigl[-\frac{\bar h_i}{y-z_i}
                 \bar\partial_x\delta^{(2)}(x-z_i)\\
 &\hspace{8mm}+\frac{\delta^{(2)}(x-z_i)}{y-z_i}\bar\partial_i\Bigr]A_X.
 \end{aligned}
 \label{eq:mixedExplicitPunctureTerms}
\end{equation}
Consequently
\begin{equation}
 (J_--J_+)_{\rm punct}=[D_y,\bar D_{\bar x}]A_X,
 \label{eq:mixedWardContactCommutator}
\end{equation}
which cancels the hard commutator in the operator difference. Away from the hard punctures the diagonal term is
\begin{equation}
 \begin{aligned}
J_-^{\rm diag}={}&\frac\pi2\Bigl[
 a_-(x)\partial_y^2\delta^{(2)}(y-x)\\
 &+(\partial_xa_-(x))\partial_y\delta^{(2)}(y-x)\\
 &+(\partial_x^2a_-(x))\delta^{(2)}(y-x)\Bigr].
 \end{aligned}
 \label{eq:mixedExplicitDiagonalTerm}
\end{equation}
The unsplit kernels in Eq.~\eqref{eq:mixedSoftLegContactKernels}, with a common extension at intersecting loci, are the fundamental expressions. The separated decomposition alone does not determine a triple-intersection contact.

For the standard positive-helicity shadow map, $\partial\mathscr S_+=-\bar\partial^3/2$ and $(v\partial+\partial v)f=\partial(vf)$. Eq.~\eqref{eq:mixedMapIntertwinerVariation} thus gives
\begin{equation*}
 \begin{gathered}
(\delta_v\mathscr S_+)f
 =\frac12\bar\partial^3(vf)-\frac12v\bar\partial^3f,\\
 \begin{aligned}
 \delta_v\mathscr S_+(x,u)
 ={}&\frac12[v(u)-v(x)]\\
 &\times\bar\partial_x^3\delta^{(2)}(x-u),
 \end{aligned}
 \end{gathered}
\end{equation*}
which is Eq.~\eqref{eq:explicitCrossShadowVariation}. The variation is distributional even though the separated kernel's holomorphic derivative vanishes.

Let $j_\pm=\pi^{-2}\int\dd^2x\dd^2y\,\bar\mu_b(x)\mu_a(y)J_\pm(x,y)$. Subtracting the positive-first response from the negative-first response gives
\begin{equation}
 \begin{aligned}
 \kappa_{\rm op}&=[\mathscr D_v,\overline{\mathscr D}_{\bar v}]A_X+j_+-j_-,\\
 \kappa_{\rm map}
 &=-j_++j_-+\mathcal O,\\
 \mathcal O
 &=-\frac1\pi\int\dd^2x\,\bar\mu_bv\partial_x\bar D_{\bar x}A_X
   +\frac1\pi\int\dd^2y\,\mu_a\bar v\bar\partial_yD_yA_X.
 \end{aligned}
 \label{eq:mixedInputOutputCancellation}
\end{equation}
Evaluating the output terms without dropping derivatives of source profiles gives
\begin{equation*}
 \begin{split}
 \mathcal O=\sum_i\bigl\{
 &-v_i(\bar\mu_{b,i}\bar\partial_i+\bar h_i\bar\partial\bar\mu_{b,i})
 +\bar v_i(\mu_{a,i}\partial_i+h_i\partial\mu_{a,i})\\
 &+(h_i-\bar h_i)\mu_{a,i}\bar\mu_{b,i}\bigr\}A_X.
 \end{split}
\end{equation*}
Combining with Eq.~\eqref{eq:mixedSmearedCommutator} leaves $\sum_iJ_i\mu_{a,i}\bar\mu_{b,i}A_X$.

Before the local frame correction, the negative-first response is
\begin{equation*}
 \mathscr D_v\overline{\mathscr D}_{\bar v}A_X
 -\sum_i\{v_i\bar\mu_{b,i}\bar\partial_i
             +\bar h_i\bar\partial(\bar\mu_bv)_i\}A_X.
\end{equation*}
Its multiplicative source-overlap coefficient is $-\bar h_i$, whereas direct differentiation of the finite frame fixes it to $-\Delta_i/2$. The required difference is $-J_i/2$; the conjugate order gives $+J_i/2$. Thus
\begin{equation}
 \begin{gathered}
\kappa_{\rm src}=-\sum_iJ_i\mu_{a,i}\bar\mu_{b,i}A_X,\\
 \begin{aligned}
 C^{\rm src}_{\bar\mu\mu}(x,y;X)
 \supset{}&-\pi^2\sum_iJ_i\delta^{(2)}(x-z_i)\\
 &\times\delta^{(2)}(y-z_i)A_X,
 \end{aligned}
 \end{gathered}
 \label{eq:transportedContactCancellation}
\end{equation}
with this representative fixed by the selected frame. It is the difference of ordered frame contributions, not the antisymmetric Hessian of an added scalar functional. The sum $\kappa_{\rm op}+\kappa_{\rm map}+\kappa_{\rm src}$ vanishes, and the symmetric response is Eq.~\eqref{eq:explicitMixedTransportHessian}. The Maurer--Cartan relation checks this completed calculation; it does not supply another copy of the already included hard commutator.

\subsubsection{Additional source-profile contact tests}
\label{subsec:additional-contact-tests}
The following local tests use the four-hard channel and separated source normalization of Section~\ref{subsec:two-order-benchmarks}.

Keep the antiholomorphic disc at $a$ and choose the holomorphic profile to equal one on the hard support patch of leg~1, zero near the other legs. Since $\bar v(\bar z_1)=-2/5-4\ii/5$, the hard commutator is $-\bar v(\bar z_1)\partial_{z_1}\mathcal F_4$ on this patch. A linear normal test profile gives
\begin{equation}
 \frac{\langle[\mathscr D_v,\overline{\mathscr D}_{\bar v}]
          \mathcal F_4,n^0\rangle_{\rm density}}{\rho_*E^3}
 =-\frac{25}{34}-\frac{25}{17}\ii.
 \label{eq:mixedPunctureNormalBenchmark}
\end{equation}
The puncture part of $j_+-j_-$ has the opposite value. The source supports are disjoint, so the local map variation and spin-frame overlap vanish; the operator ordering difference already cancels on this test.

To probe the soft diagonal, take equal overlapping profiles $\mu_a=\bar\mu_b=f_R(|z-a|^2)$ away from all hard directions, where
\begin{equation*}
 \begin{gathered}
g(u)=e^{-1/(1-u)}\mathbf1_{0\leq u<1},\\
 I_0=\int_0^1\dd u\,g(u),\qquad
 f_R(t)=\frac{g(t/R^2)}{R^2I_0}.
 \end{gathered}
\end{equation*}
Their area integral is $\pi$. The coefficient of the first normal derivative is $a_-^0=q_t^0S_-^{(0)}(t)$; its holomorphic quadratic Taylor coefficient at $a$ is $c_{20}^-=(14\ii/25)E$, with $c_{02}^+=-(14\ii/25)E$. Radial angular integration selects these coefficients, leaving
\begin{equation*}
 \mathcal R_g=
 \frac{4\int_0^1\dd u\,g^2-\int_0^1\dd u\,u^2(g')^2}{I_0^2}
 \simeq5.27430.
\end{equation*}
Hence
\begin{equation}
 \begin{gathered}
\begin{aligned}
 \frac{\kappa_{\rm op}[n^0]_{\rm density}}{\rho_*E^3}
 &=-\frac{175\ii}{68R^2}\mathcal R_g\\
 &\simeq-\frac{13.5736\,\ii}{R^2},
 \end{aligned}\\
 \kappa_{\rm map}[n^0]=-\kappa_{\rm op}[n^0].
 \end{gathered}
 \label{eq:mixedDiagonalNormalBenchmark}
\end{equation}
The radial reduction follows from
\begin{equation*}
 \begin{aligned}
&\int\dd^2w\,f(t)\\
 &\quad\times[a_-\partial^2f-(\partial a_-)\partial f+(\partial^2a_-)f]\\
 &\quad=\pi c_{20}^-\left[4\int\dd t\,f^2-\int\dd t\,t^2(f')^2\right].
 \end{aligned}
\end{equation*}
Finally let both profiles be one near leg~1 and zero near the other legs. A constant normal test profile detects the residual $J_1\mathcal F_4$ before the frame term, with
\begin{equation}
 \frac{J_1\mathcal R_4}{E^2}=\frac{625}{68},\qquad
 \frac{\kappa_{\rm src}}{\mathcal F_4}=-J_1.
 \label{eq:mixedSpinFrameBenchmark}
\end{equation}
The frame contributes $-625/68$. A test linear in the normal coordinate annihilates this term proportional to the undifferentiated delta. This gives an independent check of the contact at the hard intersection.

\subsubsection{Global domains, endpoints and mixed poles}

The standard kernels above fix the reference Ward sector. More generally, after subtracting the same principal parts, a regular difference $\mathcal R_{\rm reg}^{\rm sh}$ is a section of $\bar K_x^{\otimes2}\boxtimes K_y^{\otimes2}$. If it extends smoothly across the removed loci and satisfies
\begin{equation}
 \partial_x\mathcal R_{\rm reg}^{\rm sh}=0,\qquad
 \bar\partial_y\mathcal R_{\rm reg}^{\rm sh}=0,
 \label{eq:standardShadowHomogeneousWard}
\end{equation}
then it vanishes as a global antiholomorphic quadratic differential on the sphere after compact hard pairing. Applying this result requires the stated smooth extension and a separate contact calculation.

Relative primitives carry open connectors, closed periods and weak
cut currents when their base points or loading change. The global
formulas in Supplementary Section~S2.12
evaluate these transitions on the same physical mixed-pole coefficient.
In particular, the connector in Supplementary
Eq.~\textup{(S2.265)} acts identically on the
physical and transformed Ward currents. A quadratic Ward polynomial
has a vanishing cubic descendant within a cut chamber, while its weak
boundary current remains the one calculated in
Appendix~\ref{app:BPboundary}. General two-leg coefficients retain their
full transition data.

On a common unpinched continuation let $\mathcal G^{\BP}$ be the summed family with the same helicity-exchange prescription. Its mixed-pole ordering contribution is
\begin{equation}
 \mathcal R_{\rm pinch}
 =\left[\Res_{\Delta_+=0}\Res_{\Delta_-=0}
       -\Res_{\Delta_-=0}\Res_{\Delta_+=0}\right]
       \mathcal G^{\BP}\big|_{\rm mixed\ poles}.
 \label{eq:BPPinchRemainder}
\end{equation}
Coordinate poles commute; the physical contact has a mixed divisor that can give a nonzero jump. Metric matching therefore allows more general polar sets.

Suppose on a common distribution space
\begin{equation*}
 \mathcal A(\alpha,\beta)
 =\frac{F_0}{\alpha(\alpha+\beta)}
  +\frac{F_1}{\beta(\alpha+\beta)}
  +\frac{H(\alpha,\beta)}{\alpha+\beta}
  +\mathcal A_{\rm coord},
\end{equation*}
with total polar degree at most two and $H$ holomorphic. If $T(\alpha,\beta)$ is a holomorphic continuous operator family on that space, then
\begin{equation}
 (\Res_\beta\Res_\alpha-\Res_\alpha\Res_\beta)
 T(\alpha,\beta)\mathcal A(\alpha,\beta)=T(0,0)(F_0-F_1).
 \label{eq:mixedPoleContinuousMapLemma}
\end{equation}
Nonconstant Taylor terms of $T$ raise total degree above minus two and give no double residue. Higher weight poles can select derivatives of $T$. This argument requires holomorphy after the physical singular pullback.

For the selected prescription, the joint calculation in
Section~\ref{sec:core-constructed-matching} keeps the moving denominator,
endpoint and puncture intersections, and all relative transition currents
before taking either residue. Both orders use the same scalar finite
part and source variations. These evaluated currents determine the two
individual defects, including their symmetric part.

\subsection{Relative boundary currents and their shape variation}
\label{app:BPboundary}

\subsubsection{The obstruction to an ordinary global inverse}

In a projective coordinate neighborhood the standard shadow obeys
\begin{equation*}
 \mathscr S_+f(x)=\frac3\pi\int\dd^2u\,
                 \frac{f(u)}{(\bar x-\bar u)^4},\qquad
 \partial_x\mathscr S_+f=-\frac12\bar\partial_x^3f.
\end{equation*}
If an ordinary-distribution inverse satisfied $\partial Gf=f$ across a hard puncture, commuting weak derivatives would imply
$\partial(\mathscr S_++\frac12\bar\partial^3G)f=0$. The physical Ward tensor instead satisfies
\begin{equation}
 \begin{aligned}
\partial_x\bar D_{\bar x}A_X
 ={}&\pi\sum_i\Bigl[-\bar h_i\bar\partial_x\delta^{(2)}(x-z_i)\\
 &\hspace{8mm}+\delta^{(2)}(x-z_i)\bar\partial_i\Bigr]A_X.
 \end{aligned}
 \label{eq:BPNonzeroWardDivergence}
\end{equation}
These properties cannot all hold on one ordinary global distribution space. If $\mathcal B_\partial=1-\partial G$, the corrected identity is
\begin{equation*}
 \partial(\mathscr S_++\tfrac12\bar\partial^3G)f
 =-\frac12\bar\partial^3\mathcal B_\partial f.
\end{equation*}
A finite sum of puncture-supported delta derivatives cannot alone supply this defect: applying $\bar\partial^3$ raises its highest derivative order by three, while Eq.~\eqref{eq:BPNonzeroWardDivergence} has order at most one. Relative boundaries, nonlocal domain data or a relative meaning of the descendant are required. This is an obstruction to an ordinary-distribution globalization, not to the BP construction, whose inverse has a relative domain \cite{Banerjee:2022wht}.

\subsubsection{Cut current on a physical packet}

Take $z=x+\ii y$, $a\in\mathbb R$ and the principal logarithm. Its jump from below to above the ray $x<a$ is $2\pi\ii$. With $C_a=\Theta(a-x)\delta(y)$,
\begin{equation}
 \bar\partial\operatorname{Log}(z-a)=-\pi C_a,\qquad
 \partial\operatorname{Log}(z-a)=\frac1{z-a}+\pi C_a.
 \label{eq:BPWeakLogDerivatives}
\end{equation}
For $\epsilon_a=P(\bar z)\operatorname{Log}(z-a)$ with $\deg P\leq2$, Leibniz differentiation yields Eq.~\eqref{eq:BPExplicitCutCurrent}. It includes endpoint currents when derivatives act on $\Theta(a-x)$. A finite lower endpoint adds a constant logarithm times $P$, whose third antiholomorphic derivative is zero. Away from the endpoint the canonical transverse form is
\begin{equation}
 \begin{aligned}
 R_{C_a}[P]&:=\frac12\bar\partial^3\epsilon_a\\
 &=\frac\pi8P(x)\delta''(y)
 -\frac{\ii\pi}2P'(x)\delta'(y)\\
 &\quad-\frac{7\pi}8P''(x)\delta(y).
 \end{aligned}
 \label{eq:BPCanonicalCutCurrent}
\end{equation}
For example,
$P(x-\ii y)\delta''=P(x)\delta''+2\ii P'(x)\delta'-P''(x)\delta$
and $P'(x-\ii y)\delta'=P'(x)\delta'+\ii P''(x)\delta$; these product identities give the coefficient seven.

At hard leg~1 of the real channel, $a=z_1=1/2$, $\epsilon_1\omega_1=-4E/5$ and $\mathcal R_4=-625E^2/136$. The coefficient of the highest normal derivative in the positive soft residue is $E N_1(\bar s)/(s-a)$, where
\begin{equation*}
 \begin{aligned}
N_1(b)={}&-\frac{\epsilon_1\omega_1}{E}
 \Bigl[z_1(b-\bar z_1)^2\\
 &\quad+(1+z_1\bar z_1)(b-\bar z_1)\Bigr]\\
 ={}&\frac25(b-a)^2+(b-a).
 \end{aligned}
\end{equation*}
Pairing the normal derivative of the momentum delta with $n^0$ supplies a minus sign. In units of $\rho_*E^3$ the primitive's coefficient is therefore
\begin{equation}
 \begin{gathered}
\begin{aligned}
 P_1(b)&=-\frac{\mathcal R_4}{E^2}N_1(b)\\
 &=\frac{125}{68}b^2+\frac{375}{136}b-\frac{125}{68},
 \end{aligned}\\
 P_1''=\frac{125}{34}.
 \end{gathered}
 \label{eq:BPPhysicalWardPolynomial}
\end{equation}
Take the other puncture cuts horizontally to the left. A small rectangle about zero then intersects only the cut from leg~1 and contains no hard puncture. Let $\bar\mu(x,y)=\eta(x)\psi_b(y)$ with $\eta\in C_c^\infty(-1/8,1/8)$, $\int\dd x\,\eta=1$, $b=1/8$, and
\begin{equation*}
 \psi_b(y)=\begin{cases}
 (1+y^2/b^2)\exp\bigl(1-(1-y^2/b^2)^{-1}\bigr),&|y|<b,\\
 0,&|y|\geq b.
 \end{cases}
\end{equation*}
Its derivatives at the cut obey $\psi_b(0)=1$, $\psi_b'(0)=\psi_b''(0)=0$. Thus
\begin{equation}
 \begin{gathered}
\frac{\langle R_{C_a},\bar\mu\rangle_{n^0,\rm density}}{\rho_*E^3}
 =-\frac{875\pi}{272},\\
 -\frac1\pi\frac{\langle R_{C_a},\bar\mu\rangle_{n^0,\rm density}}{\rho_*E^3}
 =\frac{875}{272}.
 \end{gathered}
 \label{eq:BPPhysicalCutBenchmark}
\end{equation}
This density detects the ordinary-derivative extension away from hard punctures, where puncture-local counterterms pair to zero. Continuity gives a nonzero response after integration against a sufficiently localized compact tangential packet.

\subsubsection{The generic relative Ward-pole residue}

The normalization in Eq.~\eqref{eq:BPFractionalRelativeMap} follows from
\begin{equation*}
 \begin{gathered}
\Gamma(4-\Delta)\Gamma(\Delta-3)=-\frac\pi{\sin\pi\Delta},\\
 r(\Delta)=-e^{-\ii\pi\Delta}\sin\pi\Delta.
 \end{gathered}
\end{equation*}
The gamma-normalized kernel is holomorphic in its order on a fixed properly supported Volterra distribution domain; the physical amplitude's singular pullback requires separate control.

For the isolated Ward pole
$\mathcal A_{\rm pole}(\Delta;u,\bar u)=\Delta^{-1}P(\bar u)/(u-a)$,
choose finite base points $u_0=\bar u_0=1$, $a=1/2$, and fixed detours with continuous branches. With $L=x-u_0$, $c=u_0-a$,
\begin{equation}
 I_x^{\Delta+1}\frac1{u-a}
 =\frac{L^{\Delta+1}}{c\Gamma(\Delta+2)}
 {}_2F_1\left(1,1;\Delta+2;-\frac Lc\right),
 \label{eq:BPExactWardHolomorphicFactor}
\end{equation}
\begin{equation}
 I_{\bar x}^{\Delta-3}P
 =\sum_{k=0}^{2}
 \frac{P^{(k)}(\bar u_0)(\bar x-\bar u_0)^{\Delta+k-3}}
      {\Gamma(\Delta+k-2)}.
 \label{eq:BPExactWardAntiholomorphicFactor}
\end{equation}
All reciprocal gamma factors vanish linearly at zero, whereas the holomorphic factor tends to a based logarithm with locally integrable boundary values. Uniformly on compact source sets avoiding the base point, including neighborhoods of the logarithmic hard singularity,
\begin{equation}
 \Res_{\Delta=0}\mathscr Q_+^{\rm rel}(\Delta)
       \mathcal A_{\rm pole}(\Delta)=0.
 \label{eq:BPZeroWardPoleRelativeResidue}
\end{equation}
Near the base point the expansion contains
$(x-u_0)^{\Delta+m+1}(\bar x-\bar u_0)^{\Delta+k-3}$, $m\geq0$, $0\leq k\leq2$.
For conjugate local branches and a test monomial $z^p\bar z^q$, a pole at zero would require both
$m+k+p+q=0$ and $m-k+4+p-q=0$.
The vanishing sum forces every index to vanish, which is incompatible with the angular selection rule. The local base-point model therefore contributes no additional pole at the soft weight.

The difference between Eqs.~\eqref{eq:BPPhysicalCutBenchmark} and \eqref{eq:BPZeroWardPoleRelativeResidue} is the current in Eq.~\eqref{eq:BPExplicitCutCurrent}: the generic map differentiates independent complex coordinates before the boundary value, whereas ordinary weak differentiation also acts on the jump. The local endpoint residue remains valid on an unpinched complex tube.

\subsubsection{Boundary subtraction and moving contours}

For a piecewise smooth primitive, let $j_k$ be the jump of its regular $k$th antiholomorphic derivative. The current from a single weak derivative is $\mathcal N_Cj=(\ii/2)jC_a$ on the horizontal convention. Iteration gives
\begin{equation*}
 \bar\partial^3\epsilon_C
 =(\bar\partial^3\epsilon_C)_{\rm reg}
 +\mathcal N_Cj_2+\bar\partial(\mathcal N_Cj_1)
                    +\bar\partial^2(\mathcal N_Cj_0).
\end{equation*}
This proves Eq.~\eqref{eq:BPBoundaryCompletedDescendant}. For $P\operatorname{Log}$ one has $j_k=2\pi\ii P^{(k)}$, so $E_C=R_C$ and the completed representative equals the standard Ward tensor. If a deformation sweeps a region $\Omega$, then, with the corresponding orientation,
\begin{equation*}
 \epsilon_{C'}-\epsilon_C=2\pi\ii P(\bar z)\mathbf1_\Omega,
 \qquad
 E_{C'}-E_C=\pi\ii\bar\partial^3[P\mathbf1_\Omega].
\end{equation*}
The changes cancel. This proves cut independence on the quadratic Ward module. For coefficients with additional antiholomorphic poles the regular descendant need not vanish, and its transition data remain part of the definition.

In a rectangle away from endpoints, move the cut normally to $y=\varepsilon$, tapering the deformation to zero at its fixed endpoints. Differentiating the shifted form of Eq.~\eqref{eq:BPCanonicalCutCurrent} gives
\begin{equation}
 \dot R_C=-\frac\pi8P(x)\delta'''(y)
       +\frac{3\ii\pi}8P'(x)\delta''(y)
       +\frac{3\pi}8P''(x)\delta'(y).
 \label{eq:BPCutShapeDerivative}
\end{equation}
The odd source $\bar\mu_{\rm odd}=\eta(x)y\psi_b(y)$ has transverse derivative equal to one at the cut, while its value and derivatives of orders two and three vanish. Consequently
\begin{equation}
 \begin{gathered}
\frac{\langle\dot R_C,\bar\mu_{\rm odd}\rangle_{n^0,\rm density}}{\rho_*E^3}
 =-\frac{375\pi}{272},\\
 -\frac1\pi\frac{\langle\dot R_C,\bar\mu_{\rm odd}\rangle_{n^0,\rm density}}
 {\rho_*E^3}=\frac{375}{272}.
 \end{gathered}
 \label{eq:BPPhysicalShapeBenchmark}
\end{equation}
The subtraction in Eq.~\eqref{eq:BPBoundaryCompletedDescendant} has the opposite shape response. The full mixed Beltrami variation also differentiates the input, output frame and descendant.

The same conversion matters in the mixed soft-leg expression
$\partial_u[a_+(u,\bar u)/(y-u)]$. Applying $G\partial=1$ as an ordinary global identity would give
$\frac12\bar\partial_x^3[a_+(x,\bar x)/(y-x)-a_+(u_0,\bar x)/(y-u_0)]=-J_+(x,y)$,
including the weak diagonal contact. For the isolated Ward-pole model, the generic relative kernel instead has zero residue across $x=y$ away from the base point and hard intersections: Eq.~\eqref{eq:BPExactWardAntiholomorphicFactor} supplies a factor of $\Delta$, while integrating meromorphic poles of order at most two gives a locally integrable Cauchy pole or logarithm. The difference is the contact $J_+$. Its conjugate gives $J_-$. One must therefore use the same boundary operation for the generic kernels, soft-leg terms and map derivatives. This calculation fixes the boundary operation for the Ward-pole subfamily. The full physical matching defects in Eq.~\eqref{eq:physicalMixedResidueDefects} additionally use the joint estimates of Section~\ref{sec:core-constructed-matching}.

\subsection{Weight derivatives of the generic BP maps}
\label{app:nmhvWeightJets}

For the normalization in Eq.~\eqref{eq:BPGenericIntertwiners}, let
$c_1=2\gamma_E-11/6$. Logarithmic differentiation gives
\begin{equation}
 K(\Delta)=6\left[1+c_1\Delta
 +\frac12\left(c_1^2-\frac{49}{36}\right)\Delta^2
 +O(\Delta^3)\right].
 \label{eq:nmhvBPWeightNormalization}
\end{equation}
A fixed-branch kernel contributes
$L=\log(x-u)+\log(\bar x-\bar u)$ and, for physical soft energy,
$-\log(1+u\bar u)$. Multiplying by $e^{\Delta L}$ replaces $c_1$ by
$c_1+L$ in the linear and quadratic Taylor coefficients. Frame and measure factors
must be included before applying Eq.~\eqref{eq:nmhvPolarTerm}.

Using Eqs.~\eqref{eq:BPBasedFractionalIntegral} and \eqref{eq:BPFractionalRelativeMap} on the selected unpinched branches, where the integral converges,
\begin{equation*}
 \begin{aligned}
\partial_\lambda I_C^\lambda f(x)
 ={}&\frac1{\Gamma(\lambda)}\int_{x_0}^{x}\dd u\, (x-u)^{\lambda-1}\\
 &\times[\log(x-u)-\psi(\lambda)]f(u).
 \end{aligned}
\end{equation*}
At $\lambda=1$ the bracket is $\log(x-u)+\gamma_E$.

For the negative order write $L=x-x_0$, $u=x-\xi$ and
\begin{equation*}
 R_3(x,\xi)=f(x-\xi)-f(x)+\xi f'(x)
 -\frac{\xi^2}{2}f''(x)+\frac{\xi^3}{6}f'''(x).
\end{equation*}
Taylor subtraction before continuation gives
\begin{equation*}
 \begin{aligned}
 I^{-3+\eta}f={}&\frac1{\Gamma(-3+\eta)}\Biggl[\\
 &\quad\sum_{k=0}^{3}\frac{(-1)^kf^{(k)}(x)}{k!}
 \frac{L^{\eta+k-3}}{\eta+k-3}\\
 &\quad+\int_0^L\dd\xi\,\xi^{\eta-4}R_3(x,\xi)\Biggr].
 \end{aligned}
\end{equation*}
Using $1/\Gamma(-3+\eta)=-6\eta[1-\psi(4)\eta+O(\eta^2)]$ yields
$I^{-3}f=f'''$ and
\begin{equation}
 \begin{aligned}
&\left.\partial_\lambda I^\lambda f(x)\right|_{\lambda=-3} =[\log L-\psi(4)]f'''(x)\\
 &\qquad+\frac{2f(x)}{L^3}-\frac{3f'(x)}{L^2}+\frac{3f''(x)}L\\
 &\qquad-6\int_0^L\frac{R_3(x,\xi)}{\xi^4}\dd\xi.
 \end{aligned}
 \label{eq:nmhvNegativeFractionalDerivative}
\end{equation}
The integral converges for $C^4$ data on an unpinched contour and retains the base point; it is distinct from a global weak derivative across a cut. In particular $I^{-3}1=0$, whereas its order derivative is $2/L^3$.

The linear weight derivative of the positive relative map is therefore
\begin{equation}
 (\mathscr Q_+^{\mathrm{rel}})'(0)
 =\frac12\left[-\ii\pi I_x^1I_{\bar x}^{-3}
 +(\partial_\lambda I_x^\lambda)_{1}I_{\bar x}^{-3}
 +I_x^1(\partial_\lambda I_{\bar x}^\lambda)_{-3}\right].
 \label{eq:nmhvRelativeFirstJet}
\end{equation}
Higher weight derivatives follow from the same Taylor subtraction and
\begin{equation*}
 \begin{aligned}
\frac1{\Gamma(-m+\eta)}
 ={}&(-1)^mm!\,\eta\exp\Biggl[
 (\gamma_E-H_m)\eta\\
 &+\sum_{n\ge2}\frac{(-1)^{n+1}\zeta(n)-H_m^{(n)}}{n}\eta^n\Biggr],
 \end{aligned}
\end{equation*}
where $H_m^{(n)}=\sum_{j=1}^{m}j^{-n}$ and $H_m=H_m^{(1)}$.
Only finitely many such derivatives are needed after the joint polar
degree has been established.

Metric-source variation must be applied to these full maps. In a coefficient
pullback frame a conjugated positive map obeys
\begin{equation}
 \begin{aligned}
\delta_v\mathscr Q_+(\alpha)f
 ={}&\left(v\partial-\frac\alpha2\partial v\right)
    \mathscr Q_+(\alpha)f\\
 &-\mathscr Q_+(\alpha)
  \left[v\partial+\left(1+\frac\alpha2\right)\partial v\right]f.
 \end{aligned}
 \label{eq:nmhvGenericMapVariation}
\end{equation}
This conjugation includes the transported endpoints and contours. A moving denominator obeys
$\delta(D^{-p})=-p(\delta D)D^{-p-1}$; any compensating numerator zero
must be counted. The differentiated integrand and its relative boundary
conversion terms therefore need the same resolved estimates as the
unvaried family, with their own Taylor orders.

\section{Vacuum normalization and constrained response}
\label{app:vacuumresponse}

\subsection{Independent vacuum metric data}
\label{sec:weyl}

\subsubsection{Marked amplitudes and the vacuum functional}
\label{subsec:kappadef}
\label{subsec:datamap}
\label{subsec:hierarchy}

The Euler coefficient is defined by the covariant Weyl response
\begin{equation}
 \delta_\sigma W[g,0]
 =\frac{\cW}{24\pi}
 \int_\Sigma\dd^2x\sqrt g\,\sigma R
 +\delta_\sigma C_{\loc}[g].
 \label{eq:anomalydef}
\end{equation}

Appendix~\ref{sec:weylbeltrami} relates this convention to marked trace contacts and the metric central terms.

We use the standard relative Wess--Zumino functional \cite{Deser:1993yx,MazurMottola:2001,Osborn:1991gm}
\begin{equation}
 \begin{gathered}
 \begin{aligned}
 \Gamma_{\WZ}[\phi;\hg]
 ={}&\frac{1}{24\pi}\int_\Sigma\dd^2x\hsqrtg\\
 &\times\left[(\widehat\nabla\phi)^2+\hR\phi\right],
 \end{aligned}\\
 \Gamma_{\WZ}[0;\hg]=0.
 \end{gathered}
 \label{eq:wzfunctional}
\end{equation}

For $g=e^{2\phi}\hg$, expand $\phi=\sum_A'\phi_A Y_A$ in orthonormal eigenfunctions of the positive Laplacian $\hD$, with $\hD Y_A=\lambda_A Y_A$ and the constant mode omitted. Then $\partial_A\partial_B\Gamma_{\WZ}|_0=\frac{\lambda_A}{12\pi}\delta_{AB}$. Local counterterms change the Weyl coboundary while preserving the Euler coefficient.

Every marked response contains a derivative with respect to $J$. Hence the family
\begin{equation}
 W_\alpha[g,J]
 =W_0[g,J]+\alpha\bigl(\Gamma_{\Poly}[g]-\Gamma_{\Poly}[\hg]\bigr).
 \label{eq:alphacompletion}
\end{equation}

has the same marked coefficients and all their metric derivatives, while $\cW[W_\alpha]=\cW[W_0]+\alpha$. Here $\Gamma_{\Poly}$ denotes the Euler-normalized functional whose global relative definition is given in Eq.~\eqref{eq:polyakovnonlocal}. More generally, adding $P[g]-P[\hg]$ leaves every marked response unchanged. The finite reconstruction in Section~\ref{sec:marked-finite-functional-reconstruction} leaves this vacuum freedom independent of the hard sources unconstrained.

The ambiguity can be measured after a vacuum completion is specified. For the area-normalized scalar determinant, a unit-normalized spherical harmonic of eigenvalue $\lambda_\ell=\ell(\ell+1)$ has fixed-area Hessian $\cW(\lambda_\ell-2)/(12\pi)$. Local curvature terms add $(\lambda_\ell-2)^2$ times a spectral polynomial. Appendix~\ref{app:determinant} derives the determinant normalization and local curvature expansion used in the finite-mode estimator, Eq.~\eqref{eq:countertermSpectralExtraction}, and its truncation bound, Eq.~\eqref{eq:spectralRemainderBound}. Appendix~\ref{sec:induced} derives the corresponding fixed-area inverse, including the multiplier contribution and residual conformal modes. The extraction requires a derivative-order bound, as the finite-data ambiguity below demonstrates.

\subsubsection{Local curvature terms and spectral extraction}
\label{subsec:countertermanomaly}
\label{subsec:localcountertermproof}

For $C_a[g]=a\int\sqrt g\,R_g^2$, the coordinate and fixed-area Hessians are
\begin{equation}
 \begin{aligned}
 \left.\frac{\dd^2 C_a[e^{2\epsilon Y}\gamma]}{\dd\epsilon^2}\right|_0
 &=8a(\lambda^2-4\lambda+2),\\
 \left.\frac{\dd^2 C_a[e^{2\phi_\epsilon}\gamma]}{\dd\epsilon^2}\right|_0
 &=8a(\lambda-2)^2.
 \end{aligned}
 \label{eq:curvatureCountertermHessians}
\end{equation}

The common term $-32a\lambda$ illustrates why a single spherical Hessian coefficient does not isolate the Euler cocycle \cite{Deser:1993yx,MazurMottola:2001}. More generally, define $h_\ell^{\rm A}=\left.12\pi\frac{\dd^2W[e^{2\phi_\epsilon}\gamma]}{\dd\epsilon^2}\right|_0$ for $W=\cW\Gamma_{\rm rel}+C_{\loc}$ and $\ell\geq2$. The local curvature expansion in Appendix~\ref{app:determinant} gives
\begin{equation}
 \begin{gathered}
\begin{aligned}
 h_\ell^{\rm A}
 ={}&\cW(\lambda_\ell-2)\\
 &+(\lambda_\ell-2)^2p_N(\lambda_\ell),
 \end{aligned}\\
 \lambda_\ell=\ell(\ell+1),\qquad\deg p_N\leq N.
 \end{gathered}
 \label{eq:countertermNormalizedSpectrum}
\end{equation}

The degree bound follows from the retained derivative order. For distinct sampled eigenvalues $\Lambda_j=\ell_j(\ell_j+1)$, polynomial interpolation of $\frac{h^{\rm A}(\lambda)}{\lambda-2}$ at $\lambda=2$ yields
\begin{equation}
 \cW
 =\sum_{j=1}^{N+2}
 \frac{h_{\ell_j}^{\rm A}}{\Lambda_j-2}
 \prod_{k\ne j}\frac{2-\Lambda_k}{\Lambda_j-\Lambda_k}.
 \label{eq:countertermSpectralExtraction}
\end{equation}

For the curvature-squared sector this becomes
\begin{equation}
 \begin{gathered}
\cW=\frac{5}{12}h_2^{\rm A}-\frac{1}{15}h_3^{\rm A},\\
 a=\frac{1}{576\pi}
 \left(\frac{h_3^{\rm A}}{10}-\frac{h_2^{\rm A}}{4}\right)\\
 \text{for }C_{\loc}=C_a.
 \end{gathered}
 \label{eq:countertermTwoModeExtraction}
\end{equation}

For $\cW=1$ and $a=\frac1{96\pi}$, the responses $(h_2^{\rm A},h_3^{\rm A})=(20,110)$ recover both coefficients. Through $R\Delta_gR$, the corresponding combination is
\begin{equation}
 \cW=\frac{15}{28}h_2^{\rm A}
 -\frac{3}{20}h_3^{\rm A}
 +\frac{5}{252}h_4^{\rm A}.
 \label{eq:countertermThreeModeExtraction}
\end{equation}

The choice $p_1(\lambda)=1+\frac\lambda{10}$ and $\cW=1$ gives $(h_2^{\rm A},h_3^{\rm A},h_4^{\rm A})=(\frac{148}{5},230,990)$ and again recovers the Euler coefficient.

For an omitted term $b_{N+1}\lambda^{N+1}$ in $p$, the interpolation shift is fixed by its leading coefficient:
\begin{equation}
 \widehat c_{\rm W}^{(N)}-\cW
 =-b_{N+1}\prod_{j=1}^{N+2}(2-\Lambda_j).
 \label{eq:spectralTruncationBias}
\end{equation}

The two-mode and three-mode shifts are $-40b_1$ and $720b_2$. More generally, write $q(\lambda)=\frac{h^{\rm A}(\lambda)}{\lambda-2}=\cW+(\lambda-2)p_N(\lambda)+r(\lambda)$, with $r(2)=0$, $r\in C^{N+2}[2,\Lambda_{\max}]$ and $\Lambda_{\max}=\max_j\Lambda_j$. The interpolation remainder gives
\begin{equation}
 \left|\widehat c_{\rm W}^{(N)}-\cW\right|
 \leq\frac{\displaystyle\sup_{2\leq\lambda\leq\Lambda_{\max}}
 |r^{(N+2)}(\lambda)|}{(N+2)!}
 \prod_{j=1}^{N+2}(\Lambda_j-2).
 \label{eq:spectralRemainderBound}
\end{equation}

Without a derivative-order bound, any finite set of $M$ responses is compatible with a change $\delta c$ in the Euler coefficient: choose
\begin{equation}
 \delta p(\lambda)
 =-\delta c\sum_{j=1}^{M}
 \frac{1}{\Lambda_j-2}
 \prod_{k\ne j}\frac{\lambda-\Lambda_k}{\Lambda_j-\Lambda_k}.
 \label{eq:countertermFiniteDataAmbiguity}
\end{equation}

Then $\delta c(\Lambda_j-2)+(\Lambda_j-2)^2\delta p(\Lambda_j)=0$ on every sampled mode, and the polynomial is generated by local curvature terms. Thus Eq.~\eqref{eq:countertermSpectralExtraction} measures $\cW$ within the specified derivative expansion; Eq.~\eqref{eq:anomalydef} defines it on general metrics.

\subsection{Trace and central-term normalization}
\label{sec:weylbeltrami}

\subsubsection{Marked trace response}

In the selected anomaly-free marked transport, $\delta_\sigma A_X=-M_\sigma A_X$, with
\begin{align}
 M_\sigma&=\sum_i\Delta_i\sigma(z_i,\bar z_i),
 \qquad \Delta_i=h_i+\bar h_i,
 \label{eq:weylwardoperator}\\
 \langle\Theta(x)X\rangle_{\rm marked}
 &=-\sum_i\Delta_i\delta_g^{(2)}(x,z_i)A_X.
 \label{eq:localtracecontact}
\end{align}
The commutator $[D_w,M_\sigma]A_X=\sum_i\Delta_i(w-z_i)^{-1}\partial_i\sigma(z_i,\bar z_i)A_X$ gives derivative contacts at the hard insertions. Including the variations of the covariant delta function, source frame and metric parametrization yields $\delta_\phi T^{\met}=-\pi\delta_\mu(\sqrt g\,\Theta^{\met})$ and its conjugate. The operators $M_\sigma$ commute, and a Wess--Zumino-consistent anomaly contributes a symmetric metric Hessian. A local curvature term changes this Hessian through
\begin{equation}
 \delta_\sigma\int\dd^2x\sqrt g\,R^2
 =\int\dd^2x\sqrt g\,\sigma
 \left(-2R^2+4\Delta_gR\right),
 \label{eq:r2variation}
\end{equation}
while preserving the Euler anomaly class.

\subsubsection{Metric central terms}
\label{sec:central}

When the trace and stress responses derive from a common locally covariant functional, Polyakov normalization gives
\begin{equation}
 \cW=\frac{c_L+c_R}{2},
 \qquad \Kphi=\frac{\cW}{12\pi}
 \quad (W=\log Z).
 \label{eq:kappacentralrelation}
\end{equation}
Here $c_L,c_R$ multiply the fourth-order stress OPE poles and $\Kphi$ is the anomaly-sector kinetic coefficient~\cite{Nguyen:2021BoundaryActions}; covariant local coboundaries preserve the Euler class~\cite{MazurMottola:2001}. The gravitational anomaly is proportional to $c_L-c_R$, so a parity-symmetric metric sector has $c_L=c_R=\cW$. The convention $\Gamma=-W$ reverses the kinetic sign. Vacuum division sets $W_{\trn}[g,0]=0$ and its vacuum metric coefficients to zero; a microscopic stress algebra additionally requires its metric coupling. Equation~\eqref{eq:kappacentralrelation} applies order by order when the common functional exists, with quantum tests using the loop-corrected soft and stress Ward data~\cite{He:2017fsb,Pasterski:2022djr,Donnay:2020lur}.

\subsection{Determinant normalization and local curvature response}
\label{app:determinant}

The determinant and its local variations use the Polyakov--Alvarez relation and heat-kernel expansion \cite{Polyakov:1981rd,Alvarez:1983,Vassilevich:2003xt,Dowker:1994Polyakov}.

For $\nu$ real scalar fields on a closed surface, with the constant scalar mode removed and common determinant units, define
\begin{equation}
 W_{\det}[g]
 =-\frac{\nu}{2}\log\Det{}'\!\left(\frac{\Delta_g}{\mu^2}\right),
 \qquad \Delta_g=-\nabla_g^2.
 \label{eq:scalardeterminant}
\end{equation}
The standard Polyakov--Alvarez relation, including the zero-mode normalization \cite{Dowker:1994Polyakov}, is
\begin{equation}
 \begin{aligned}
 &W_{\det}[e^{2\phi}\hg]-W_{\det}[\hg]
 =\nu\Gamma_{\WZ}[\phi;\hg]\\
 &\quad-\frac{\nu}{2}\log\frac{A_{e^{2\phi}\hg}}{A_{\hg}}
 +W_{\rm norm}^{\loc}[\phi;\hg],
 \end{aligned}
 \label{eq:polyakovalvarez}
\end{equation}
where $\Gamma_{\WZ}$ is defined in Eq.~\eqref{eq:wzfunctional} and $W_{\rm norm}^{\loc}$ contains local covariant measure terms. The area factor is a global normalization of the scalar zero mode. Multiplication of the scalar partition function by $A_g^{\nu/2}$ removes this factor and gives the unit relative functional used here,
\begin{equation}
 \begin{gathered}
\Gamma_{\Poly}[g]-\Gamma_{\Poly}[\hg]
 :=\Gamma_{\rm rel}[g;\hg]\\
 =-\frac12\log\!\left[
 \frac{A_{\hg}\Det{}'\Delta_g}{A_g\Det{}'\Delta_{\hg}}
 \right],\\
 \Gamma_{\rm rel}[e^{2\phi}\hg;\hg]=\Gamma_{\WZ}[\phi;\hg].
 \end{gathered}
 \label{eq:polyakovnonlocal}
\end{equation}
The reference-dependent ratio has the additive Weyl cocycle and extends over a smooth sphere-metric family by diffeomorphism covariance. Its scalar field content fixes $\Delta\cW=\Delta c_L=\Delta c_R=\nu$. In this paper we use it as a specified vacuum completion; a gravitational ensemble fixes its own boundary operator and measure.

The zero-mode distinction matters when metric derivatives are taken before imposing the area constraint. On the nonconstant orthonormal harmonics defined in Section~\ref{subsec:kappadef}, Eq.~\eqref{eq:polyakovalvarez} gives
\begin{equation}
 \left.\frac{\partial^2W_{\det}}
 {\partial\phi_A\partial\phi_B}\right|_0
 =\frac{\nu}{12\pi}\lambda_A\delta_{AB}
 -\frac{2\nu}{A_{\hg}}\delta_{AB}
 +H_{AB}^{\loc},
 \qquad A,B\ne0.
 \label{eq:determinantmodehessian}
\end{equation}
The term proportional to $A_{\hg}^{-1}$ comes from the area logarithm. Although this logarithm is constant on an area-preserving path, the constrained constant Weyl mode also contributes through the linear Euler variation. This produces the factor $\lambda-2$ in Eq.~\eqref{eq:fixedareadeterminanttest}. The local Hessian $H_{AB}^{\loc}$ can contain both linear and higher powers of $\lambda_A$; its spectral coefficients are separated from the Euler coefficient by the curvature expansion below.

For comparison, the inverse Laplacian with its constant mode projected out defines
\begin{equation}
 \begin{gathered}
\begin{aligned}
 F_\perp[g]
 ={}&\frac{1}{96\pi}\int_\Sigma\dd^2x\sqrt g\,\\
 &\times(R_g-\overline R_g)\Delta_{g,\perp}^{-1}(R_g-\overline R_g),
 \end{aligned}\\
 \overline R_g=\frac{1}{A_g}\int_\Sigma\dd^2x\sqrt g\,R_g.
 \end{gathered}
 \label{eq:curvaturebilinearcomparison}
\end{equation}
Its fixed-area response contains the rational spectral factor calculated in Eq.~\eqref{eq:fixedareabilineartest}. Thus the reference-dependent difference $\Gamma_{\rm rel}[g;\hg]-F_\perp[g]+F_\perp[\hg]$ carries shape dependence in addition to constant-mode normalization. Eq.~\eqref{eq:polyakovnonlocal} supplies the completed functional entering the anomaly relations.

\subsubsection{Quadratic local curvature terms}

We now derive the polynomial form needed by the harmonic extraction. On the unit round sphere, the background curvature is $R=2$ and all of its covariant derivatives vanish. In two dimensions the Riemann tensor is determined by $R$. At fixed area, both the constant term in a local curvature expansion and the term linear in $R$ integrate to constants. Consequently, $\int\sqrt g\,f(R)$ has constrained Hessian $4f''(2)(\lambda-2)^2$ along a normalized harmonic. On the unconstrained coordinate path $\phi=\epsilon Y$, the additional area variation gives $4[f(2)-2f'(2)]$. Setting $f(R)=aR^2$ recovers Eq.~\eqref{eq:curvatureCountertermHessians}.

For a finite sum of covariant curvature contractions with bounded derivative order, integration by parts reduces the quadratic term on the path in Eq.~\eqref{eq:fixedareaspherepath} to
\begin{equation}
 \begin{aligned}
&C_{\loc}[g_\epsilon]-C_{\loc}[\gamma] =\int_{S^2}\dd^2x\sqrt\gamma\,
 (R_{g_\epsilon}-2)\,\\
 &\qquad\times\mathcal P_N(\Delta_\gamma)(R_{g_\epsilon}-2)
 +O(\epsilon^3),\\
 &\deg\mathcal P_N\leq N.
 \end{aligned}
 \label{eq:localCurvatureQuadraticPolynomial}
\end{equation}
Rotational covariance makes the scalar differential operator a polynomial in $\Delta_\gamma$; commuting derivatives changes its coefficients through the constant background curvature. Since $R_{g_\epsilon}-2=2\epsilon(\lambda-2)Y+O(\epsilon^2)$, the Hessian of Eq.~\eqref{eq:localCurvatureQuadraticPolynomial} is $8(\lambda-2)^2\mathcal P_N(\lambda)$. Hence $p_N=96\pi\mathcal P_N$ in Eq.~\eqref{eq:countertermNormalizedSpectrum}. The terms $a_r\int\sqrt g\,R\Delta_g^rR$ supply $8a_r\lambda^r(\lambda-2)^2$ for $r\geq0$, while nonlinear powers of $R$ change the coefficients of the same polynomial. This reduction establishes the derivative-order input used in Eq.~\eqref{eq:countertermSpectralExtraction}.

\subsubsection{Fixed-area spherical response}
\label{subsec:fixedAreaResponse}

Let $\gamma$ be the unit round sphere and let $Y$ be a real harmonic satisfying $\Delta_\gamma Y=\lambda Y$, $\int\sqrt\gamma\,Y=0$ and $\int\sqrt\gamma\,Y^2=1$. The area-preserving path is
\begin{equation}
 \begin{gathered}
\begin{aligned}
 \phi_\epsilon={}&\epsilon Y-\frac12\log\!\Biggl[\frac{1}{4\pi}\\
 &\qquad\times\int\dd^2x\sqrt\gamma\,e^{2\epsilon Y}\Biggr]\\
 ={}&\epsilon Y-\frac{\epsilon^2}{4\pi}+O(\epsilon^3),
 \end{aligned}\\
 A_{e^{2\phi_\epsilon}\gamma}=4\pi.
 \end{gathered}
 \label{eq:fixedareaspherepath}
\end{equation}

Its quadratic constant term contributes through the linear Euler variation. Substitution into Eq.~\eqref{eq:wzfunctional} gives
\begin{equation}
 \begin{gathered}
\begin{aligned}
 &\Gamma_{\rm rel}[e^{2\phi_\epsilon}\gamma;\gamma] =\frac{\epsilon^2}{24\pi}(\lambda-2)+O(\epsilon^3),
 \end{aligned}\\
 \left.\frac{\dd^2W_{\det}}{\dd\epsilon^2}\right|_0
 =\frac{\nu}{12\pi}(\lambda-2)\\
 \text{at fixed area},
 \end{gathered}
 \label{eq:fixedareadeterminanttest}
\end{equation}

Here $W_{\det}$ is the independently specified ensemble of $\nu$ real scalars in Eq.~\eqref{eq:scalardeterminant}; local terms are held fixed. The $\ell=1$ kernel is the residual conformal orbit. For comparison, the projected curvature bilinear of Eq.~\eqref{eq:curvaturebilinearcomparison} gives
\begin{equation}
 \begin{gathered}
F_\perp[e^{2\phi_\epsilon}\gamma]
 =\frac{\epsilon^2}{24\pi}\frac{(\lambda-2)^2}{\lambda}+O(\epsilon^3),\\
 \Gamma_{\rm rel}-F_\perp
 =\frac{\epsilon^2}{12\pi}\frac{\lambda-2}{\lambda}+O(\epsilon^3).
 \end{gathered}
 \label{eq:fixedareabilineartest}
\end{equation}

At $\ell=2,3$, its coefficients are $\frac83,\frac{25}3$, whereas the relative determinant gives $4,10$. The difference measures the shape-dependent zero-mode and reference contribution \cite{Dowker:1994Polyakov}.

\subsection{Stationary soft fields and the fixed-area inverse}
\label{sec:induced}

For a saddle $S_C[g,C_*,0]=0$, the classical functional $\Gamma_{\rm cl}[g]=S[g,C_*(g),0]$ has Hessian
\begin{equation}
 \Gamma_{{\rm cl},gg}=S_{gg}-S_{gC}S_{CC}^{-1}S_{Cg}
 \label{eq:softSchurComplement}
\end{equation}
on an invertible complement to the soft zero modes. If $S[g,0,0]=S_C[g,0,0]=0$ throughout the metric family, the saddle $C_*=0$ has zero classical metric Hessian. A curvature source contributes through $S_{Cg}$; for a projected scalar Laplacian, its source $R-\overline R$ produces the bilinear in Eq.~\eqref{eq:curvaturebilinearcomparison}. The connected vacuum functional receives the Gaussian contribution $-\tfrac12\log\Det{}'P_g$, with its zero-mode normalization and local measure terms.

The supertranslation memory kernel is fourth order~\cite{Nguyen:2021dpa}; on the unit sphere, $P_\gamma=\Delta_\gamma(\Delta_\gamma-2)$ acts on $\ell\geq2$, excluding the translation modes $\ell=0,1$. Its determinant additionally requires a covariant continuation and zero-mode measure. On a differentiable positive elliptic family with constant-rank projected domain, $\delta\log\Det{}'P_g=\FP_{s=0}\Tr{}'(P_g^{-s-1}\delta P_g)$; the reference memory spectrum alone leaves these vacuum data unspecified.

For the Euler-normalized completion with area potential $\mu_L A_g$, use
\begin{equation}
 \mathscr F_{\rm A}[\phi;\Lambda_{\rm A}]
 =\cW\Gamma_{\rm rel}[e^{2\phi}\gamma;\gamma]
 +\mu_L A_{e^{2\phi}\gamma}
 -\Lambda_{\rm A}\bigl(A_{e^{2\phi}\gamma}-4\pi\bigr).
 \label{eq:fixedAreaVariationalFunctional}
\end{equation}
Stationarity at $\phi=0$ fixes $\Lambda_{\rm A}=\mu_L+\cW/(24\pi)$. The full Hessian, including the multiplier term, is
\begin{equation}
 \begin{gathered}
\cD_{\rm A}=\Delta_\gamma-2,\\
 H_{\rm A}=\frac{\cW}{12\pi}\cD_{\rm A}\\
 \text{on the area tangent space}.
 \end{gathered}
 \label{eq:fixedAreaLiouvilleOperator}
\end{equation}
The area constraint removes $\ell=0$; fixing the conformal frame removes the $\ell=1$ kernel. With $\mathsf P_{\rm A}$ projecting onto $\ell\geq2$, the Green operator satisfies
\begin{equation}
 \bigl(\mathsf P_{\rm A}\cD_{\rm A}\mathsf P_{\rm A}\bigr)_x
 G_{\rm A}(x,y)=\mathsf P_{\rm A}(x,y),
 \qquad G_{\rm A}=\mathsf P_{\rm A}G_{\rm A}\mathsf P_{\rm A}.
 \label{eq:fixedAreaLiouvilleGreen}
\end{equation}
Its eigenvalues are $(\lambda_\ell-2)^{-1}$, giving susceptibility $12\pi/[\cW(\lambda_\ell-2)]$ for $\cW\ne0$. The area potential drops out on the constraint surface. The shift $-2$ follows from the multiplier fixed by the Euler variation and would be lost by projecting only after unconstrained differentiation. This gives the variational interpretation of Eq.~\eqref{eq:fixedareadeterminanttest}; $\Gamma=-W$ has the opposite Hessian sign.

\bibliographystyle{JHEP}
\bibliography{reference}

\end{document}